\pdfoutput=1
\documentclass[12pt]{article}
\usepackage{cite}
\usepackage{dsfont}
\usepackage{graphicx}
\usepackage{latexsym}
\usepackage{mathrsfs}
\usepackage[overload]{textcase}

\font\grande=cmr9.5 scaled \magstep4
\font\medio=cmr9.5 scaled \magstep2
\outer\def\beginsection#1\par{\medbreak\bigskip
      \message{#1}\leftline{\bf#1}\nobreak\medskip
\vskip-\parskip
      \noindent}
\vfuzz=\maxdimen

\begin{document}
\bibliographystyle{unsrt}
\titlepage
\vspace{1cm}
\begin{center}
{\grande Ultralight scalars and invisible gravitons}
\vspace{1.5 cm}\\
Massimo Giovannini\footnote{e-mail address: massimo.giovannini@cern.ch}\\
\vspace{1cm}
{{\sl Department of Physics, CERN, 1211 Geneva 23, Switzerland }}\\
\vspace{0.5cm}
{{\sl INFN, section of Milan-Bicocca, 20126 Milan, Italy}}
\vspace*{1cm}
\end{center}
\vskip 0.3cm
\centerline{\medio  Abstract}
\vskip 0.5cm
\noindent
Unless the tensor to scalar ratio is small enough, the non-adiabatic mode associated with the ultralight bosons may be dangerously comparable with the adiabatic spectrum of curvature inhomogeneities. We point out here that this statement depends on the timeline of the post-inflationary expansion rate. Provided the decelerated evolution is slower than radiation for a sufficiently long period, the relative weight of the non-adiabatic and adiabatic components does not conflict with the direct constraints coming from the large-scale observations. Conversely the presence of ultralight bosons excludes the presence of long stages expanding faster than radiation. The synthesis of the light nuclei and the spectrum of the relic gravitons at high frequencies set the relevant constraints on the duration of the decelerated stages, on the delayed dominance of radiation and on the mass range of the ultralight scalars.
\vspace{5mm}
\vfill
\newpage

\renewcommand{\theequation}{1.\arabic{equation}}
\setcounter{equation}{0}
\section{Introduction and motivations}
\label{sec1}
The adiabatic paradigm of structure formation (see, for instance, \cite{ONEa,ONEb}) stipulates that the large-scale inhomogeneities observed by ground-based and satellite experiments are Gaussian prior to matter-radiation equality while the entropic (or non-adiabatic) fluctuations must always be subdominant \cite{TWOa,TWOb}. Both physical ingredients are concurrently present when a stage of decelerated expansion is complemented by an early inflationary evolution driven by a single inflaton field \cite{ONEa,ONEb} and this (minimal) version of the adiabatic paradigm has been repeatedly tested in the last thirty years. Indeed, the first uncontroversial evidences of its soundness came from the cross-correlations between the temperature and the polarization anisotropies of the Cosmic Microwave background (CMB) accurately measured for the first time by the WMAP experiment \cite{THREEa,THREEb,THREEc,THREEd,THREEe}. The current form of the adiabatic paradigm does not only constrain the entropic (or non-adiabatic) inhomogeneities of the primeval plasma, but also the tensor modes of the geometry. In fact, as a consequence of its inflationary origin, the large-scale spectrum of curvature fluctuations  is complemented by a diffuse background of relic gravitons \cite{FOURa,FOURb,FOURc,FOURd,FOURe,FOURf,FOURg,FOURh,FOURi,FOURl}. In the framework of the adiabatic paradigm the constraints on the tensor component are expressed in terms of $r_{T}$ denoting the ratio between the scalar and tensor power spectra  at a conventional pivot wavenumber $k_{p} = 0.002 \, \mathrm{Mpc}^{-1}$. The recent determinations either suggest $r_{T} < {\mathcal O}(0.06)$ or, more stringently, $r_{T} < {\mathcal O}(0.03)$ \cite{FIVEa,FIVEb,FIVEc,FIVEd,FIVEe}. 

Although the value of $r_{T}$ is customarily related to the profile of the inflaton potential, 
a physical reduction of the tensor to scalar ratio does also depend upon the timeline of the decelerated evolution: the total number of $e$-folds is comparatively larger when the Universe expand at a rate that is slower than radiation and the increased number of $e$-folds (elapsed since the crossing of the CMB scales) ultimately suppresses $r_{T}$ \cite{SIXa}. If the increment in the number of $e$-folds is sufficiently large the value of $r_{T}$ becomes small enough to make the tensor modes invisible by the large-scale observations. By contrast, when the post-inflationary expansion rate is faster than radiation the number of $e$-folds gets reduced and the value of $r_{T}$ becomes comparatively larger: this logic has been used to justify an increase of $r_{T}$ for monomial potentials \cite{SIXb,SIXc} since the Bicep-2 data release (eventually
contaminated by foreground emissions) suggested a value of the tensor to scalar ratio larger than the one suggested by monomial potentials in conjunction with a post-inflationary timeline dominated by radiation \cite{SIXd}. In spite of the undeniable successes, it is fair to say that the current paradigm does not univocally determine the timeline of the post-inflationary expansion rate that is ordinarily assumed to be dominated by radiation between the end of inflation and the equality time.

A modified post-inflationary evolution (possibly leading to a larger number of $e$-folds) is tenable if the large-scale inhomogeneities are adiabatic but it is not  clear what may happen when non-adiabatic fluctuations participate in the dynamics of the primeval plasma. In this investigation, for the sake of concreteness, the attention is focused on the non-adiabatic modes induced by the massive pseudoscalar particles \cite{SEVENa,SEVENb,SEVENc,SEVENd,SEVENe}. It is actually well known that the coherent oscillations around the minimum of the potential of a cosmic 
axion background lead to various effects \cite{SEVENe} and when the pseudoscalar field becomes dominant it can act as a dark matter background. A particularly interesting mass range for the axion implies that its mass $m = {\mathcal O}(10^{-22}) \, \, \mathrm{eV}$ \cite{SEVENf}. Since the de Broglie wavelength becomes astrophysically relevant for $m < {\mathcal O}(10^{-20})\,\, \mathrm{eV}$ the small-scale structures can be suppressed if this ultralight scalar is the dominant (or predominant) component of the dark matter. If the field is extremely light, it has not started oscillating by recombination and behaves more like dark energy than matter, shifting the expansion history and gravitational potentials, requiring the field to behave as matter by recombination gives roughly $m > {\mathcal O}(10^{-27})\,\, \mathrm{eV}$ \cite{FIVEa,FIVEb,FIVEc,FIVEd,FIVEe}, the limits coming 
the Lyman-$\alpha$ would imply $m \geq {\mathcal O}(10^{-22})\,\, \mathrm{eV}$ \cite{SEVENe}. 

Unless the tensor to scalar ratio is sufficiently small, the spectrum of the non-adiabatic modes induced by the ultralight bosons may be comparable with the adiabatic solution \cite{SEVENf,SEVENg}. As we are going to see, also this statement depends on the post-inflationary history of the plasma. In particular, provided the expansion rate is slower than radiation after inflation (but prior to the synthesis of the light nuclei), the relative weights of the non-adiabatic and adiabatic inhomogeneities is naturally suppressed by so avoiding the direct conflicts with the constraints coming from the observations of the temperature and polarization anisotropies of the CMB. Conversely, when the post-inflationary evolution is dominated by radiation (or if it expands faster than radiation) the adiabatic and the non-adiabatic modes at matter-radiation equality are comparable within one order of magnitude. This relative weight can only be suppressed if $r_{T}$ is itself smaller than the current bounds. If follows that an increase in the total number of $e$-folds can simultaneously suppress $r_{T}$ and the non-adiabatic component associated with the curvature inhomogeneities; this will eventually make the adiabatic solution dominant, as required by the tenets of the adiabatic paradigm illustrated above \cite{ONEa,ONEb}. 

The post-inflationary evolution is not only constrained by the non-adiabatic 
inhomogeneities and by the low-frequency data on the tensor modes (always related to $r_{T}$) but it is also affected by the tensor modes of the geometry at small scales [i.e. $k \gg k_{p}$ or $\nu \gg \nu_{p} = {\mathcal O}(\mathrm{aHz})$] since the high frequency gravitons set important 
constraints on the expansion rate, on the duration of the post-inflationary 
stages and, indirectly, on the total number of $e$-folds. The sensitivities of operating detectors \cite{LIGO0,LIGO0a,LIGO0b,LIGO1,LIGO1a,LIGO1b} are notoriously insufficient to measure the diffuse backgrounds of relic gravitons but in the future new detectors might cover different frequencies \cite{LIGO3} even beyond the so-called audio band ranging between few Hz and $10$ kHz.  The gravitational waves produced thanks to the variation of the space-time curvature should then become an object of future empirical investigations even at high frequencies while 
at intermediate frequencies (in the nHz range) the backgrounds of relic gravitons could be observed by the pulsar timing arrays \cite{NANO1,NANO2,PPTA1,PPTA2,PPTA2a} that are now primarily focussed  on the diffuse astrophysical signals. We actually know that every variation of the expansion rate produces shots of gravitons with given averaged multiplicities and specific statistical properties. If these spectra will ever be detected the timeline of the expansion rate might be directly probed without the need of postulating a particular post-inflationary paradigm before the time of big bang nucleosynthesis whose striking successes are a signature of radiation dominance for typical temperatures smaller than ${\mathcal O}(10)$ MeV. It turns out that, in the present context, the most relevant frequency range involves the region between the MHz and the THz; it can be actually argued that the largest frequency of cosmic gravitons must be smaller than ${\mathcal O}(\mathrm{THz})$ \cite{EIGHT1}. 
 
The layout of this investigation is, in short, the following. In sec. \ref{sec2} the governing equations  are introduced with particular attention to the adiabatic and non-adiabatic inhomogeneities. In sec. \ref{sec3} the different post-inflationary histories are illustrated in specific physical situations with the purpose of demonstrating that the total number of $e$-folds (elapsed since the exit of the CMB scales) depends upon the successive expanding stages after inflation.
Section \ref{sec4} discusses the relative weights of the adiabatic and of the non-adiabatic 
modes when the post-inflationary evolution includes different phases that expand either faster or slower than radiation.  After briefly addressing the constraints from the tensor modes at high frequencies, in sec. \ref{sec5} all the concurrent limits are analyzed in the perspective of ultralight scalars. The concluding remarks are collected in sec. \ref{sec6}. To clarify the notations and to elucidate some derivation, some useful complementary considerations have been collected in the appendices \ref{APPA} and \ref{APPB}.

\renewcommand{\theequation}{2.\arabic{equation}}
\setcounter{equation}{0}
\section{The evolution of the system}
\label{sec2}
The total action of the problem consists 
of the gravitational contribution $S_{g}$ supplemented by the terms accounting for the dynamics 
of the inflaton (indicated by $S_{\varphi}$) and of the ultralight scalar (denoted by $S_{\sigma}$):  
\begin{equation}
S = S_{g} + S_{\varphi} + S_{\sigma} + S_{m}.
\label{EVS1}
\end{equation}
In Eq. (\ref{EVS1}) $S_{m}$ stands for the action of the matter sources 
that can be eventually present in the late time Universe. At early times (i.e. during the accelerated stage of expansion) the dynamics is determined, in practice, by the sum of $S_{g}$ and $S_{\varphi}$:
\begin{equation}
S_{g} + S_{\varphi} = - \frac{1}{2 \ell_{P}^2} \int \,d^{4} x\, \sqrt{-g}\,\,R \, + \int d^{4} x \, \sqrt{-g} \,\,\biggl[ \frac{1}{2} g^{\alpha\beta} \partial_{\alpha} \varphi 
\partial_{\beta} \varphi - V(\varphi) \biggr],
\label{EVS2}
\end{equation}
while the contribution to $\sigma$ (subleading both during inflation and during the post-inflationary evolution) may eventually dominate around the decoupling time 
if the dark matter interpretation is enforced \cite{SEVENf,SEVENg}.
In Eq. (\ref{EVS2}) $\ell_{P} = \sqrt{8 \pi G}$, $R$ indicates the Ricci scalar and $V(\varphi)$ is the inflaton potential whose specific properties will be specified later on\footnote{Furthermore, in what follows, the signature of  $g_{\mu\nu}$ (with determinant $g$) is mostly minus $(+, \, -,\, -,\,-)$ and we shall often trade the Planck length $\ell_{P}$ for the reduced Planck mass 
$\overline{M}_{P} = M_{P}/\sqrt{8\,\pi}$ according to $\ell_{P} = 1/\overline{M}_{P}$. }. With the same notations the  action for the ultralight scalar can be written as  
\begin{equation}
S_{\overline{\sigma}} = \int \, d^{4} x \, \, \sqrt{-g} \biggl[ \frac{M^2}{2} \, g^{\alpha\beta} \partial_{\alpha} \overline{\sigma} 
\partial_{\beta} \overline{\sigma} - \overline{m}^4 \, \overline{U}(\overline{\sigma})\biggr],
\label{EVS3}
\end{equation}
where $M$ is the Peccei-Quinn scale \cite{SEVENa,SEVENb,SEVENc,SEVENd,SEVENe} whose range of variation (always smaller than $\overline{M}_{P}$) falls in the window $10^{15} \,\, \mathrm{GeV} < M < 10^{18} \,\,\mathrm{GeV}$. In Eq. (\ref{EVS3}) $\overline{U}(\overline{\sigma})$ is a dimensionless function  
of $\overline{\sigma}$. When $\overline{U}(\overline{\sigma}) = \overline{U}(\overline{\sigma}+ 2 \pi)$ the action is invariant 
for $\overline{\sigma} \to \overline{\sigma} + 2 \pi$.  If we choose $  \overline{U}(\overline{\sigma}) = (1 - \cos{\overline{\sigma}})$ 
we have that $\overline{U}(\overline{\sigma}) = \overline{\sigma}^2/2$ for $|\overline{\sigma}| <1$. We can finally redefine 
the field (i.e. $\sigma = M\,  \overline{\sigma}$) and the mass term (i.e. $m = \overline{m}^2/M$) and the canonical form of the action 
becomes:
\begin{equation}
S_{\sigma} =  \int d^{4} x \, \sqrt{-g} \,\,\biggl[ \frac{1}{2} g^{\alpha\beta} \partial_{\alpha} \sigma 
\partial_{\beta} \sigma - U(\sigma) \biggr],
\label{EVS4}
\end{equation}
where $U(\sigma) = m^2 \sigma^2/2$. The typical range of the mass term will be $10^{-27} \, \mathrm{eV} < m < 10^{-20} \mathrm{eV}$. Some of the considerations discussed later on do not depend on the 
particular shape of the potential and this is why $U(\sigma)$ is kept generic in some of the forthcoming discussions. Finally the matter part of the action ($S_{m}$ in Eq. (\ref{EVS1})) takes the form of a perfect relativistic fluid with associated energy-momentum tensor $T_{\mu}^{\,\,\nu} = (p + \rho) u_{\mu}\, u^{\nu} - p \delta_{\mu}^{\,\,\mu}$ where $p$ and $\rho$ denote, respectively, the pressure and the energy density.

\subsection{Evolution of the homogeneous components}
The components of the system described by the 
action (\ref{EVS1}) are not simultaneously dominant and their relative importance 
defines the relevant physical regimes. Following the tenets of the concordance scenario 
we consider the evolution of the homogeneous components of the various fields in a conformally flat background metric $\overline{g}_{\mu\nu} = a^2(\tau) \eta_{\mu\nu}$ where $\tau$ indicates the conformal time coordinate, $a(\tau)$ is the scale factor and $\eta_{\mu\nu}$ is the Minkowski metric. This choice is actually motivated by the current data that are compatible with a spatially 
flat background geometry but they are also consistent with the early inflationary stage 
where the intrinsic (spatial) curvature is exponentially suppressed in comparision 
with the extrinsic (Hubble) curvature \cite{ONEa,ONEb}. The evolution equations of the background geometry can then be written in terms of the total energy density and pressure (i.e. $\rho_{t}$ and $p_{t}$) whose explicit expression is given, respectively, by:
\begin{eqnarray}
\rho_{t} = \rho_{\varphi} + \rho_{\sigma} + \rho, \qquad p_{t} = p_{\varphi} + p_{\sigma} + p.
\label{BB1}
\end{eqnarray}
During the inflationary stage $\rho_{t} = \rho_{\varphi}$ and $p_{t} = p_{\varphi}$ since 
$(\rho_{\sigma} + \rho) \ll \rho_{\varphi}$ and $(p_{\sigma} + p) \ll p_{\varphi}$.
In the post-inflationary stage the coherent oscillations of $\varphi$ are still dominant, at least 
for some time before the radiation stage is reached at a typical scale $H_{r}$.
It is quite plausible that radiation must prevail already prior to the onset of big-bang nucleosyhthesis 
where the expansion rate is severely constrained by the relative abundances of the light nuclear elements \cite{ONEa,ONEb}. This means that, in the model-independent perspective adopted here $H_{r} \geq {\mathcal O}(10^{-44})\, \overline{M}_{P}$ i.e. $H_{r} \geq {\mathcal O}(10^{-16}) \, \mathrm{eV}$. 
By definition 
$\rho_{\varphi}$ and $p_{\varphi}$ are 
\begin{equation}
\rho_{\varphi} = \frac{\varphi^{\prime\, 2}}{2\, a^2} + V(\varphi), \qquad p_{\varphi} = \frac{\varphi^{\prime\, 2}}{2\, a^2} - V(\varphi),
\label{BB2}
\end{equation}
where the prime indicates throughout a derivation with respect to the cosmic time coordinate. With the very same notations the energy density and the pressure of the light scalar are 
$\rho_{\sigma} = \sigma^{\prime\, 2}/(2\,a^2) + U(\sigma)$ and $p_{\sigma} = \sigma^{\prime\, 2}/(2\, a^2) - U(\sigma)$. Although the dominant component of the system varies, the evolution equations of the geometry can always be written in terms 
of the global variables $\rho_{t}$ and $p_{t}$ of Eq. (\ref{BB1})
\begin{equation}
3\, {\mathcal H}^2 = a^2\, \ell_{P}^2\,\, \rho_{t},\qquad 6\,{\mathcal H}^{\prime} = - a^2\, \ell_{P}^2\,\,(\rho_{t} + 3 p_{t}),
\label{BB3}
\end{equation}
where ${\mathcal H}= (\ln{a})'$ is the comoving Hubble rate while the physical Hubble rate is $ H= {\mathcal H}/a$. The equations collected in Eq. (\ref{BB3}) notoriously imply the covariant conservation of the total energy-momentum tensor of the sources, i.e.  $\rho_{t}' + 3 {\mathcal H} (\rho_{t} + p_{t}) =0$. 
The energy-momentum tensors of  the inflaton, of 
the light scalar and of the fluid sources are all covariantly conserved; so, for instance the covariant 
conservation of the light scalar implies 
$\rho_{\sigma}^{\prime} + 3 {\mathcal H} ( \rho_{\sigma} + p_{\sigma})=0$ which is equivalent 
to the usual Klein-Gordon equation written in the case of a conformally flat background geometry:
\begin{equation}
\sigma'' + 2 {\mathcal H} \sigma' +  a^2\,U_{,\,\sigma} =0,
\label{BB5}
\end{equation}
where $U_{,\,\sigma} = \partial_{\sigma} U$ indicates the derivative of the potential with respect to $\sigma$. As we are going to see in specific situations, depending on the convenience it is possible either to formulate 
the evolution of the system in terms of $\rho_{t}$ and $p_{t}$ or 
to consider separately the various components. 

\subsection{Curvature inhomogeneities}
Eeven if it is often more practical to analyze the problem in a specific gauge,
for the present purposes we like to start with the gauge-invariant evolution of curvature inhomogeneities 
defined in terms of the Bardeen potentials $\Phi$ and $\Psi$ \cite{EIGHT2,EIGHT3,EIGHT4}; these variables are swiftly introduced in appendix \ref{APPA} (see in particular Eq. (\ref{APA4})) together with the related technical considerations. The evolution of $\Phi$ and $\Psi$ depends upon the gauge-invariant fluctuation of the total energy density and pressure.
In particular from the $(00)$ perturbed component of the Einstein equations we obtain 
the Hamiltonian constraint
\begin{equation}
 \nabla^2 \Psi - 3 {\mathcal H} ( \Psi^{\prime} + {\mathcal H} \Phi) = \frac{\ell_{P}^2}{2}\,a^2\, \delta^{(g)} \rho_{t},
 \label{PP4}
 \end{equation}
 where $\delta^{(g)} \rho_{t}$ indicates the (gauge-invariant) fluctuation of the total energy density 
 (see Eq. (\ref{APA4a}) and the related discussion in appendix \ref{APPA}). The gauge-invariant 
 pressure fluctuation $ \delta^{(g)} p_{t}$ enters instead the dynamical equation coming from the spatial components of the perturbed Einstein's equations:
\begin{equation}
 \Psi^{\prime\prime} + {\mathcal H} ( \Phi^{\prime} + 2 \Psi^{\prime}) + ( {\mathcal H}^2 + 2 {\mathcal H}^{\prime}) \Phi  + \frac{1}{3} \nabla^2 ( \Phi - \Psi) = \frac{\ell_{P}^2}{2} a^2 \delta^{(g)} p_{t}.
 \label{PP5}
 \end{equation}
 The difference between the two Bardeen potentials depends upon the anisotropic stress $\Pi_{t}$
 since $\nabla^2 ( \Phi - \Psi) = 3 \ell_{P}^2 \Pi_{t}/2$, where we used that $\partial_{i} \partial_{j} \Pi^{i\, j} = \nabla^2 \Pi_{t}$.  Moreover, when a certain system contains diverse physical components the total pressure fluctuation is not only given by the variation of the energy density; more specifically we have that  $\delta^{(g)} p_{t} = c_{s\,t}^2 \delta^{(g)} \rho_{t} + \delta p_{nad}$ where $\delta p_{nad}$ is the non-adiabatic pressure fluctuation. When the relevant wavelengths are larger than the Hubble radius prior to matter radiation equality the $\delta p_{nad}$ leads to the standard mixed initial data that are discussed in the literature; depending on how we count the non-adiabatic (or entropic) initial data involve four or five independent solutions of the Einstein-Boltzmann hierarchy at early times \cite{NINE1,NINE2,NINE3,NINE4,NINE5}. In what follows, however, we are not going to deal with the non-adiabatic initial conditions of the Einstein-Boltzmann hierarchy but we shall rather consider the way non-adiabatic fluctuations are produced at much earlier times during post-inflationary epoch. If we now multiply Eq. (\ref{PP4}) by $c_{s\,t}^2 $ and subtract the obtained equation from Eq. (\ref{PP5}) we obtain a complicated looking expression 
\begin{equation}
\Psi^{\prime\prime} + {\mathcal H}[ \Phi^{\prime} + ( 2 + 3 c_{s\,t}^2 ) \Psi^{\prime}] + 
[ {\mathcal H}^2 ( 1 + 3 c_{s\,t}^2) + 2 {\mathcal H}'] \Phi - c_{s\,t}^2 \nabla^2 \Psi + \frac{1}{3}\nabla^2( \Phi - \Psi) = \frac{\ell_{P}^2}{2} a^2 \delta p_{nad}.
\label{PP6}
\end{equation}
Equation (\ref{PP6}) can now be rewritten in terms of the curvature inhomogeneities on comoving orthogonal hypersurfaces (conventionally indicated by ${\mathcal R}$) that are traditionally 
 employed to set the initial conditions of CMB anisotropies prior to matter-radiation equality when the relevant wavelengths are still larger than the Hubble radius (see e.g. \cite{ONEa} and also \cite{THREEa,THREEb,THREEc}) ; 
\begin{equation}
{\mathcal R}^{\prime} = - \frac{{\mathcal H}}{p_{t} + \rho_{t}} \delta p_{nad} + 
\frac{ 2 {\mathcal H}}{3 \ell_{P}^2 a^2 (p_{t} + \rho_{t})} \nabla^2(\Phi - \Psi) - \frac{2 c_{s\, t}^2 {\mathcal H}}{\ell_{P}^2 (p_{t} + \rho_{t})} \nabla^2 \Psi.
\label{PP7}
\end{equation}
In terms of the Bardeen's potentials (see also appendix \ref{APPA}) the curvature inhomogeneities 
are given by:
\begin{equation}
{\mathcal R} = - \Psi - {\mathcal H}( {\mathcal H} \Phi + \Psi^{\prime})/({\mathcal H}^2 - {\mathcal H}^{\prime})
\label{PP7a}
\end{equation}
We can now recall that the anisotropic stress $\Pi^{i\,j}$ is related 
to the Laplacian of the difference between the two Bardeen's potentials as 
$ \nabla^2 ( \Phi - \Psi) = 3\, \ell_{P}^2 \,a^2\, \Pi_{t}/2$ where we use the notation 
$\nabla^2 \Pi_{t} = \partial_{i} \, \partial_{j} \, \Pi^{i\,j}$; on this basis 
Eq. (\ref{PP7}) becomes
\begin{equation}
{\mathcal R}^{\prime} = \Sigma_{{\mathcal R}}  - \frac{2 a^2 }{\ell_{P}^2 {\mathcal H} \, z_{t}^2 } \nabla^2\Psi, \qquad 
\Sigma_{{\mathcal R}} = \frac{{\mathcal H} \, \Pi_{t}}{p_{t} + \rho_{t}} - \frac{{\mathcal H} \, \delta p_{nad}}{p_{t} + \rho_{t}},
\label{PP8}
\end{equation}
where we introduced the background function $z_{t} = a^2 \sqrt{p_{t} + \rho_{t}}/[ c_{s\,t}^2 \, {\mathcal H}]$. At this point we take the derivative of both sides of Eq. 
(\ref{PP7}) and obtain 
\begin{equation}
\bigl( {\mathcal R}^{\prime} - \Sigma_{{\mathcal R}}\bigr)^{\prime} = \biggl(2 {\mathcal H} - 2 \frac{z_{t}^{\prime}}{z_{t}} - \frac{{\mathcal H}^{\prime}}{{\mathcal H}}\biggr) \bigl( {\mathcal R}^{\prime} - \Sigma_{{\mathcal R}}\bigr) - \frac{2 a^2 }{\ell_{P}^2 \, z_{t}^2 \, {\mathcal H}} \nabla^2 \Psi^{\prime}.
\label{PP9}
\end{equation}
To eliminate $\nabla^2 \Psi^{\prime}$ we use  Eq. (\ref{PP7a}) and trade $\Phi$ for the anisotropic stress $\Pi_{t}$; at the end of this lengthy but straightforward procedure we obtain the following equation 
\begin{equation}
\bigl( {\mathcal R}^{\prime} - \Sigma_{{\mathcal R}}\bigr)^{\prime}  + 2 \frac{z_{t}^{\prime}}{z_{t}} \bigl( {\mathcal R}^{\prime} - \Sigma_{{\mathcal R}}\bigr) - c_{s\,t}^2 \nabla^2 {\mathcal R} = \frac{3 a^4}{z_{t}^2} \Pi_{t}.
\label{PP10}
\end{equation}
In the large-scale limit the anisotropic stress is negligible and the Laplacians are also 
subleading; therefore the effective evolution of the non-adiabatic mode reduces 
to ${\mathcal R}^{\prime} - \Sigma_{{\mathcal R}} =0$ \cite{NINE1,NINE2,NINE3,NINE4,NINE5}. Finally, when $\Sigma_{{\mathcal R}} \to 0$ and 
$\Pi_{t} \to 0$ we recover the evolution of the adiabatic mode, in particular during inflation.

\subsection{The early inflationary evolution}
It is well known \cite{ONEa,ONEb}  that during inflation  $V(\varphi)$ prevails over all the other components entering $\rho_{t}$ and $p_{t}, $ i.e. $ \rho_{t} \equiv \rho_{\varphi}$ and $\rho_{t} + p_{t} \equiv \rho_{\varphi} + p_{\varphi} =  \varphi^{\prime\, 2}/a^2 \ll V(\varphi)$.
Because of the dominance of $V(\varphi)$ the evolution equations of the homogeneous components (see Eq. (\ref{BB3})) assume a simpler form where the second derivatives of the scale factor and of the inflaton are neglected 
\begin{equation}
3\,H^2 = \ell_{P}^2 V(\varphi), \qquad 3 H \, \dot{\varphi} + V_{\, \varphi} =0.
\label{INF2}
\end{equation}
As usual, in Eq. (\ref{INF2}) the overdot denotes a derivation with respect to the cosmic time coordinate\footnote{The cosmic and the conformal time coordinates  are  related as $ a(\tau) d\tau = d\, t$. This means that also the time derivatives in the two parametrizations have a simple algebraic connection; so for instance $\dot{\varphi}= \varphi^{\prime}/a$ where the prime 
indicates the derivation with respect to the conformal time coordinate $\tau$.}  $t$ and, as already stressed after Eq. (\ref{BB3}), $H = {\mathcal H}/a$; the same notations are used throughout 
the whole investigation. Equations (\ref{INF2}) notoriously imply that during the stage of accelerated expansion the so-called slow-roll parameters must always be smaller than $1$, i.e. 
\begin{equation}
\epsilon = - \dot{H}/H^2 = (\overline{M}_{P}^2/2)(V_{\, \varphi}/V)^2 < 1, \qquad 
\eta = \ddot{\varphi}/(H\, \dot{\varphi}) <1, \qquad \overline{\eta}= \overline{M}_{P}^2\,\,(V_{,\varphi\varphi}/V),
\label{INF3}
\end{equation}
where  $\overline{\eta}$ is actually a linear combination of $\epsilon$ and $\eta$ [i.e. $\overline{\eta} = (\epsilon - \eta$)]. The inflationary expansion is concluded when $\epsilon(t_{e}) \to 1$ where $t_{e}$ formally indicates the moment at which $\dot{H} \to - H^2$; furthermore, in the context of 
single-field scenarios considered here, the condition $\dot{H} = - H^2$ demands that $V(\varphi_{e}) = \dot{\varphi}_{e}^2$. When inflation ends all the three slow-roll parameters 
 become ${\mathcal O}(1)$ and the total duration of inflation is notoriously the same both in the 
 cosmic and in the conformal time parametrizations: 
 \begin{equation}
 N_{t}= \int_{t_{b}}^{t_{e}} H \, d t =  \int_{\tau_{b}}^{\tau_{e}} {\mathcal H} \, d \tau, 
 \label{INF4}
 \end{equation}
where $t_{b}$ and $\tau_{b}$ denote the beginning of the inflationary expansion. In connection with Eq. (\ref{INF4}) we remark that the 
total number of $e$-folds $N_{t}$ is unfortunately {\em not} a physically meaningful variable 
since the values of $t_{b}$ (or $\tau_{b}$) cannot be observationally assessed. 
This is why it is more relevant to deal with the number of $e$-folds elapsed since the wavenumber $k$ crossed the comoving expansion rate $a\, H$ (i.e. $k \simeq a_{k} H_{k}$)
\begin{equation}
 N_{k} =  \int_{\tau_{k}}^{\tau_{e}} {\mathcal H} \, d \tau, \qquad k \tau_{k} ={\mathcal O}(1),  
 \label{INF4a}
 \end{equation}
where $\tau_{k}$ indicates the crossing time of the $k$-mode \cite{ONEa} (see also \cite{NINE6}). Since the different $k$-modes do not cross the comoving expansion rate all at the same time, Eq. (\ref{INF4a}) stipulates that $N_{k}$ is not the same for different wavenumbers. 
The wavelengths that were still larger than the Hubble radius between matter-radiation equality and photon decoupling turn out to be the most relevant ones since their corresponding (comoving)
wavenumbers are directly probed (at the present time) by the current CMB experiments \cite{FIVEa,FIVEb,FIVEc,FIVEd,FIVEe}.  In short these wavenumbers  
will be referred to as the CMB scales: from a quantitative viewpoint the CMB scales are defined here as $k = {\mathcal O}(k_{p})$ where $k_{p} =0.002\, \mathrm{Mpc}^{-1}$ is the pivot scale at which the power spectra of 
curvature inhomogeneities are assigned (see, in this respect, also the introductory considerations 
of sec. \ref{sec1}).  During inflation the evolution of curvature inhomogeneities 
follows from Eq. (\ref{PP10}) in the limit $\delta p_{nad} \to 0$, $\Pi_{t} \to 0$ and $z_{t} \to z_{\varphi} 
= a \, \varphi^{\prime}/{\mathcal H}$. Furthermore, at the crossing time,
the power spectrum of curvature inhomogeneities must be evaluated  for $\tau \to \tau_{k}$
\begin{equation}
P_{{\mathcal R}}(k,\,\tau_{k}) = \frac{k^3}{2 \pi^2} \bigl| {\mathcal R}_{k}(\tau_{k})\bigr|^2 = \frac{k^{2}}{4 \pi^2 z^2_{\varphi}(\tau_{k})}, \qquad {\mathcal R}_{k}(\tau_{k}) = \frac{e^{i k \tau_{k}}}{\sqrt{2 k} \, z_{\varphi}(\tau_{k})}.
\label{INF5} 
\end{equation}
Depending upon the convenience, Eq. (\ref{INF5}) can be rephrased in various ways (either in terms of the inflationary 
potential or directly through the Hubble rate at $\tau_{k} = 1/k$); for the present ends a sensible parametrization is given by 
\begin{equation}
P_{{\mathcal R}}(k) = \frac{1}{\pi \, \epsilon_{k}} \biggl(\frac{H_{k}}{M_{P}}\biggr) \equiv {\mathcal A}_{{\mathcal R}}, \qquad k = {\mathcal O}(k_{p}),
\label{INF6}
\end{equation}
where we indicated, with shorthand notation\footnote{With the same shorthand notation we are going to introduce $\epsilon_{k} = \epsilon(\tau_{k})$, $\overline{\eta}_{k} = \overline{\eta}(\tau_{k})$, $\eta(\tau_{k}) = \eta_{k}$ and so on and so forth.}, $P_{{\mathcal R}}(k) = P_{{\mathcal R}}(k,\tau_{k})$.  Because Eq. (\ref{INF6}) 
holds for $k = {\mathcal O}(k_{p})$, ${\mathcal A}_{{\mathcal R}}$ estimates the amplitude of the power spectrum for $k \to k_{p}$. However, when $ k \neq {\mathcal O}(k_{p})$ 
the power spectrum of curvature inhomogeneities is customarily written, for 
phenomenological purposes \cite{FIVEa,FIVEb,FIVEc,FIVEd,FIVEe}, as $P_{{\mathcal R}}(k) = 
{\mathcal A}_{{\mathcal R}}\, (k/k_{p})^{n_{s} -1}$ where $n_{s} = 1 - 6 \epsilon_{k} + 2 \overline{\eta}_{k}$ is the scalar spectral index whose explicit expression actually depends upon the specific inflationary scenario through the slow-roll parameters of Eq. (\ref{INF3}). A relevant quantity (already mentioned in sec. \ref{sec1}) is constructed from the quotient between the tensor and the scalar power spectra and it is customarily referred to as the tensor to scalar ratio, i.e.
 $r_{T}(k,\tau)= P_{T}(k,\tau)/P_{{\mathcal R}}(k,\tau)$. When evaluated at the time the CMB scales crossed the comoving horizon $r_{T}(k) =r_{T}(k, \tau_{k})$ 
turns out to be proportional to $\epsilon_{k}$
 \begin{equation}
r_{T}(k)= 8 \ell_{P}^2\, a^2_{k}/z_{\varphi}^2(\tau_{k}) = 8\ell_{P}^2 \, (\dot{\varphi}_{k}^2/H_{k}^2) \to 16 \, \epsilon_{k}.
\label{INF7}
\end{equation}
As already discussed in sec. \ref{sec1}, the value of $r_{T}= r_{T}(k)$ is currently constrained for $k = {\mathcal O}(k_{p})$ by the observational data implying, roughly speaking, that $r_{T} \leq {\mathcal O}(0.03)$ \cite{FIVEa,FIVEb,FIVEc,FIVEd,FIVEe}. In the conventional approach the limits on $r_{T}$ are customarily viewed as constraints on the first derivative (with respect to $\varphi$) of the inflationary potential since, according to the consistency relation of Eq. (\ref{INF7}), $r_{T}(k) \propto \epsilon_{k}=  (\overline{M}_{P}^2/2)(V_{\, \varphi}/V)_{k}^2$. 
As an example, in the case of monomial potentials $r_{T}(k) \propto {\mathcal O}(1/N_{k})$ 
while for the plateau-like potential the tensor to scalar ratio gets smaller $r_{T} \propto {\mathcal O}(1/N_{k}^2$). The spirit of this investigation is, however, slightly different: while there is no doubt that plateau-like potentials are more consistent with the current data than the monomials \cite{FIVEa,FIVEb,FIVEc,FIVEd,FIVEe}, the minute value of $r_{T}$ is not only related to the 
{\em shape} of $V(\varphi)$  but also to the actual determination of $N_{k}$ that is affected by the post-inflationary evolution: as specifically discussed in sec. \ref{sec3},  $N_{k}$ changes depending upon the physical features of the post-inflationary evolution. 

A final technical remark is in order. Although, for the present purposes, the shape of the inflationary potential is not central, for the sake of concreteness we consider here the class of inflationary potentials that can be parametrized in the following form:
\begin{equation}
V(\varphi) = M_{inf}^{4}\, v(\overline{\varphi}), \qquad \overline{\varphi} = \varphi/\overline{M}_{P},
\label{INF8}
\end{equation}
where $M_{inf}$ indicates the scale of the $V(\varphi)$ while $\overline{\varphi}$ and $v(\overline{\varphi})$ are both dimensionless. In the inflationary limit (i.e. $\overline{\varphi} \gg 1$) the potential develops a plateau ${\mathcal O}(M_{inf}^4)$ (i.e. $v(\overline{\varphi}) \to 1$) and in the opposite regime  $v(\overline{\varphi}) = \overline{\varphi}^{2 q}$ for $\overline{\varphi} \ll 1$. Different analytic expressions with these properties exist and some of them have been 
studied in detail in \cite{SIXa};  the forthcoming considerations do not rely 
on the explicit forms of $v(\overline{\varphi})$ however the following two examples 
\begin{equation}
v_{1}(\overline{\varphi}) = (1 - \exp{- \beta \overline{\varphi}})^{2 q}, \qquad v_{2}(\overline{\varphi}) = \frac{\beta^{p}\,\, \overline{\varphi}^{\,2 q}}{[ 1 + \beta^2 \,\,\overline{\varphi} ^{\,\frac{4 q}{p}}]^{\frac{p}{2}}},
\label{INF9}
\end{equation}
will be considered hereunder for illustration with the caveat that more contrived expressions of $v(\overline{\varphi})$ (obeying the above requirements on $v(\overline{\varphi})$) can be always concocted.

\subsection{Evolution of the ultralight scalar during inflation}
During the inflationary stage governed  Eq. (\ref{INF2}) the background of the $\sigma$-field does not affect the dynamics since $\sigma$ remains ${\mathcal O}(M)$ even after the conclusion of inflation (but before the oscillations of $\sigma$ taking place for $H= H_{m} = {\mathcal O}(m)$). Although the mass of the light scalar is much smaller than the inflationary expansion rate (i.e. $\mu = m/H_{1} \simeq m/H_{k} \ll 1$) the corresponding fluctuations are still amplified with a quasi-flat spectrum and with an amplitude ${\mathcal O}(H_{k})$. In fact, throughout the inflationary stage the fluctuation of $\sigma$ (i.e. $\chi_{\sigma} = \delta \sigma$) obeys, in Fourier space, the following (approximate) equation: 
\begin{equation}
\overline{\chi}_{\sigma}^{\prime\prime} + \biggl[ k^2 + a^2 \, H^2 \, \mu^2 - a^2 H^2 ( 2 - \epsilon)\biggr] \overline{\chi}_{\sigma}=0,
\label{INF10}
\end{equation}
where $\overline{\chi}_{\sigma}(k,\tau) = a(\tau)\, \chi_{\sigma}(k,\tau)$. 
While during the inflationary epoch the fluctuations 
of $\sigma$ and $\varphi$ (i.e. $\delta \sigma = \chi_{\sigma}$ and $\delta \varphi = \chi_{\varphi}$) remain decoupled (since $\mu \ll 1$), after the end of inflation (when $\epsilon(\tau_{e}) \to 1$) the fluctuations of the ultralight scalar and of the inflaton are coupled. In sec. \ref{sec4} and in appendix \ref{APPA}) the coupled evolution of $\chi_{\sigma}$ and $\chi_{\varphi}$ is analyzed with greater details. Here we are just going to focus on the most relevant aspects of the physical result. First of all, because the slow-roll parameters affect the definition of the conformal time 
coordinate $\tau$, we also have that $\tau$ depends on $\epsilon$
\begin{equation}
\tau = \int dt/a(t) = - (a\, H)^{-1} + \epsilon \int \,d a/(a^2 H),
\label{INF11}
\end{equation}
where the second equality follows after integration by parts; from Eq. (\ref{INF11}) 
we then deduce $a\, H = - 1/[ (1- \epsilon) \tau]$ and Eq. (\ref{INF10}) becomes:
\begin{equation}
\overline{\chi}_{\sigma}^{\prime\prime} + \biggl[ k^2 + \frac{\mu^2}{(1 - \epsilon)^2 \tau^2} - \frac{( 2 - \epsilon)}{( 1 - \epsilon)^2 \tau^2}\biggr] 
\overline{\chi}_{\sigma} =0.
\label{INF12}
\end{equation}
If $m = {\mathcal O}(10^{-22})\, \mathrm{eV}$ and $H_{1} = {\mathcal O}(10^{-5})\, M_{P}$ we have 
that $\mu^2 = {\mathcal O}(10^{-90})$ and the contribution of the corresponding term can be safely neglected in the evolution of $\chi_{\sigma}$ and in the determination of the corresponding 
power spectrum
\begin{equation}
P_{\sigma}(k,\tau) = \frac{k^3}{2 \pi^2 a^2} \bigl|\overline{\chi}_{\sigma}(k,\tau)|^2 = {\mathcal A}_{\sigma}(\nu_{\sigma}) H^2 \, \biggl(\frac{k}{a \, H}\biggr)^{3 - 2 \nu_{\sigma}},
\label{INF13}
\end{equation}
where ${\mathcal A}_{\sigma}(\nu_{\sigma}) = 2^{2\nu_{\sigma} - 3}\, \Gamma^2(\nu_{\sigma})/\pi^3$ (see also Eq. (\ref{NAD8i}) of appendix \ref{APPA} and discussions therein). Since 
the solution of Eq. (\ref{INF12}) is obtained via Hankel functions \cite{TEN1,TEN2}, $\nu_{\sigma} = (3 - \epsilon)/[2 ( 1-\epsilon)] \simeq 3/2 - 2 \epsilon + {\mathcal O}(\epsilon^2)$  is the corresponding Bessel index. Given that the power spectrum is quasi-flat, the pre-factor is estimated in the limit $\nu_{\sigma} \to 3/2$, i.e. ${\mathcal A}_{\sigma} \to 1/(4 \pi^2)$. Recalling that, at the crossing time, $k = a\, H= {\mathcal H}$, when the modes $k = {\mathcal O}(k_{p})$ cross the expansion rate the power spectrum is given by $P_{\sigma}(k,\tau_{k}) = H_{k}^2/(4 \pi^2)$. Because $H_{k}^2< {\mathcal O}(M^2)$ the corresponding fluctuations are (predominantly) Gaussian \cite{FIVEa,FIVEb,FIVEc,FIVEd,FIVEe}. This also means, as as often propounded in qualitative discussions, that in Fourier space the fluctuation of the light fields are, during inflation, of the order of $H_{k}/(2\pi)$. 

\subsection{The conventional late-time evolution}
\begin{figure}[ht!]
\begin{center}
\includegraphics[width=9cm,height=7cm]{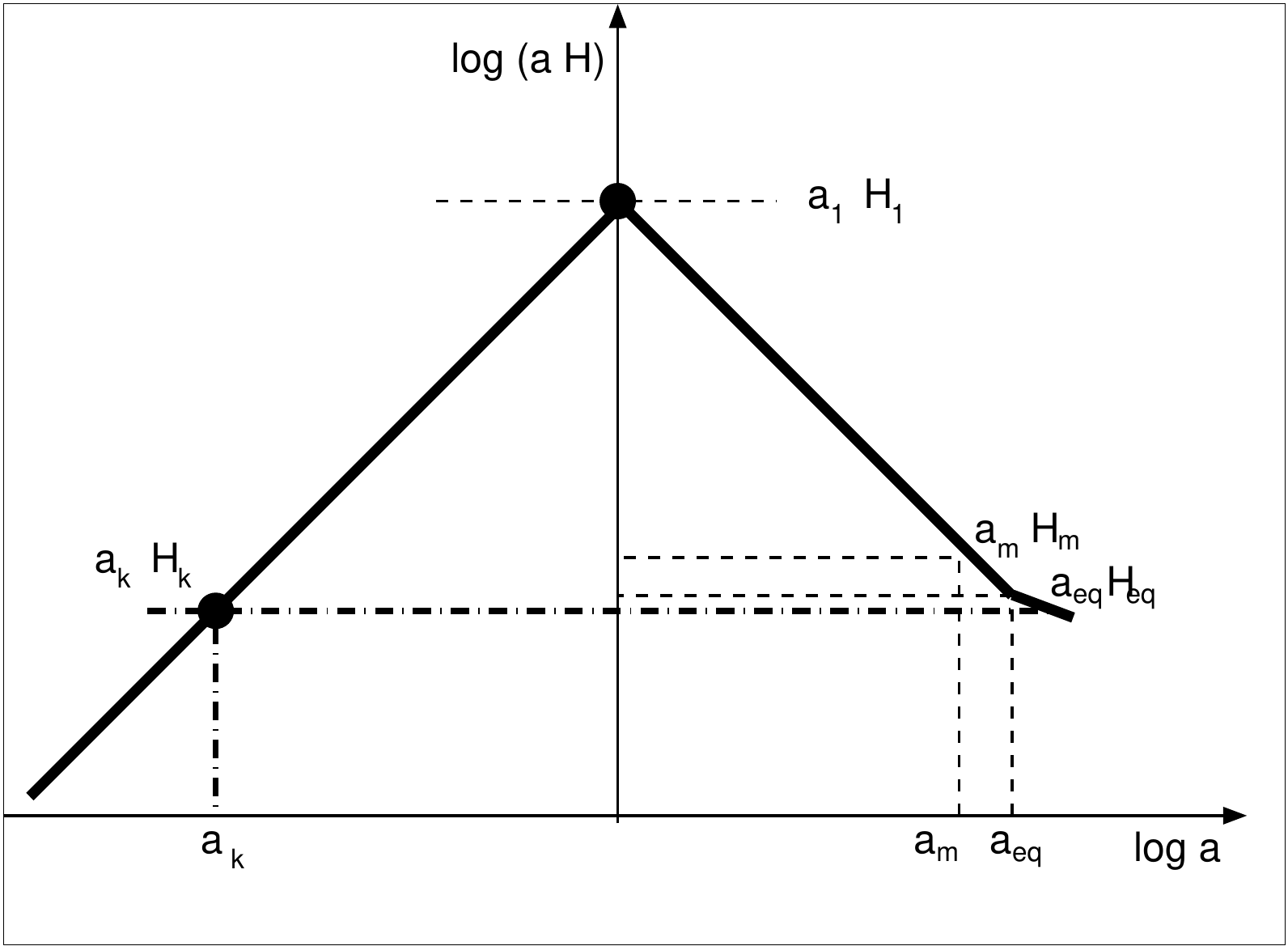}
\end{center}
\caption{\label{FIGURE1} The common logarithm of the (comoving) expansion rate 
$a \, H$ is illustrated as a function of the common logarithm of the scale factor. While during the inflationary stage 
(see the leftmost region of the cartoon) $a\, H$ is proportional to the scale factor, in the 
radiation stage $a\,H$ scales instead as $a^{-1}$.  When 
$k = a_{k} H_{k}$ the (comoving) wavenumber crosses the (comoving) expansion rate. In this conventional timeline the CMB wavenumbers 
$k = {\mathcal O}(k_{p})$ crossed the expansion rate nearly $60$ $e$-folds before the end of inflation and reentered after matter radiation equality (but prior to photon decoupling).}
\end{figure}

The conventional timeline of the post-inflationary expansion rate is illustrated in Fig. \ref{FIGURE1} where, on the vertical axis, the (common) logarithm 
of the comoving expansion rate is plotted against the (common) logarithm of the scale factor. 
The inflationary stage corresponds to the leftmost region of Fig. \ref{FIGURE1} where 
$ H\, a$ increases linearly with $a$. With a dark blob the crossing region of the scales $k = {\mathcal O}(k_{p})$ (i.e. $k = a_{k}\, H_{k}$) has been also indicated. The CMB scales eventually cross again the expansion rate between matter-radiation equality and photon decoupling (see the rightmost region on the cartoon). All in all, in the standard lore of Fig. \ref{FIGURE1} the evolution after inflation is almost suddenly dominated by radiation and the total energy density and pressures are:
 \begin{equation}
\rho_{t} = \sigma^{\, \prime\, 2}/(2 a^2) + U(\sigma) + \rho_{r}, \qquad 
p_{t} = \sigma^{\, \prime\, 2}/(2 a^2) - U(\sigma) + p_{r}.
\label{AINF1}
\end{equation}
Because more general evolutions are considered in sec. \ref{sec3}, it is now mandatory to discuss, for immediate comparison, the evolution of $\sigma$ in the timeline of Fig. \ref{FIGURE1}. Thanks to the inflationary initial conditions the potential of $\sigma$ prevails against the corresponding kinetic energy while the radiation background exceeds both $\sigma^{\prime\, 2}/(2 a^2)$ and $U(\sigma)$;
more specifically we have $(\rho_{r} + p_{r}) \gg \sigma^{\, \prime\, 2}/a^2$ and $\rho_{r} \gg U_{\sigma} \gg \sigma^{\, \prime\, 2}/(2 \,a^2)$. Moreover, while the evolution
of $\sigma$ reads in general
\begin{equation}
\sigma^{\prime\prime} + 2 {\mathcal H} \sigma^{\prime} + a^2 U_{\sigma} =0,
\label{AINF2}
\end{equation}
owing to the hierarchy between $\sigma^{\prime\, 2}$ and $U(\sigma)$ 
$\sigma$ obeys, right after the end of inflation, the 
following approximate equation
\begin{equation}
\sigma^{\prime} = - \frac{a^2}{5 {\mathcal H}} \,\, U_{,\,\sigma}, \qquad H > H_{m} = {\mathcal O}(m),
\label{AINF3}
\end{equation}
that is valid, as indicated,  before the onset of the oscillating stage of $\sigma$ (i.e. for $H> H_{m}$).
Once Eq. (\ref{AINF3}) is inserted into Eq. (\ref{AINF2}), the following 
approximate expression is obtained
\begin{equation}
\frac{1}{5} \frac{{\mathcal H}'}{{\mathcal H}^2} \frac{\partial U}{\partial\sigma} a^2 - \frac{4}{5} \frac{\partial U}{\partial \sigma} a^2 + a^2 \frac{\partial U}{\partial \sigma} - \frac{1}{5 {\mathcal H}} \frac{\partial^2 U}{\partial \sigma^2} a^2 \sigma^{\prime} =0.
\label{AINF4}
\end{equation}
The last term at the left hand side of Eq. (\ref{AINF4}) contains three derivatives of the potential (since $\sigma^{\prime}$ is proportional to $U_{,\,\sigma}$); this 
term is actually negligible while the other contributions cancel when ${\mathcal H}^{\prime} = - {\mathcal H}^2$ and radiation prevails against the other sources. Although Eq. (\ref{AINF4}) implies that $\sigma$ remains ${\mathcal O}(M)$, the Hubble rate steadily decreases and eventually it becomes of order $m$.
For $H < H_{m} = {\mathcal O}(m)$, $\sigma$ undergoes a stage of coherent 
oscillations where $(\sigma^{\prime}/a)^2 \propto a^{-3}$ as it follows from Eq. (\ref{AINF2}) written in the cosmic time parametrization
(i.e. $\ddot{\sigma} + 3 H \dot{\sigma} + m^2 \sigma=0$); after 
rescaling the field amplitude the following equation is readily deduced without much effort
\begin{equation}
(a^{3/2} \, \sigma)^{\cdot\cdot} + [ m^2 - (9 H^2/4 + 3\dot{H}/2)]\,(a^{3/2} \, \sigma)=0.
\label{AINF5}
\end{equation}
Again since $\dot{H} = - 2 H^2$,  for $m^2 \gg H^2$ the rescaled variable $ a^{3/2} \, \sigma$ 
is given as a combination of trigonometric functions with argument $m\, t$.
According to the parameters selected in Fig. \ref{FIGURE1} the oscillatory stage occurs approximately when $a_{1}\,H_{1}/(a_{m} \, H_{m}) = {\mathcal O}(10^{23})$
where we considered, as an example, $m= {\mathcal O}(10^{-22})\, \mathrm{eV}$ (to be compared 
with $H_{eq} = {\mathcal O}(10^{-28}) \mathrm{eV}$). Within the timeline 
of Fig. \ref{FIGURE1} the critical density parameter 
of the light scalar (i.e. $\Omega_{\sigma}(a) = \rho_{\sigma}/[ 3\,H^2\, \overline{M}_{P}^2 ]$) becomes, at the equality time\footnote{For the numerical estimates we recall that $H_{eq} = 2.026 \times 10^{-28} (h_{0}^2 \Omega_{M0}/0.134)^2 \, [h_{0}^2 \Omega_{R0}/(4.15\times10^{-5})]^{-3/2} \, \mathrm{eV}$ where $\Omega_{M0}$ and $\Omega_{R0}$ are the current 
critical fractions of dusty matter and radiation; the previous figure also corresponds to $H_{eq}/M_{P} = 1.67\times 10^{-56}$ for the same typical values of the parameters. Note also that $H_{0} = 1.23 \times 10^{-61} (h_{0}/0.7)$ where $h_{0}$ indicates throughout the current indetermination in the value of the Hubble rate.},
\begin{equation}
\Omega_{\sigma}(a_{eq}) = \frac{8\pi}{3} \, \biggl(\frac{M}{M_{P}}\biggr)^2 \, 
\sqrt{\frac{m}{H_{eq}}} = 0.3 \biggl(\frac{M}{10^{17} \mathrm{GeV}}\biggr)^2 
\biggl(\frac{m}{10^{-22}\, \mathrm{eV}}\biggr)^{1/2}.
\label{AINF6}
\end{equation}
which is why ultralight scalars are customarily regarded as potential dark 
matter candidates \cite{SEVENf,SEVENg}. From Eqs. (\ref{INF12})--(\ref{INF13}) it follows that  $\chi_{\sigma}(k,\tau)$ (i.e. the Fourier mode of the fluctuation of the ultralight scalar) gets amplified during inflation with a quasi-flat spectrum and, 
although the associated power spectra, do not contribute to the curvature inhomogeneities during inflation, they will also be present when $\sigma$ starts oscillating and potentially induce further sources of curvature inhomogeneities in the large-scale limit. This contribution is controlled 
by $\delta p_{nad}$ appearing in Eqs. (\ref{PP7})--(\ref{PP8}); moreover, since at 
large scales the sources of anisotropic stress are negligible we have that $\Phi= \Psi$. The explicit form of non-adiabatic pressure fluctuations is  
\begin{equation}
 \delta p_{nad} =  (1/3 - c_{s\,t}^2) \delta \rho_{r} 
+ \Psi ( c_{s\,t}^2 -1) \frac{\sigma^{\prime\, 2}}{a^2} + \frac{\sigma^{\prime}\,\chi_{\sigma}^{\prime}}{a^2} ( 1 - c_{s\,t}^2) - U_{,\sigma}\chi_{\sigma} ( 1 + c_{s\,t}^2).
\label{AINF7}
\end{equation}
The total sound speed of the system can be written as
\begin{equation}
 c_{s\,t}^2 =  \frac{ 4\, \rho_{r} \, a^2 + 9\, \sigma^{\prime\, 2} +
6 \, a^2 \,(\sigma^{\prime} /{\mathcal H}) U_{,\sigma}}{ 3(4 \rho_{r}\, a^2 + 3  \sigma^{\prime\, 2})}.
\label{AINF8}
\end{equation}
From Eq. (\ref{AINF8}) we argue that if the field $\sigma$ is subdominant during a radiation stage $ c_{s\,t}^2 \to 1/3$ and the last term at the right hand side of Eq. (\ref{AINF7}) gives the leading contribution to the non-adiabatic pressure inhomogeneity, 
 $\delta p_{nad} \simeq - 4 U_{,\sigma} \chi_{\sigma}/3$. In the case of a generic 
 barotropic fluid with equation of state $p = w \, \rho$ replaces the radiation stage we have $c_{s\,t} \to \sqrt{w}$ 
 and the explicit form of $\delta p_{nad}$ is obtained from Eq. (\ref{AINF7}) by replacing 
 the $1/3$ with $w$. The approximate form of $\delta p_{nad}$ becomes then 
 \begin{equation}
  \delta p_{nad} = - U_{,\sigma}\chi_{\sigma} ( 1 + c_{s\,t}^2) = - U_{,\sigma}\chi_{\sigma} ( 1 + w).
  \label{AINF9}
 \end{equation}
 When Eq. (\ref{AINF9}) is inserted into Eqs. (\ref{PP8})--(\ref{PP9}) the explicit form of the evolution equation of the curvature inhomogeneities is obtained. But since it is appropriate to present first the connection between the different post-inflationary histories and the suppression of the tensor to scalar ratio, the analysis of the solutions of the evolution equations involving $\delta p_{nad}$ is postponed to sec. \ref{sec4}. 

\renewcommand{\theequation}{3.\arabic{equation}}
\setcounter{equation}{0}
\section{Post-inflationary histories and the suppression of $r_{T}$}
\label{sec3}
The evolution of the light scalars and the suppression of the tensor modes 
of the geometry for $k = {\mathcal O}(k_{p})$ are both determined by the
 post-inflationary timelines that will now be scrutinized in greater detail. 
If the expansion rate is faster 
than radiation during a certain portion of the decelerated stage after inflation, the number of $e$-folds elapsed since the crossing of the CMB scales ($N_{k}$ in Eq. (\ref{INF4}))
gets comparatively smaller; the opposite is true if the expansion 
rate gets slower than radiation during a certain post-inflationary timeline. 
\subsection{Generic features of the post-inflationary evolution}
To examine and motivate the implications of the previous two observations
we first remark that the 
the ratio between $a_{1}\, H_{1}$ and $a_{m}\, H_{m}$  not only depends upon
the mass range of the $\sigma$ field but also upon the maximal curvature scale $H_{1}$. If we recall the shape of Fig. \ref{FIGURE1} and 
the observations of Eqs. (\ref{INF6})--(\ref{INF7}) we can 
compute $H_{1}$ which is of the order of $H_{k}$ (i.e. 
the Hubble rate evaluated at the crossing of the CMB scales).
\begin{figure}[ht!]
\begin{center}
\includegraphics[width=9cm,height=7cm]{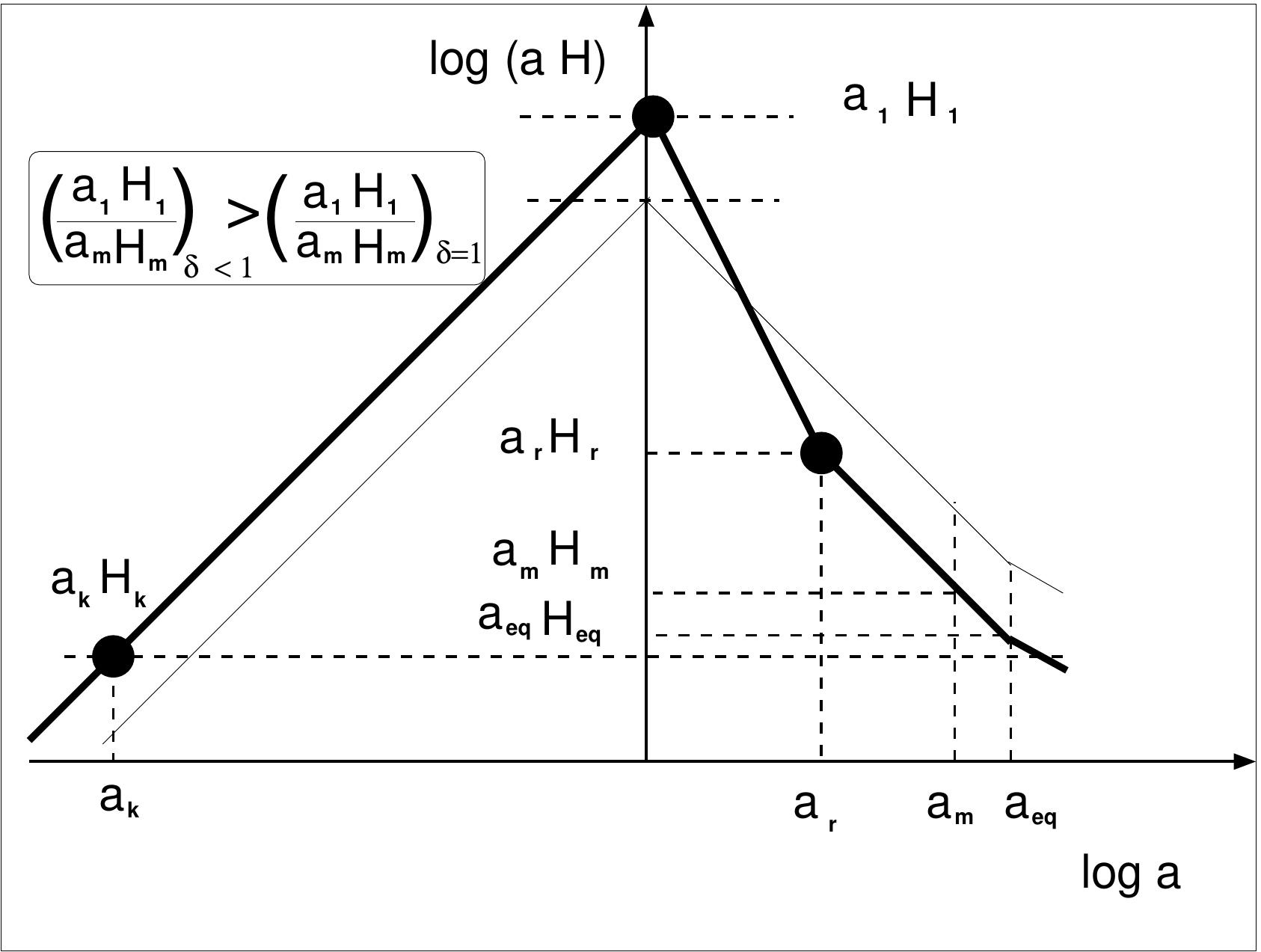}
\end{center}
\caption{\label{FIGURE2} The timeline of the comoving expansion 
rate is illustrated when the post-inflationary evolution initially evolves at a rate that is slower 
than radiation (i.e. $\delta < 1$ in the language of Eq. (\ref{POS2})). The thin line below the thicker profile corresponds, for comparison, to the timeline of Fig. \ref{FIGURE1}. Because the overall slope of the comoving expansion rate gets modified its maximal value increases and the number of $e$-folds elapsed since the crossing of the CMB scales gets larger. For the purposes of this particular cartoon we assumed $\delta = 1/2$ and $H_{r} = {\mathcal O}(10^{-13})$; we actually remind that we must always require that $H_{r} \geq 10^{-16} \mathrm{eV}$, as implied by the constraints coming from the synthesis of the light nuclei.}
\end{figure}
With these premises, a simple estimate shows that:
\begin{equation}
\frac{a_{1} H_{1}}{a_{m} \, H_{m}} = \frac{\tau_{m}}{\tau_{1}} = {\mathcal O}(10^{23}),
\label{POS1}
\end{equation}
for $m = {\mathcal O}(10^{-22})\, \mathrm{eV}$. Let us now suppose, as illustrated in Fig. \ref{FIGURE2}, that the end of inflation (occurring at an approximate scale $H_{1}$)
{\em does not} immediately precedes the radiation stage. Between the end 
of inflation and the beginning of radiation the comoving expansion rate 
evolves as $a \, H \propto a^{-\delta}$ where $\delta$ accounts for the expansion 
rate in the intermediate post-inflationary stage preceding the radiation epoch. In this situation the estimate of Eq. (\ref{POS1}) gets modified as follows:
 \begin{equation}
 \frac{a_{1} H_{1}}{a_{m} H_{m}} = \sqrt{\frac{H_{1}}{H_{m}}} \biggl(\frac{H_{r}}{H_{1}}\biggr)^{\frac{\delta -1}{2(\delta+1)}}.
 \label{POS2}
 \end{equation}
 When $\delta \to 1$ the expansion rate coincides with radiation and 
the result of Eq. (\ref{POS2}) becomes exactly the one qualitatively anticipated in Eq. (\ref{POS1}) which is now written in more explicit terms as a function of the relevant parameters:
\begin{equation}
\frac{a_{1} H_{1}}{a_{m} H_{m}} = 5.15\times 10^{22} \biggl(\frac{m}{10^{-22}\, \mathrm{eV}}\biggr)^{-1/2} \biggl(\frac{{\mathcal A}_{{\mathcal R}}}{2.41\times10^{-9}}\biggr)^{1/4}  \biggl(\frac{r_{T}}{0.03}\biggr)^{1/4}.
\label{POS3}
\end{equation}
If the expansion rate is slower than radiation (i.e. $\delta<1$ in Eq. (\ref{POS2})) we must have $a_{1} H_{1}/(a_{m} \, H_{m}) 
> {\mathcal O}(10^{23})$; this happens since $H_{r} < H_{1}$ in Eq. (\ref{POS2}) and this is 
the physical situation illustrated in Fig. \ref{FIGURE2} where, for the sake of concreteness, we took $\delta = 1/2$. As 
anticipated the ratio between $a_{1} \, H_{1}$ and the other typical scales 
(like $a_{m} H_{m}$) gets larger than in the case where $\delta \to 1$. The increase 
is given by the ratio appearing in Eq. (\ref{POS2}) which becomes, for $\delta \to 1/2$, $(H_{r}/H_{1})^{-1/6}$. The value of $H_{r}$ should be larger than the scale 
associated with big-bang nucleosynthesis taking place when the temperature 
ranges between $0.1$ and $1$ MeV. This requirement implies the lower bound $H_{r} > 10^{-16} \mathrm{eV}$; in Fig. \ref{FIGURE2} we selected $H_{r} = 10^{-13} \, \mathrm{eV}$ 
so that, approximately, $(H_{r}/H_{1})^{-1/6} = {\mathcal O}(10^{6})$. 
\begin{figure}[ht!]
\begin{center}
\includegraphics[width=9cm,height=7cm]{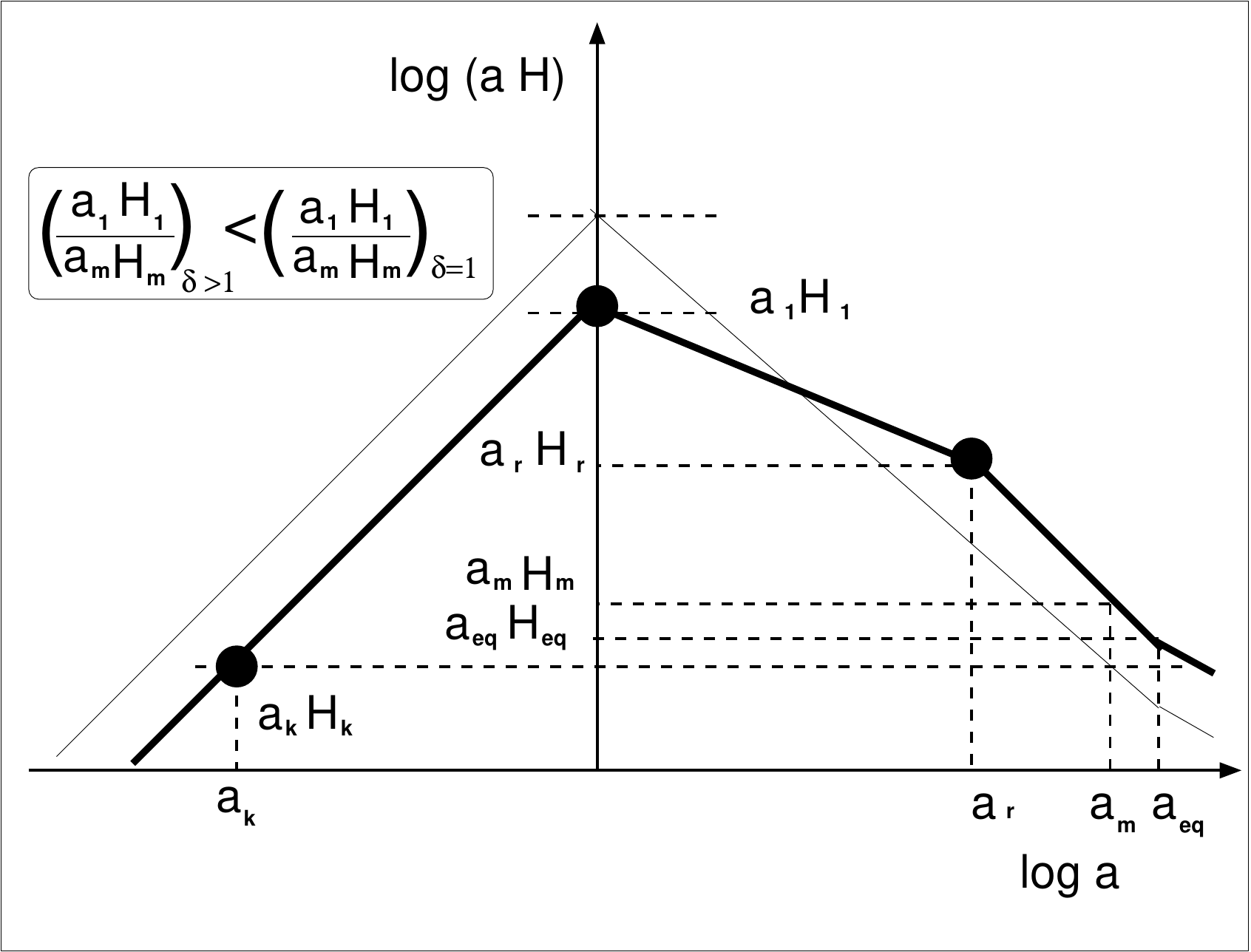}
\end{center}
\caption{\label{FIGURE3} The timeline of the comoving expansion 
rate is illustrated when the post-inflationary evolution initially evolves at a rate faster
than radiation [i.e. $\delta > 1$ in the language of Eq. (\ref{POS2})]. The thin line corresponds, for comparison, to the conventional case of Fig. \ref{FIGURE1}. Not only $a_{1} H_{1}/(a_{m} \, H_{m})$ gets reduced in comparison with Fig. \ref{FIGURE2} but also the number of $e$-folds elapsed since the crossing of the CMB scale is smaller than in the case of Figs. \ref{FIGURE1} and \ref{FIGURE2}.  All the parameters are the same as in Fig. \ref{FIGURE2} except for the value of $\delta$ that now has chosen to be $\delta \to 2$, always for the sake of illustration.}
\end{figure}

The same argument leading to Eq. (\ref{POS2}) remains valid when the expansion rate is faster than radiation, as illustrated in the cartoon of Fig. \ref{FIGURE3}. In this dynamical situation, however, 
$a_{1} H_{1}/(a_{m} \, H_{m}) < {\mathcal O}(10^{23})$, i.e. smaller than in the case where the post-inflationary evolution is dominated by radiation; this happens because $\delta > 1$ and, at the same time, $H_{r} <H_{1}$ in Eq. (\ref{POS2}). In Fig. \ref{FIGURE3} this remark is complemented by illustrating a post-inflationary expansion rate  faster than radiation (for the sake of concreteness we took $\delta \to 2$). If $\delta > 1$ the factor $(H_{r}/H_{1})^{(\delta -1)/[2 (\delta +1)]}$ of Eq. (\ref{POS2}) reduces the estimate valid for $\delta \to 1$ (see also Eq. (\ref{POS3})). Again 
if we take the same value selected before (i.e. $H_{r} = 10^{-13} \mathrm{eV}$)
we obtain $(H_{r}/H_{1})^{1/6} = {\mathcal O}(10^{-6})$ implying a 
reduction of roughly six orders of magnitude of the ratio $a_{1} H_{1}/(a_{m} \, H_{m})$ 
in comparison with the case $\delta \to 1$. Instead of ${\mathcal O}(10^{23})$ 
(as it follows from Eq. (\ref{POS3})) we shall have that $a_{1} H_{1}/(a_{m} \, H_{m})={\mathcal O}(10^{17})$. The same conclusion holding for $a_{m} H_{m}$ also holds for other typical scales 
like $a_{eq}\, H_{eq}$. What matters for the present purposes 
is that this class of observations directly affect the number of $e$-folds 
elapsed since the crossing of the CMB scales, as we are going to see 
in the next subsection. 

\begin{figure}[!ht]
\centering
\includegraphics[height=6.5cm]{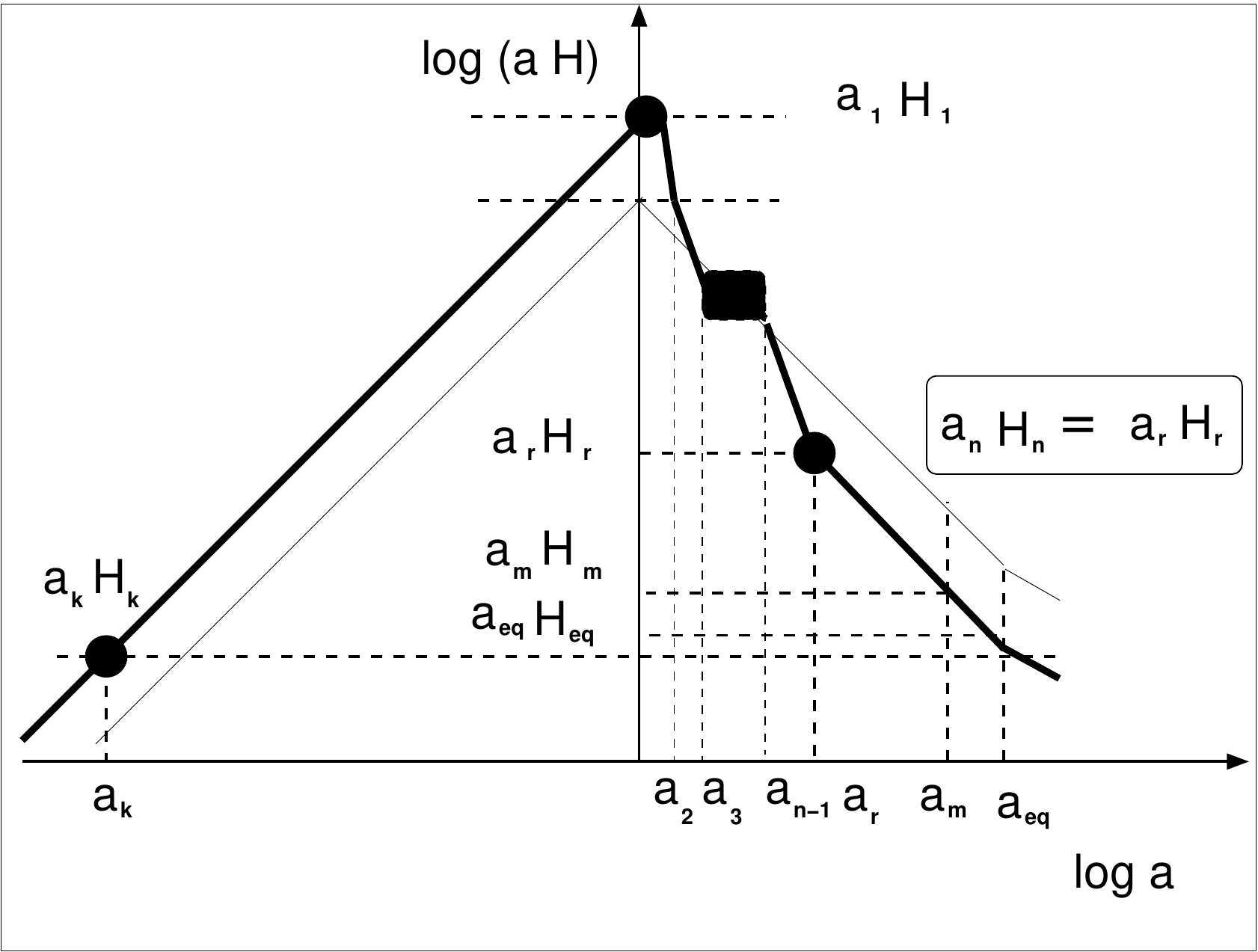}
\includegraphics[height=6.5cm]{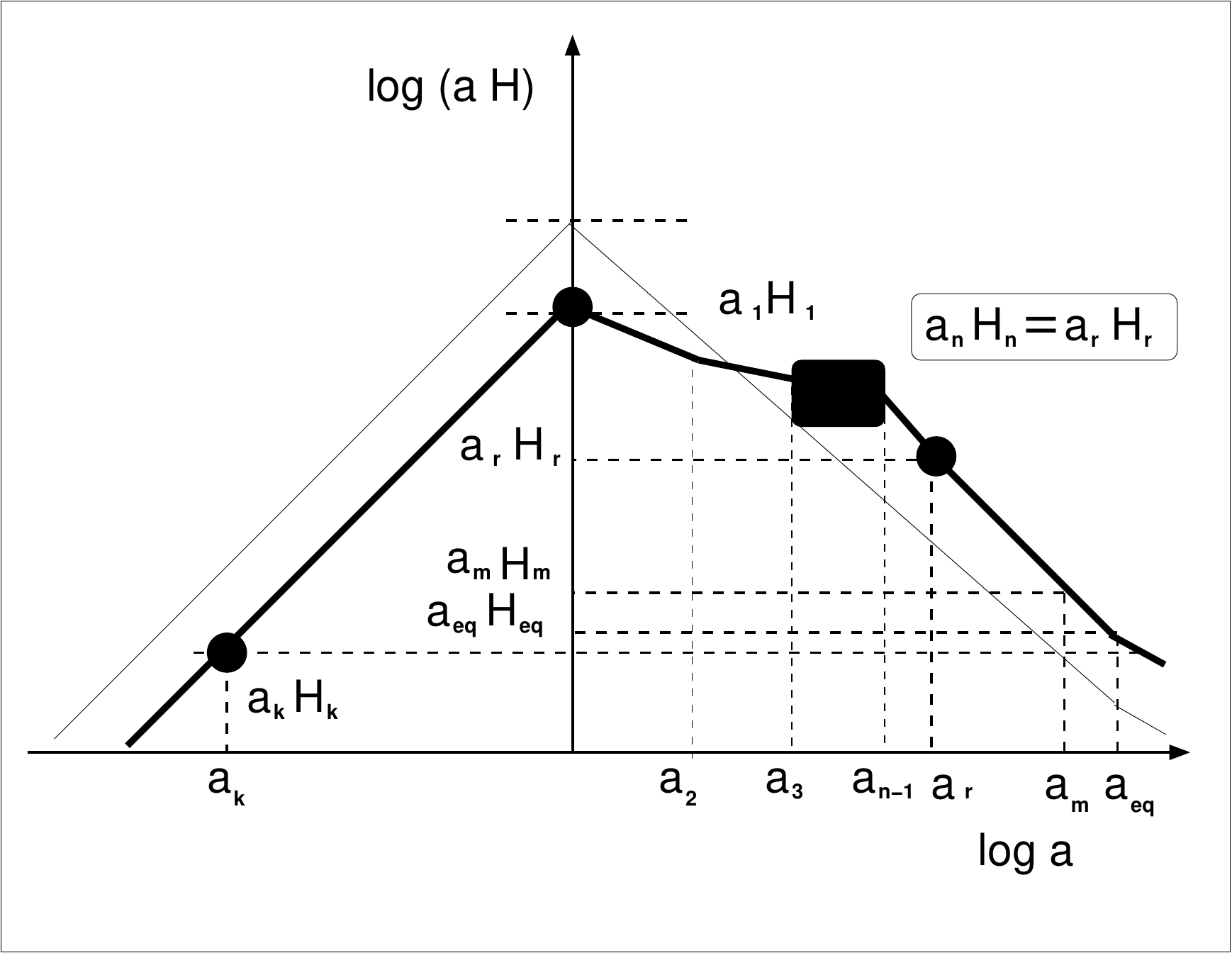}
\caption[a]{In both plots we illustrate the common logarithm of $a\,H$. As before the left portion of each cartoon corresponds to the inflationary evolution while the post-inflationary epoch consists of  $n-1$ successive stages of expansion characterized by the rates $\delta_{i}$. In the right plot the overall evolution is faster than radiation. In the same averaged sense the timeline of the left plot is slower than radiation. The black rectangles account for possible presence of further sub-phases. In the present conventions $a_{n}$ coincides with $a_{r}$ (i.e. the beginning of the standard radiation-dominated evolution). As already stressed post-inflationary stage is bounded by the curvature scale of big-bang nucleosynthesis so that $H_{r} \geq 10^{-44} \, M_{P} = {\mathcal O}(10^{-16}) \mathrm{eV}$. }
\label{FIGURE4}      
\end{figure}

From the technical viewpoint, in Eq. (\ref{POS2}) we considered a {\em single} post-inflationary stage expanding at an arbitrary rate and preceding the radiation phase. This hypothesis can be easily relaxed and when  many expanding stages precede the radiation dominance, Eq. (\ref{POS2}) depends upon all the successive stages of expansion. We can therefore imagine a succession of $n$ different stages each of them characterized by a specific expansion rate $\delta_{i}$ with $i=1,\, .\,.\,.\,n$. During each of the stages the expansion rate gets reduced so that we can construct the combinations $H_{i+1}/H_{i} < 1$. The general form of Eq. (\ref{POS2}) then becomes
\begin{equation}
\frac{a_{1} H_{1}}{a_{m} H_{m}} = \sqrt{\frac{H_{1}}{H_{m}}} \prod_{i=1}^{n-1} \,\, \biggl(\frac{H_{i+1}}{H_{i}} \biggr)^{\frac{\delta_{i}-1}{2(\delta_{i} + 1)}}.
\label{POS4}
\end{equation}
 The product appearing in Eq. (\ref{POS4}) runs from $1$ to $(n-1)$ 
 simply because the beginning of the $n$th stage coincide with the radiation epoch; in other 
 words $H_{n} = H_{r}$. When all the $\delta_{i}$ are equal, Eq. (\ref{POS4}) 
 coincides with Eq. (\ref{POS2}) since, in this situation
 \begin{equation}
  \prod_{i=1}^{n-1} \,\, \biggl(\frac{H_{i+1}}{H_{i}} \biggr)^{\frac{\delta_{i}-1}{2(\delta_{i} + 1)}}
 = \biggl[\prod_{i=1}^{n-1} \,\, \biggl(\frac{H_{i+1}}{H_{i}} \biggr)\biggr]^{\frac{\delta -1}{2(\delta +1)}}
 = \biggl(\frac{H_{r}}{H_{1}}\biggr)^{\frac{\delta -1}{2(\delta +1)}}.
 \label{POS5}
 \end{equation}
In Fig. \ref{FIGURE4} the possibility of a phase with multiple expanding stages is illustrated.
The black rectangles account for the possible presence of various sub-phases. As stressed 
in connection with Eq. (\ref{POS4}) the $n$th expansion rate coincides with the one of the radiation 
stage, i.e. $H_{n} \equiv H_{r}$. 

\subsection{The number of $e$-folds and the post-inflationary timeline}
According to Figs. \ref{FIGURE2} and \ref{FIGURE3} the number of $e$-folds 
can be either larger or smaller than the conventionally quoted figure ${\mathcal O}(60)$. To 
make this statement more quantitative we now estimate number of $e$-folds elapsed since the crossing of the CMB scales $k= {\mathcal O}(k_{p})$ with particular attention to the 
dynamical situations illustrated in Figs. \ref{FIGURE2} and \ref{FIGURE3}. Since, by definition, 
the crossing time is given by $k = {\mathcal O}(a_{k}\, H_{k})$, we can express  the crossing condition in terms of the number of $e$-folds elapsed since the crossing 
of $k= {\mathcal O}(k_{p})$ (see Eq. (\ref{INF4})):
\begin{equation}
\biggl(\frac{k}{a_{0}\, H_{0}}\biggr) \biggl(\frac{a_{0} \, H_{0}}{a_{eq}\, H_{eq}}\biggr)  \biggl(\frac{a_{eq} \, H_{eq}}{a_{r}\, H_{r}}\biggr) \biggl(\frac{a_{r} \, H_{r}}{a_{1}\, H_{1}}\biggr) 
\biggl(\frac{a_{1} \, H_{1}}{a_{e}\, H_{e}}\biggr) \biggl(\frac{a_{e} \, H_{e}}{a_{k}\, H_{k}}\biggr) = {\mathcal O}(1).
\label{EFOL1}
\end{equation}
In Eq. (\ref{EFOL1}) $a_{e}$ denotes the end of inflation while $a_{1}$ is the onset of the 
post-inflationary evolution. As in the case of the observational estimates 
we shall be assuming that $a_{e} = a_{1}$. In contrast with the standard approach 
however the onset of the radiation dominance is conceptually distinct 
from the end of inflation. Finally in Eq. (\ref{EFOL1}) $H_{0}$ is the current 
value of the Hubble rate and we shall be using the convention $a_{0} =1$: in this way 
comoving and physical wavenumbers coincide at the present time.
In case a single expanding stage different from radiation 
takes place between $a_{1}$ and $a_{r}$ the crossing condition of eq. (\ref{EFOL1}) is:
\begin{equation}
\biggl(\frac{k}{a_{0} \, H_{0}} \biggr) \frac{\sqrt{H_{0}/M_{P}}}{( 2 \,\pi \, \Omega_{R\,0}\, \epsilon_{k} \, {\mathcal A}_{{\mathcal R}})^{1/4}} \frac{ \sqrt{H_{1}/H_{k}}}{{\mathcal C}(g_{s}, g_{\rho})} \, \, \biggl(\frac{H_{r}}{H_{1}}\biggr)^{- \frac{\delta -1}{2(\delta + 1)}}
\,\, e^{N_{k}} = {\mathcal O}(1).
\label{EFOL2}
\end{equation}
where the factor ${\mathcal C}(g_{s}, g_{\rho})= \bigl(g_{s,\, eq}/g_{s,\, r}\bigr)^{1/3} \, \bigl(g_{\rho,\, r}/g_{\rho,\,eq}\bigr)^{1/4}$ accounts for the evolution of the relativistic species 
between the onset on the radiation epoch and the equality time\footnote{In particular $g_{\rho}$ is the number of relativistic degrees of freedom in the plasma while $g_{s}$ denotes the effective number of relativistic degrees of freedom appearing in the entropy density. In the simplest situation we would have $g_{s,\, r}= g_{\rho,\, r} = 106.75$ (corresponding to the degrees of freedom of the standard model of particle interactions) and $g_{s,\, eq}= g_{\rho,\, eq} = 3.94$; this means that ${\mathcal C}(g_{s}, \, g_{\rho})= 0.75$.}. From Eq. (\ref{EFOL2}) 
we can then obtain the number of $e$-folds elapsed since the crossing 
of the CMB scales $k = {\mathcal O}(k_{p})$ and the result is:
\begin{eqnarray}
N_{k} &=& 59.5  + \frac{1}{4} \ln{\biggl(\frac{r_{T}}{0.03}\biggr)} +\frac{1}{4} \ln{\biggl(\frac{{\mathcal A}_{{\mathcal R}}}{2.41\times 10^{-9}}\biggr)}  +   \frac{1}{4} \ln{\biggl(\frac{h_{0}^2 \, \Omega_{R0}}{4.15\times 10^{-5}}\biggr)}
 \nonumber\\
&+&  \ln{{\mathcal C}(g_{s}, \, g_{\rho})} - \ln{\bigl(k/k_{p}\bigr)} + \frac{(\delta -1)}{2(\delta + 1)} \ln{(H_{r}/H_{1})}.
\label{EFOL3}
\end{eqnarray}
After careful inspection of the various terms determining $N_{k}$, the last 
contribution appearing in Eq. (\ref{EFOL3}) is potentially the largest one. 
The swiftest numerical estimate follows by setting $k \to k_{p}$
and by considering all the other parameters fixed to their typical values.
The expansion scale of big bang 
nucleosynthesis (BBN in what follows) is approximately $H_{bbn} = {\mathcal O}(10^{-44}) \, M_{P}$ whereas 
the inflationary expansion rate is $ H_{k} \simeq H_{1} \simeq \sqrt{ \pi {\mathcal A}_{{\mathcal R}} \, r_{T}}/4$, as explained in Eq. (\ref{INF6}). With these caveats Eq. (\ref{EFOL3}) suggests: 
\begin{equation}
N_{k} = {\mathcal O}(60) - {\mathcal O}(45) \biggl(\frac{\delta -1}{\delta+1}\biggr),
\label{EFOL4}
\end{equation}
where we have chosen the lowest possible value for $H_{r}$, i.e. $H_{r} = {\mathcal O}(H_{bbn})$. 

Let us now consider few concrete situations. Assuming the sources for the evolution of the geometry are given by perfect fluids (with barotropic equation of state) we have  $\delta =2/(3 w +1)$ and  $\delta_{max} \to 2$ since $w \to 0$ is the minimal value of $w$. In this case, from Eq. (\ref{EFOL4}) $N_{k} = {\mathcal O}(60)  - {\mathcal O}(15)= {\mathcal O}(45)$. Conversely, when $\delta < 1$ the post-inflationary expansion rate between $H$ and $H_{r}$ is instead slower than radiation and $N_{k} > {\mathcal O}(60)$. Again, if the background is driven by 
perfect  fluids, $\delta_{min} =1/2$ and it corresponds to a stiff plasma $w\to 1$ where the sound speed and the speed of light coincide. In this second case 
 Eq. (\ref{EFOL4}) implies that $N_{k} = {\mathcal O}(60) + {\mathcal O}(15)$. All in all, from the 
results of Eqs. (\ref{EFOL2})--(\ref{EFOL3})  we can conclude that
\begin{equation}
{\mathcal O}(45) \leq N_{k} \leq {\mathcal O}(75).
\label{EFOL4a}
\end{equation}
The same reasoning that 
leads to Eq. (\ref{EFOL4a}) can be repeated for different classes of sources driving the background evolution. For instance the post-inflationary expansion rate can correspond to a stage dominated by 
an oscillating scalar field with an approximate potential $\varphi^{2 q}$ near 
the origin \cite{ELEVENa,ELEVENb,ELEVENc}. Grossly speaking this is 
the dynamical framework suggested by the class of potentials already discussed in Eqs. (\ref{INF8})--(\ref{INF9}). As we are going to see in sec. \ref{sec4}, also in this situation the results 
summarized by Eq. (\ref{EFOL4a}) apply. The indetermination affecting $N_{k}$ also influences $N_{c}$, i.e. the critical number of $e$-folds required to fit inside the current Hubble patch the redshifted value of the event horizon at the onset of inflation. If $H_{b} \simeq H_{1}$ is the Hubble rate at the onset of inflation,  $N_{c}$ follows from the requirement  $H_{b}^{-1} (a_{0}/a_{b}) \simeq H_{0}^{-1}$ measuring the maximal number of $e$-folds presently accessible to large-scale observations ($N_{max}$ in what follows) \cite{NINE6}. The value of $N_{c}$ is determined by fitting the (redshifted) inflationary event horizon inside the current Hubble patch, i.e. $H_{b}^{-1} (a_{0}/a_{i}) \simeq H_{0}^{-1}$ where $H_{b}$ denotes the expansion rate in the initial stages of inflation.The values of $N_{k}$ and $N_{max}$ are related as:
\begin{equation}
N_{k} = N_{c} - \ln{\biggl(\frac{k}{a_{0}\, H_{0}}\biggr)} -\ln{\frac{H_{k}}{H_{1}}}.
\label{EFOL4b}
\end{equation}

A natural extension of Eq. (\ref{EFOL3}) involves the presence of multiple post-inflationary stages parametrized by different values of the expansion rate conventionally denoted by $\delta_{i}$ with $i =1,\,.\,.\,n$. This is the timeline described 
by the cartoons of Fig. \ref{FIGURE4}; we then have that the last term appearing at the right hand side of Eq. (\ref{EFOL3}) must be replaced by the 
combination 
\begin{equation}
\frac{1}{2}\sum_{i}^{n-1} \, \biggl(\frac{\delta_{i} -1}{\delta_{i} + 1}\biggr) \, \ln{(H_{i+1}/H_{i})},
\label{EFOL4c}
\end{equation}
where, as in Eqs. (\ref{POS4})--(\ref{POS5}), $(H_{i+1}/H_{i})<1$ while the values of the various $\delta_{i}$ can be either smaller or larger 
than $1$. The initial stage of the post-inflationary evolution begins after the end of inflation 
(i.e.  $H \simeq H_{f} = H_{1}$) while the $n$-th stage conventionally coincides with the standard radiation-dominated evolution i.e. $a_{r} = a_{n}$ and $\delta_{n} = 1$. As already explained in connection with Eqs. (\ref{POS4})--(\ref{POS5}),  $H_{r} >10^{-44} \, M_{P}$ implying that the big bang nucleosynthesis takes place when radiation is already dominant. 
During the $i$-th stage of the sequence $a \, H \propto a^{- 1/\delta_{i}}$; 
since, by construction, $a_{n} = a_{r}$ we also have that $H_{n}= H_{r}$; this means that 
$\xi_{n-1} = H_{n}/H_{n-1} = H_{r}/H_{n-1}$. It finally follows  that the 
product of all the $\xi_{i}$ coincides with $H_{r}/H$, namely
\begin{equation}
\prod_{i=1}^{n-1} \xi_{i} = \xi_{1}\,\,\xi_{2}\,\,.\,.\,.\,\,\xi_{n-2} \,\, \xi_{n-1} = \xi_{r} = H_{r}/H < 1.
\label{EFOL4d}
\end{equation}
The final expression of $N_{k}$ can also be written in the following manner
\begin{equation}
N_{k} = \overline{N}_{k} + \Delta N_{k}, \qquad  \Delta N_{k} \equiv \Delta N_{k}(\delta_{i}, H_{1},H_{r}) = \frac{1}{2}\sum_{i}^{n-1} \, \biggl(\frac{\delta_{i} -1}{\delta_{i} + 1}\biggr) \, \ln{(H_{i+1}/H_{i})},
\label{EFOL4e}
\end{equation}
where $\overline{N}_{k}$ is the number of $e$-folds elapsed since the crossing of the CMB scales when the post-inflationary evolution is purely dominated by radiation (see Fig. \ref{FIGURE1}) while $\Delta N_{k}$ indicates the contribution 
of the post-inflationary evolution when the plasma is not dominated by radiation. By definition in a radiation stage (i.e. when $\delta_{i} =1$ for all the $i$)  $\Delta N_{k}$ vanishes. 
As a final remark the transition to radiation is the period where the entropy and the radiation is produced. Since it takes place when all the large-scale inhomogeneities of observational 
interest are outside the horizon the different approaches to the reheating dynamics are not expected to affect the large-scale power spectra \cite{ONEa,ONEb}.  While it is true that  
energy is probably lost during the transition to radiation,  in the case 
of single-field inflationary models this might be, after all, not such a bad approximation since 
the combined action of the reheating and of the preheating dynamics leads to 
a process that is almost sudden. This is the approximation where the various observational 
collaborations estimate $\overline{N}_{k}$ \cite{FIVEa,FIVEb,FIVEc,FIVEd,FIVEe}.
 
 \subsection{Potential-independent suppression of $r_{T}$}
Since the values of $N_{k}$ are determined by the post-inflationary evolution, as anticipated the suppression of $r_{T}(k)$  for $k = {\mathcal O}(k_{p})$ does not uniquely depend upon the inflaton potential. This is the theme addressed in the remaining part of this section. Let us first remark that the general expression of $N_{k}$ (see Eqs. (\ref{INF4})--(\ref{INF4a}))  can also be phrased directly in terms of the inflaton potential as
\begin{eqnarray}
N_{k} = \frac{1}{\overline{M}_{P}^2} \int_{\varphi_{e}}^{\varphi_{k}} \biggl(\frac{V}{V_{,\, \varphi}}\biggr) \,\,d \varphi \equiv \int_{\overline{\varphi}_{e}}^{\overline{\varphi}_{k}} \biggl(\frac{v}{v_{,\,\overline{\varphi}}}\biggr) \,\, d \overline{\varphi},
\label{NEF1}
\end{eqnarray}
where $\varphi_{k} = \varphi(\tau_{k})$ and $\overline{\varphi}_{k} = \overline{\varphi}(\tau_{k})$ 
are the values of the field when the scale $k$ crosses the comoving Hubble radius while the second expression refers to the parametrizations of the potential already introduced in Eqs. (\ref{INF8})--(\ref{INF9}).  As pointed out before, also in Eq. (\ref{NEF1}) $\varphi_{e}$ refers 
to the end of inflation [see Eq. (\ref{INF3}) and discussion thereafter];  moreover the requirement $\epsilon_{e} =1$ demands $\dot{\varphi}_{e}^2 = 2 \,\epsilon_{e} \, V_{e}/[3 - \epsilon_{e}]$. With these specifications Eq. (\ref{NEF1}) unambiguously determines the connection between $\overline{\varphi}_{k}$ and $N_{k}$.  Since for $\tau\to \tau_{k}$ also the slow-roll parameters of Eq. (\ref{INF3}) depend upon the powers of $\overline{\varphi}_{k}$, we can accurately obtain the 
scaling of the slow-roll parameters and of $r_{T}(k) = r_{T}(k,\tau_{k})$ with the 
number of $e$-folds $N_{k}$ (see, in this respect, Eq. (\ref{INF7})). Depending upon the properties 
of the potentials the slow-roll parameters and the tensor to scalar ratio 
exhibit different scalings as a function of the number of $e$-folds elapsed since 
the crossing of the CMB scales. For instance, in the case of the monomial 
potentials \cite{SIXb,SIXc} (see also \cite{TWELVEa,TWELVEb}) it is well known 
that, barring for some numerical factors, $\epsilon_{k}$, $\overline{\eta}_{k}$ and 
$r_{T}(k)$ all scale as $1/N_{k}$. If we now recall Eq. (\ref{EFOL4e}) we see 
that 
\begin{equation}
r_{T}(k) = 16 \epsilon_{k} = {\mathcal O}\biggl(\frac{1}{\overline{N}_{k} + \Delta N_{k}}\biggr),
\label{SCAL1}
\end{equation}
where $\overline{N}_{k}$ is the number of $e$-folds computed in the conventional 
(radiation-dominated) timeline of Fig. \ref{FIGURE1} while 
$\Delta N_{k} = \Delta N_{k}(\delta_{i}, H_{1}, H_{r})$ is the contribution 
of the post-inflationary stage. If the expansion rate after inflation is, in average, 
faster than radiation we generically have $\Delta N_{k} < 0$ so that 
$r_{T}(k)$ is {\em less} suppressed than in the conventional situation where the post-inflationary timeline is dominated by radiation\footnote{This effect has been actually used 
in Refs. \cite{SIXb,SIXc} to address the Bicep-2 anomaly \cite{SIXd} in the 
framework of the monomial potentials.}.  In the case of plateau-like potentials, always 
barring for the numerical factors, $r_{T}(k) = {\mathcal O}(1/N_{k}^2)$. 
More complicated scalings can be envisaged in the case of hill-top potentials \cite{TWELVEc,TWELVEd,TWELVEe,TWELVEf,TWELVEg}. In all these 
situations, as suggested in Eq. (\ref{SCAL1}), the value of $N_{k}$ depends upon $\Delta N_{k}$ which is positive (negative) when the post-inflationary evolution, in an averaged sense, is slower (faster) than radiation. Let us finally consider, for the sake of concreteness, the examples of Eq. (\ref{INF9}) where the potentials scale as $1/N_{k}^2$; more specifically 
for $v_{1}(\overline{\varphi})$ we obtain:
\begin{equation}
r_{T}(k)= \frac{8}{\beta^2 \, (\overline{N}_{k} + \Delta N_{k})^2}, \qquad \overline{\eta}_{k} 
= - \frac{1}{(\overline{N}_{k} + \Delta N_{k})}.
\label{SCAL2}
\end{equation}
In Eq. (\ref{SCAL2}) the suppression does not only 
depend upon the shape and upon the parameters of the potential but also upon the post-inflationary evolution encoded in $\Delta N_{k}$. We also remark that $\overline{\eta}_{k}$ does not scale as $1/N_{k}^2$ and this is exactly what it is needed to reproduce the adiabatic spectral index [ see Eqs. (\ref{INF6}) and (\ref{NAD8g})]; more specifically we have
\begin{equation}
n_{s}(k) = 1 - \frac{3}{\beta^2 \, (\overline{N}_{k} + \Delta N_{k})^2} - \frac{2}{(\overline{N}_{k} + \Delta N_{k})}.
\label{SCAL3}
\end{equation}
As already mentioned, according to current data, the upper limits on $r_{T}(k)$ range between ${\mathcal O}(0.06)$ and ${\mathcal O}(0.03)$ while the spectral index is 
accurately determined in the range $0.9644(8) < \overline{n}_{s} < 0.9653(2)$ with a central value corresponding to $0.9649$ \cite{FIVEa,FIVEb,FIVEc,FIVEd,FIVEe}. 
The potentials $v_{1}(\overline{\varphi})$ and $v_{2}(\overline{\varphi})$ mentioned 
in Eq. (\ref{INF9}) are consistent with the current limits on $r_{T}(k)$ and on $n_{s}(k)$ 
\cite{SIXa}. In the presence of a post-inflationary stage expanding at a rate slower 
than radiation they lead to value of $r_{T}(k)$ which is even smaller 
than the one obtained in the conventional situation. In this way the low frequency gravitons 
are even more invisible than in the conventional situation.
Let us finally move to the second potential appearing in Eq. (\ref{INF9});  
for the potential $v_{2}(\overline{\varphi})$ the slow-roll parameters are then given by
\begin{equation}
\epsilon_{k} = \frac{ 2 q^2 \, \beta^{- \frac{2 p}{p + 2 q}}}{[ 4 q ( p + 2 q) (\overline{N}_{k} + \Delta N_{k})/p]^{\frac{p + 4 q}{p + 2 q}}}, \qquad  \overline{\eta}_{k} = - \frac{ p + 4 q}{2 ( p + 2 q) \, (\overline{N}_{k} + \Delta N_{k})}.
\label{SCAL4}
\end{equation}
Once more, from Eq. (\ref{SCAL4}) we obtain 
\begin{eqnarray}
r_{T}(k) &=& \frac{32 \, q^2 \,  \beta^{ - 2/(1+ 2q/p)}}{[ 4\, q\, ( p + 2 q)\, (\overline{N}_{k} + \Delta N_{k})/p]^{(p+ 4 q)/(p+ 2 q)}}, 
\label{SCAL5}\\
n_{s}(k) &=& 1 - \frac{12 q^2 \, \beta^{ - 2/(1+ 2q/p)}}{[ 4\, q\, ( p + 2 q)\, (\overline{N}_{k} + \Delta N_{k})/p]^{(p+ 4 q)/(p+ 2 q)}}- \frac{p +4 q}{(p + 2\, q) \, (\overline{N}_{k} +\Delta N_{k})}.
\label{SCAL6}
\end{eqnarray}
Since the CMB scales $k = {\mathcal O}(k_{p})$ correspond to $\nu = {\mathcal O}(\mathrm{aHz})$ of the graviton spectrum
we can also argue that, depending upon the values of $\Delta N_{k}$, the spectral 
energy density of the gravitons (see sec. \ref{sec5}) is comparatively more suppressed at low frequencies when $\Delta N_{k}>0$. In this case the gravitons 
are practically invisible in the low-frequency range but they are very energetic 
for frequencies between the MHz and the THz, as we are going to discuss in 
the first part of sec. \ref{sec5}.

\renewcommand{\theequation}{4.\arabic{equation}}
\setcounter{equation}{0}
\section{The adiabatic and the entropic modes}
\label{sec4}
\subsection{Global fluid approach}
After clarifying the nature of the post-inflationary evolution and its direct 
 connection with the suppression of the tensor to scalar ratio we are now 
 going to compute the non-adiabatic\footnote{Sometimes the non-adiabatic solutions are referred to as isocurvature modes. This is actually a misnomer \cite{TWELVEh} (see also \cite{TWELVEh2}) since it seems to imply the absence 
 of curvature inhomogeneities that are instead present. In some cases this terminology makes more sense when the relativistic inhomogeneities are treated in the synchronous coordinate system that is however known for the presence of spurious gauge modes \cite{TWELVEi,TWELVEl} and that will not be specifically employed in this analysis.} curvature inhomogeneities induced by the ultralight scalar without assuming that the post-inflationary evolution is dominated by radiation as it happens in Fig. \ref{FIGURE1}.
In the large-scale limit the relevant equation follows from Eq. (\ref{PP7}) after neglecting the spatial gradients. Furthermore 
the dominant component of $\delta p_{nad}$ (i.e. the non-adiabatic pressure fluctuations) is deduced from Eqs. (\ref{AINF7})--(\ref{AINF8}) and (\ref{AINF9}) so that the evolution of ${\mathcal R}$ eventually becomes:
\begin{equation}
{\mathcal R}^{\prime} = \frac{{\mathcal H}}{p_{t} + \rho_{t}} (1 + c_{s\,t}^2) \,\,U_{,\, \sigma}\,\, \chi_{\sigma}.
\label{NAD1}
\end{equation}
Equation (\ref{NAD1}) describes the evolution of the curvature inhomogeneities 
induced by a light scalar when the post-inflationary evolution is dominated by a 
generic fluid background with sound speed $c_{s\,t}^2$. The growth rate of ${\mathcal R}$ can be actually estimated from Eq. (\ref{NAD1}) and, up to a numerical factor, this strategy reproduces the results by different analyses where the light scalar evolves in the background provided by the oscillations of the inflaton during a decelerated stage of expansion. Furthermore Eq. (\ref{NAD1}) is not specific of a radiation background and it can be employed  during a decelerated stage evolving either faster or slower than radiation. To deduce the evolution of the homogeneous component of $\sigma$ in the backgrounds discussed in sec. \ref{sec3}, it is also convenient to generalize Eq. (\ref{AINF3}) (valid in a radiation epoch) to the case where the fluid evolution is 
not dominated by radiation. Focussing, for simplicity, on the timelines illustrated in the cartoons 
of Figs. \ref{FIGURE2} and \ref{FIGURE3} the analog of Eq. (\ref{AINF3}) is:
\begin{equation}
\sigma^{\prime} = - \frac{\delta \,a^2}{{\mathcal H} \, (4 \delta + 1)} U_{,\, \sigma}, \qquad {\mathcal H} \, a^{1/\delta} = H\, a^{1 + 1/\delta} = \mathrm{constant}.
\label{NAD2}
\end{equation}
When $\delta \to 1$, Eq. (\ref{NAD2}) reproduces Eq. (\ref{AINF3}); moreover Eq. (\ref{NAD2}) correctly describes 
the evolution that precedes the oscillatory stage and it implies that $\sigma$ remains 
${\mathcal O}(M)$ at least until $H = {\mathcal O}(m)$. When $H > m$ 
the same is true for $\chi_{\sigma}$ whose amplitude 
remains frozen to $H_{k}/(2 \pi)$. Equation (\ref{NAD2}) can then be easily solved and the result is: 
\begin{equation}
{\mathcal R}(k, a) = {\mathcal R}_{ad}(k) + \frac{{\mathcal B}(\delta)}{\overline{M}_{P}^2} \chi_{\sigma}(k) \, U_{,\,\sigma} \, \biggl(\frac{a}{{\mathcal H}}\biggr)^2, \qquad {\mathcal B}(\delta) = \frac{\delta}{6 (\delta +1)}.
\label{NAD3}
\end{equation}
In Eq. (\ref{NAD3}) the integration constant corresponds to ${\mathcal R}_{ad}(k)$ since ${\mathcal R}_{ad}^{\prime} =0$ for typical wavelengths larger than the Hubble radius. The second contribution in Eq. (\ref{NAD3}) (also indicated in what follows by ${\mathcal R}_{nad}(k, a)$) is the non-adiabatic term induced by $\delta p_{nad}$. The 
result of Eq. (\ref{NAD3}) follows from Eq. (\ref{NAD2}) after changing the integration variable from $\tau$ to $a$ and by taking into account that ${\mathcal H}\, a^{1/\delta}$ is constant for $H >m$. In the limit $\delta \to 1$ Eq. (\ref{NAD3}) reproduces the result valid in the radiation epch.

\subsection{Interacting scalars and their fluctuations}
The strategy leading to Eq. (\ref{NAD3}) is also appropriate when the 
 ultralight scalar evolves in the background provided by another heavier 
 scalar field $\varphi$ that coincides, in the present case, with the  inflaton. To verify this point the same
calculation leading to the results of Eqs. (\ref{NAD2})--(\ref{NAD3}) will now be repeated by directly considering two interacting scalar fields.  We start by remarking that the gauge-invariant result of Eq. (\ref{NAD3}) can be analyzed in the most convenient coordinate system. When the inflaton $\varphi$ and the light scalar $\sigma$ are simultaneously 
present, it is quite practical to consider the gauge where the curvature 
inhomogeneities only depend upon the fluctuations of the two minimally coupled scalars, namely\footnote{A number of specific discussions related to the gauge (\ref{NAD4a}) are reported in the appendix \ref{APPA} [see, in particular Eq. (\ref{APA8}) and the discussions therein].}
\begin{equation}
{\mathcal R} = - \frac{{\mathcal H}(\varphi^{\prime}\, \chi_{\varphi} + \sigma^{\prime} \, \chi_{\sigma})}{\varphi^{\prime \, 2} + \sigma^{\prime\, 2}}, \quad \chi_{\varphi} = \delta_{s}^{(1)} \varphi, \quad \chi_{\sigma} = \delta_{s}^{(1)}\sigma,
\label{NAD4a}
\end{equation}
where $\delta_{s}^{(1)}$ denotes the first-order scalar fluctuation of the corresponding 
quantity. As shown in appendix \ref{APPA}, in the coordinate 
system leading to Eq. (\ref{NAD4a}) the dynamics of the inhomogeneities can be effectively reduced to a pair of coupled equations involving solely $\chi_{\varphi}$ and $\chi_{\sigma}$ while the explicit dependence upon the metric fluctuations is formally eliminated. Barring 
for some relevant technical detail reported in appendix \ref{APPA} (see, in particular, Eqs. (\ref{APA9}) and (\ref{APA12})--(\ref{APA16})), the dynamics of $\chi_{\varphi}(\vec{x},\tau)$ and $\chi_{\sigma}(\vec{x}, \tau)$ is determined from the following system:
\begin{eqnarray}
 &&\chi_{\varphi}'' + 2 {\mathcal H} \chi_{\varphi}' - \nabla^2 \chi_{\varphi} + {\mathcal Q}_{\varphi\varphi} \chi_{\varphi} + 
  {\mathcal Q}_{\varphi\sigma} \chi_{\sigma}=0,
 \label{NAD4}\\
  &&\chi_{\sigma}'' + 2 {\mathcal H} \chi_{\sigma}' - \nabla^2 \chi_{\sigma} + {\mathcal Q}_{\sigma\sigma} \chi_{\sigma} + 
  {\mathcal Q}_{\sigma\varphi} \chi_{\varphi}=0,
  \label{NAD5}
 \end{eqnarray}
where the three coefficients ${\mathcal Q}_{\varphi\varphi}$, ${\mathcal Q}_{\sigma\sigma}$ and ${\mathcal Q}_{\varphi\sigma} = {\mathcal Q}_{\sigma\varphi}$ only depend on the background evolution and they are:
\begin{eqnarray}
&& {\mathcal Q}_{\varphi\varphi} = a^2 V_{,\,\varphi\varphi} + \ell_{P}^2\, \biggl[ 2 \, a^2\, V_{,\,\varphi}\, (\varphi^{\prime}/{\mathcal H}) + \varphi^{\prime\,2} \bigl(2 +  {\mathcal H}^{\prime}/{\mathcal H}^2 \bigr)\biggr],
\label{NAD6}\\
&&   {\mathcal Q}_{\sigma\sigma} = a^2 U_{,\,\sigma\sigma} + \ell_{P}^2\, \biggl[ 2 \, a^2\, 
U_{,\,\sigma} \, (\sigma^{\prime}/{\mathcal H}) 
 + \sigma^{\prime\, 2}\,\bigl(2 +  {\mathcal H}^{\prime}/{\mathcal H}^2 \bigr)\biggr],
\label{NAD7}\\
&& {\mathcal Q}_{\varphi\sigma}= 
\ell_{P}^2\, \biggl[  a^2\, U_{,\,\sigma}\,(\varphi^{\prime}/{\mathcal H}) 
 +  a^2\, V_{,\,\varphi} \bigl(\sigma^{\prime}/{\mathcal H}\bigr)
 + \varphi^{\prime} \sigma^{\prime} \bigl(2 +  {\mathcal H}^{\prime}/{\mathcal H}^2 \bigr)\biggr].
\label{NAD8}
\end{eqnarray}
During inflation (corresponding 
to the leftmost region of Figs. \ref{FIGURE1}, \ref{FIGURE2} and \ref{FIGURE3})
 $m/H \simeq {\mathcal O} (10^{-45})$ for 
$m = {\mathcal O}(10^{-22})\, \mathrm{eV}$ and $ \sigma = {\mathcal O}(M)$; furthermore, in the same limit, the scale 
of the potential $M_{inf}$ is, in practice, of the order of $\sqrt{H M_{P}}$. Given these hierarchies the only relevant term in Eq. (\ref{NAD4}) is ${\mathcal Q}_{\varphi\varphi}$ that can be explicitly rewritten as:
\begin{eqnarray}
{\mathcal Q}_{\varphi\varphi} &=& a^2 \, H^2 \biggl[ \frac{V_{,\,\varphi\varphi}}{H^2} + 2 \frac{V_{,\,\varphi}}{H^2\, \overline{M}_{P}^2} \biggl(\frac{\dot{\varphi}}{H}\biggr) + \biggl( 3 + \frac{\dot{H}}{H^2}\biggr) \frac{\dot{\varphi}^2}{H^2 \, \overline{M}_{P}^2}\biggr]
\nonumber\\
&=& a^2 \, H^2 [ 3 \overline{\eta} - 12 \epsilon + 2 ( 3 - \epsilon)]
\label{NAD8a}
\end{eqnarray}
where, as before, the overdot indicates a derivation with respect to the cosmic time coordinate. The second equality 
is Eq. (\ref{NAD8a}) follows from Eqs. (\ref{INF2}) and from the definition of the slow-roll parameters $\epsilon$ and $\eta$ already introduced in Eq. (\ref{INF3}). From Eqs. (\ref{NAD4})--(\ref{NAD5}) and (\ref{NAD8}) also the adiabatic spectrum of the curvature inhomogeneities is readily deduced; to avoid digressions this analysis is reported in the final part of appendix \ref{APPA}.

While during the inflationary stage the potential of $V(\varphi)$ dominates and the Universe accelerates, the limit $V\to 0$ is directly comparable with the result of Eq. (\ref{NAD3}) since, in this case,  ${\mathcal H}^{\prime} \to - 2 {\mathcal H}^2$ and the background geometry decelerates. So, for $V\to 0$  Eq. (\ref{NAD6}) demands ${\mathcal Q}_{\varphi\varphi}\to 0$ while Eqs. (\ref{NAD7})--(\ref{NAD8}) imply, respectively:
\begin{eqnarray}
{\mathcal Q}_{\sigma\sigma} = a^2 U_{,\,\sigma\sigma} + 2 \ell_{P}^2 a^2 U_{,\,\sigma} \, (\sigma^{\prime}/{\mathcal H}), \qquad
{\mathcal Q}_{\varphi\sigma} =\ell_{P}^2\, a^2\, U_{,\,\sigma}\,(\varphi^{\prime}/{\mathcal H}).
\label{NAD910}
\end{eqnarray}
The second contributions to ${\mathcal Q}_{\sigma\sigma}$ is negligible; recalling in fact Eq. (\ref{NAD2}) 
we have that $\sigma^{\prime}$ is proportional to $U_{,\,\sigma}$ and, for this reason, 
the second contribution is proportional to $(m/M_{P})^2 = {\mathcal O}(10^{-100})$ 
for $m = {\mathcal O}(10^{-22})\, \mathrm{eV}$. The system of Eqs. (\ref{NAD4})--(\ref{NAD5}) 
becomes therefore 
\begin{eqnarray}
&& \chi_{\sigma}^{\prime\prime} + 2 {\mathcal H} \chi_{\sigma}^{\prime} + a^2 U_{,\,\sigma\sigma} - \nabla^2 \chi_{\sigma} =0,
\label{NAD11}\\
&& \chi_{\varphi}^{\prime\prime} + 2 {\mathcal H} \chi_{\varphi}^{\prime} - \nabla^2\chi_{\varphi} + \ell_{P}^2 \, a^2 \, U_{,\,\sigma}
(\varphi^{\prime}/{\mathcal H}) \chi_{\sigma} =0.
\label{NAD12}
\end{eqnarray}
Once the evolution of $\chi_{\varphi}$ and $\chi_{\sigma}$ is determined from Eqs. 
(\ref{NAD11})--(\ref{NAD12}) the curvature inhomogeneities follow from Eq. (\ref{NAD4a}).
To compare the non-adiabatic component deduced in Eq. (\ref{NAD3}) with 
the solution of Eqs. (\ref{NAD11})--(\ref{NAD12}) the Laplacians are negligible  
since Eq. (\ref{NAD3}) applies in the large-scale limit. In this situation 
 $\chi_{\sigma}$ is constant while $\chi_{\varphi}$ is determined 
by direct integration of Eq. (\ref{NAD12}):
\begin{equation}
\chi_{\varphi}(k, a) = - \frac{U_{\,,\sigma} \, (\varphi^{\prime}/{\mathcal H})}{36 \, \overline{M}_{P}^2} \, \chi_{\sigma}(k)\, \frac{a^2}{{\mathcal H}^2}.
\label{NAD14}
\end{equation}
If Eq. (\ref{NAD14}) is inserted into Eq. (\ref{NAD4a}) the non-adiabatic contribution to the curvature inhomogeneities 
is readily obtained and it is: 
\begin{equation}
{\mathcal R}_{nad} = - \frac{{\mathcal H}}{\varphi^{\prime}} \chi_{\varphi} - \frac{{\mathcal H}\, \sigma^{\prime}}{\varphi^{\prime\, 2}} \chi_{\sigma}= \frac{\chi_{\sigma}(k)}{18 \, \overline{M}_{P}^2} U_{\,,\sigma} \frac{a^2}{{\mathcal H}^2},
\label{NAD15}
\end{equation}
where we used that $\varphi^{\prime\, 2} \gg \sigma^{\prime\, 2}$ and also that, in the 
limit $V\to 0$, $\varphi^{\prime\, 2} = 6 \overline{M}_{P}^2 \, {\mathcal H}^2$.
It is relevant to note that both terms appearing in Eq. (\ref{NAD4a}) 
contribute since, according to Eq. (\ref{NAD2}),  $\sigma^{\prime}= a^2 \, U_{,\,\sigma}/(6 \, {\mathcal H})$ for $\delta \to 1/2$.
The result of Eq. (\ref{NAD15}) coincides, as expected, with the second term of Eq. (\ref{NAD3}) when $\delta \to 1/2$ since ${\mathcal B}(1/2)= 1/18$. This means that 
the two approaches have in fact the same physical content.

\subsection{Rate of growth when the oscillations dominate}
After inflation the kinetic and the potential terms are comparable and both relevant for the dynamics while the inhomogeneities of the light scalar $\sigma$ provide the dominant contribution to the non-adiabatic curvature fluctuations. As already suggested in Eq. (\ref{INF8}) the potential of $\varphi$ will be written as  $V(\varphi) = M_{inf}^4 \, v(\overline{\varphi})$ 
but instead of considering the limit $\overline{\varphi} \gg 1$ (where $v(\overline{\varphi}) \to 1$ and $\overline{\varphi} = \varphi/\overline{M}_{P}$) the focus will be on the post-inflationary evolution $v(\overline{\varphi}) = \overline{\varphi}^{2\,q}$. 
From Eq. (\ref{BB2}) $\rho_{\varphi} = \dot{\varphi}^2/2 + V(\varphi)$ and it can be argued that, during the oscillating stage, $\dot{\rho}_{\varphi} \simeq 0$ \cite{ELEVENa,ELEVENb,ELEVENc}. 
Actually the covariant conservation of the energy-momentum tensor of $\varphi$ demands $\dot{\rho}_{\varphi} + 3 H \dot{\varphi}^2 =0$; however, since during oscillations $3 \, H \dot{\varphi}^2 \ll \dot{\rho}_{\varphi}$, we conclude that $\dot{\rho}_{\varphi}\to 0$. 
The constant value of the energy density is then identified with $V_{m} = V(\varphi_{max})$ indicating the maximum of the potential during the oscillatory stage; in this way we obtain $\rho_{\varphi} = V_{m} $ and also $\dot{\varphi}^2 = 2 (V_{m} - V)$. The expansion rate, the potential, the kinetic term are all oscillating but what matters 
for the present ends are the scaling properties of the various dynamical quantities. For this purpose we can smooth out the oscillations by defining the appropriate averaged quantities. For instance the analog of the slow-roll parameter 
now reads, in average,
\begin{equation}
\overline{\epsilon} = -  \frac{\langle \dot{H}\rangle}{\langle H^2\rangle} \equiv \frac{3\,\,\langle \dot{\varphi}^2 \rangle }{\langle \dot{\varphi}^2 \rangle  + 2 \langle V \rangle},
 \label{OSC1}
 \end{equation}
 where the second equality simply follows from the background equations when $\varphi$ dominates;
in Eq. (\ref{OSC1}) $\langle .\,.\,.\,\rangle$  indicates the averaging over the oscillations between 
$0$ and $t_{m}= t_{max}$ (or $\varphi_{m}= \varphi_{max}$) when $V$ coincides with $V_{m}$.
 Since $\dot{\varphi}^2 = 2 (V_{m} - V)$,  it is possible to rewrite Eq. (\ref{OSC1}) as\footnote{Note that if the integration variable is changed from $t$ to $\varphi$ (or from $t/t_{m}$ to $y = \varphi/\varphi_{m}$)
 the second equality in Eq. (\ref{OSC1a}) follows immediately.}:
\begin{equation}
\overline{\epsilon} = 3 \,\,\frac{\int_{0}^{t_{m}}( V_{m} - V) \,\, dt}{2 \int_{0}^{t_{m}}V_{m\,\, d\,t}} \equiv 
\frac{3 \int_{0}^{1} \, d\, x\,\,\sqrt{ 1 - y^{2\,q}}}{ \int_{0}^{1} \,\,dx/\sqrt{ 1 - y^{2\,q}}},
\label{OSC1a}
\end{equation}
where $y= \varphi/\varphi_{m}$. 
The integrals appearing in Eq. (\ref{OSC1a}) can be performed exactly and the overall result is $\overline{\epsilon} = 3\,q/(q+1)$. From the connection between conformal and cosmic time coordinates,  the scaling properties of all the relevant dynamical quantities can be easily derived. Let us first consider ${\mathcal H}$ and 
$\overline{\varphi}^{\prime}$:
\begin{equation}
{\mathcal H} = {\mathcal H}_{\ast} (a/a_{\ast})^{(1- 2q)/(q+1)},\qquad \overline{\varphi}^{\prime} =  \overline{\varphi}_{\ast}^{\prime} \,(a/a_{\ast})^{(1-2q)/(q+1)},
\label{OSC2}
\end{equation}
implying that $\overline{\varphi}^{\prime}/{\mathcal H} = \overline{\varphi}_{\ast}^{\prime}/{\mathcal H}_{\ast}$. Recalling that $a \, H= {\mathcal H}$ we can also obtain $H = H_{\ast} (a/a_{\ast})^{- 3q/(q+1)}$
with $\dot{\varphi}/H = \dot{\varphi}_{\ast}/H_{\ast}$. From the preceding results 
the scaling law of $\overline{\varphi}$ can also be derived and it is given by  $\overline{\varphi} = \overline{\varphi}_{\ast} (a/a_{\ast})^{-3/(q+1)}$; with the same strategy  we obtain
 \begin{equation}
 V_{\,,\varphi} \propto (M_{inf}^{4}/\overline{M}_{P}) (a/a_{1})^{- 3\frac{(2 q -1)}{(q+1)}}, \qquad 
 V_{\,,\varphi\,\varphi} \propto (M_{inf}^{4}/\overline{M}_{P}^2) (a/a_{1})^{- 6\frac{(q -1)}{(q+1)}}.
\label{OSC3}
\end{equation}
The results of Eqs. (\ref{OSC1})--(\ref{OSC1a}) and (\ref{OSC2})--(\ref{OSC3}) are directly corroborated by the numerical integration of the system. It  turns out that the cosmic time coordinate is slightly more convenient, however, after solving numerically the system, we can directly 
plot ${\mathcal H} = a H$ as a function of the scale factor for different values of $q$ to check the validity of the 
results derived above.
\begin{figure}[!ht]
\centering
\includegraphics[width=8cm,height=9cm]{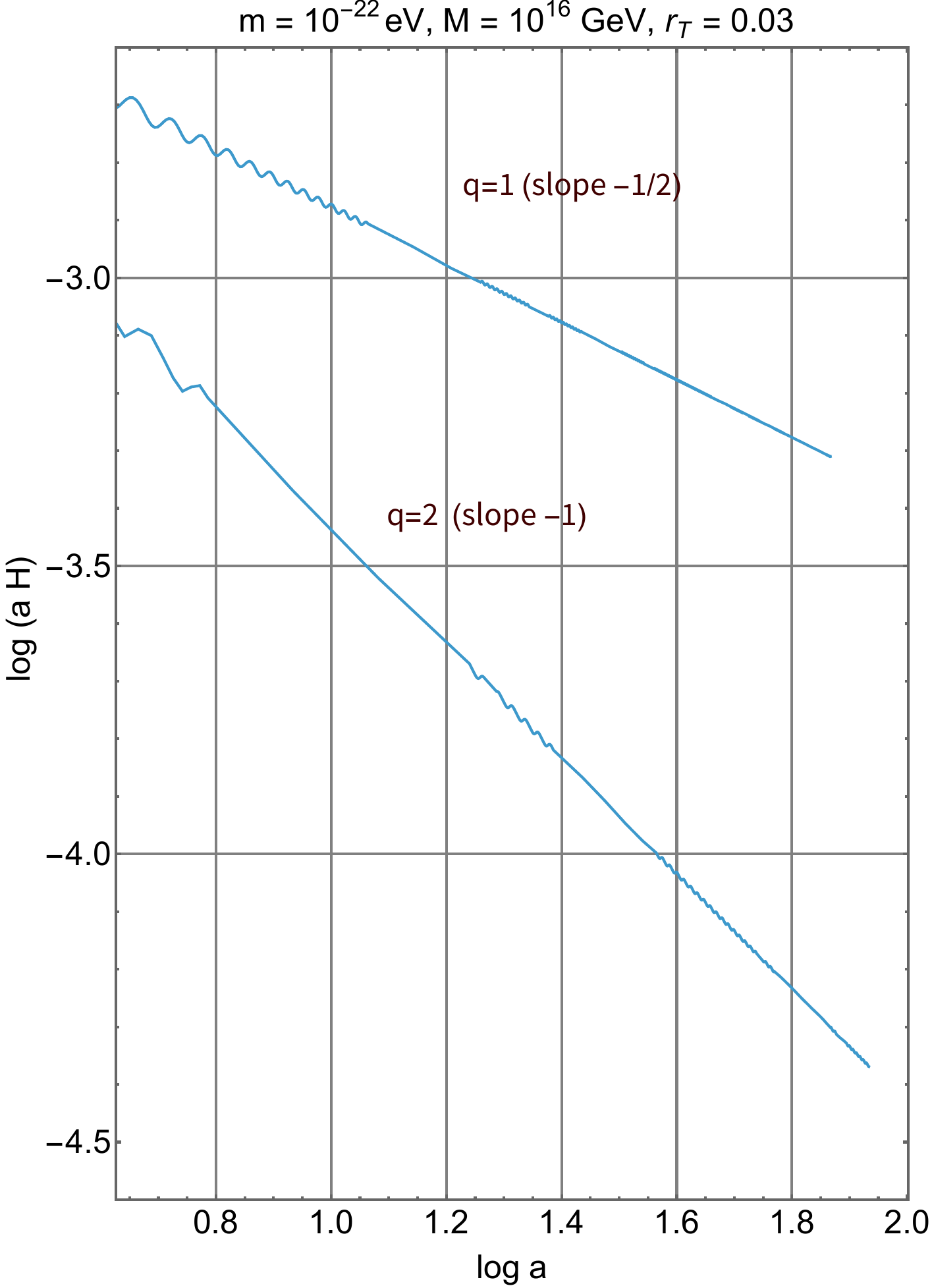}
\includegraphics[width=8cm,height=9cm]{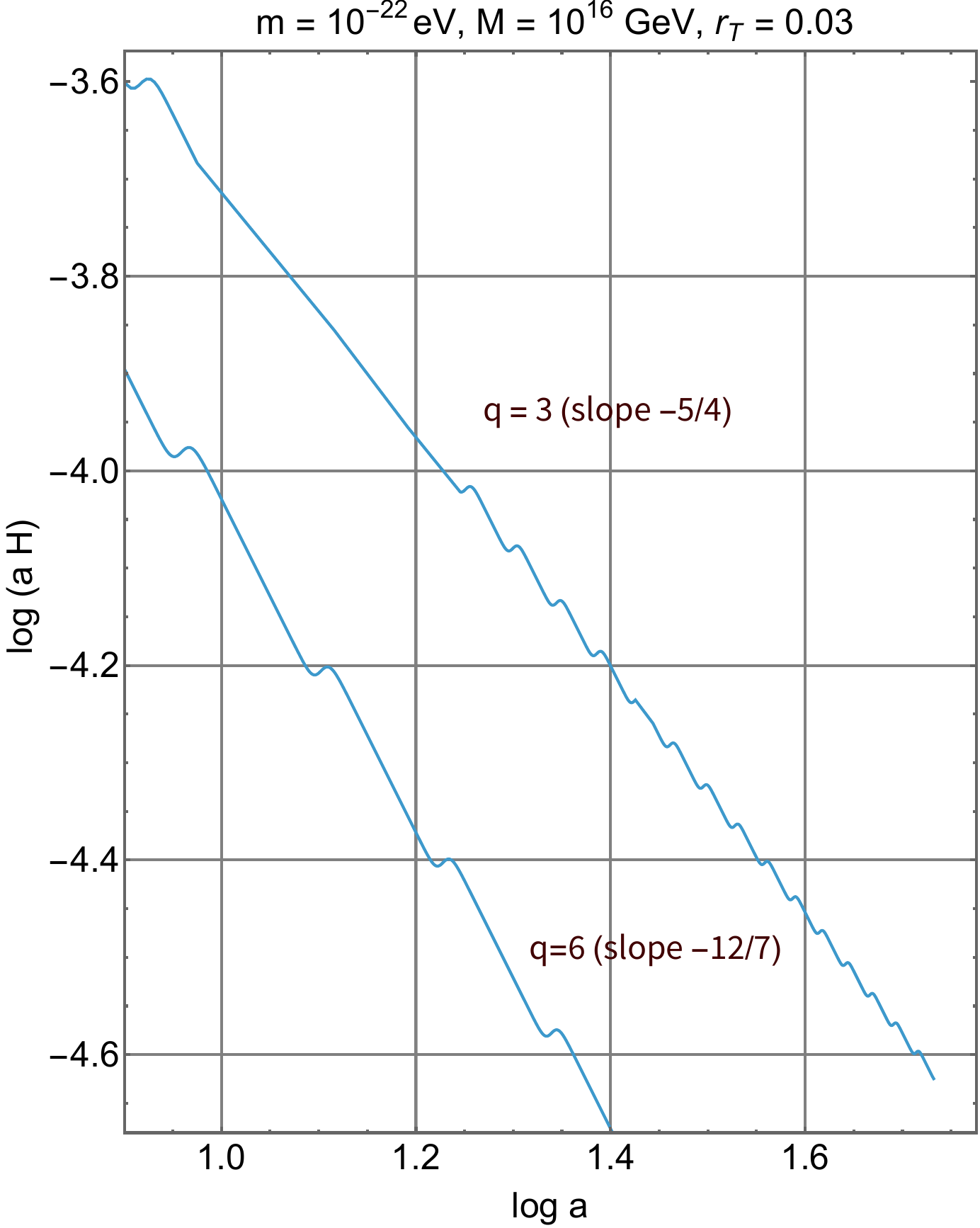}
\caption[a]{The scaling of ${\mathcal H} = a H$ is illustrated for different values of $q$. Common logarithms
are employed on both axes. Barring for the wiggles caused by the oscillations of $\varphi$, the various curves are in fact modulated straight lines with computable slopes that 
follow from the results of Eq. (\ref{OSC2}). The analytic and the numerical results 
coincide: in both plots the analytic values of the slopes slopes have been indicated 
and they perfectly match the numerical result. }
\label{FIGURE5}      
\end{figure} 
In Fig. \ref{FIGURE5} we illustrate this analysis for different values of $q$. The numerical 
results agree with the slopes deduced from Eq. (\ref{OSC2});  the numerical integration 
is sensitive to the values of $r_{T}$ since the tensor to scalar ratio ultimately fixes $M_{inf}$. 
With a similar method the other scalings of Eqs. (\ref{OSC1})--(\ref{OSC2}) can be deduced numerically 
and the obtained results also agree with the analytical expectations. Inserting Eqs. (\ref{OSC1})--(\ref{OSC1a}) and (\ref{OSC2})--(\ref{OSC3}) into Eqs. (\ref{NAD4})--(\ref{NAD5}) and (\ref{NAD6})--(\ref{NAD7}) we can identify the dominant and sub-dominant terms. This step eventually leads to a simpler form of the system since the only terms that count are the ones that are unsuppressed for $a \gg a_{\ast}$. With this strategy the analog 
of Eq. (\ref{NAD12}) now becomes:
\begin{equation}
 \chi_{\varphi}^{\prime\prime} + 2 {\mathcal H} \chi_{\varphi}^{\prime} - \nabla^2\chi_{\varphi} + \ell_{P}^2 \, \biggl[ a^2 \, U_{,\,\sigma}
(\varphi^{\prime}/{\mathcal H}) + ( 2 + {\mathcal H}^{\prime}/{\mathcal H}^2) \varphi^{\prime}\, \sigma^{\prime}\biggr] \chi_{\sigma} =0.
\label{OSC3a}
\end{equation}
The integration of Eq. (\ref{OSC3}) in the large-scale limit leads to a result that is qualitatively similar to (but quantitatively different from) Eq. (\ref{NAD14}). Since 
$\chi_{\sigma}(k,\tau) = \chi_{\sigma}(k) \simeq H_{k}/(2\pi)$, in the large-scale limit Eq. (\ref{OSC3}) becomes:
\begin{equation}
\partial_{\tau} \biggl[ a^2 \, \partial_{\tau} \chi_{\varphi}(k,\tau)\biggr] =  - a^2\,\biggl[ a^2 \, U_{,\,\sigma}
(\varphi^{\prime}/{\mathcal H}) + ( 2 + {\mathcal H}^{\prime}/{\mathcal H}^2) \varphi^{\prime}\, \sigma^{\prime}\biggr] \frac{\chi_{\sigma}(k)}{\overline{M}_{P}^2}.
\label{OSC3b}
\end{equation}
If we integrate twice the previous equations and neglect the integration constant which 
is fixed by the adiabatic solution we have
\begin{equation}
\chi_{\varphi}(k, a) = - \overline{{\mathcal B}}(q)\,\,\frac{U_{,\,\sigma} \, (\varphi^{\prime}/{\mathcal H}^2)}{ \overline{M}_{P}^2} \, \chi_{\sigma}(k)\, \frac{a^2}{{\mathcal H}},\qquad \overline{{\mathcal B}}(q)= \frac{(1+q)^2}{9(2 q+1)^2}.
\label{OSC4}
\end{equation}
We can now use Eq. (\ref{OSC4}) into the expression of the curvature 
inhomogeneities (see Eq. (\ref{NAD4a})) evaluated in the limit $\varphi^{\prime\,2} \gg \sigma^{\prime\, 2}$ 
and the result is:
\begin{equation}
{\mathcal R}_{nad} = \frac{\chi_{\sigma}(k)}{\overline{M}_{P}^2} \, \overline{{\mathcal D}}(q)U_{,\, \sigma} \frac{a^2}{{\mathcal H}^2},\qquad \overline{{\mathcal D}}(q)= \frac{(4 q+1)\, (q+1)^2}{18 \, q\, (2 q +1)^2}.
\label{OSC5}
\end{equation}
As in the case of Eq. (\ref{NAD15}) both terms appearing in Eq. (\ref{NAD4a}) 
contribute to the final result of Eq. (\ref{OSC5}).
Since the physical situation corresponds to $q>1$ we can expand $\overline{{\mathcal D}}(q)$ 
in powers of $1/q$ and eventually obtain
\begin{equation}
\overline{{\mathcal D}}(q) = \frac{1}{18}\biggl[ 1 + \frac{7}{4 q} + {\mathcal O}(q^{-2})\biggr].
\label{OSC6}
\end{equation}
Equation (\ref{OSC6}) reproduces the result of Eq. (\ref{NAD4a}) to leading order in $1/q$; this implies that 
the results obtained with the global fluid approach are accurate to the level ${\mathcal O}(0.1)$ which is more
than sufficient for the actual determination of the constraints simultaneously involving, the non-adiabatic 
mode, the dark-matter implications and the post-inflationary evolution (see, in this respect, sec. \ref{sec5}).

\subsection{Dark matter and constraints from non-adiabatic modes}
From the timeline illustrated in Fig. \ref{FIGURE1} the critical 
density parameter of the $\sigma$ field is given by Eq. (\ref{AINF6}) that shall now be generalized to the situation where the radiation background is not always dominant during the post-inflationary evolution, as already suggested in Figs. \ref{FIGURE2}, \ref{FIGURE3} and \ref{FIGURE4}. In particular 
for a single post-inflationary stage preceding the radiation dominance the result (\ref{AINF6}) becomes\footnote{As in Eq. (\ref{EFOL2}) 
${\mathcal C}(g_{s}, g_{\rho})= \bigl(g_{s,\, eq}/g_{s,\, r}\bigr)^{1/3} \, \bigl(g_{\rho,\, r}/g_{\rho,\,eq}\bigr)^{1/4}$; $g_{\rho}$ is the number of relativistic degrees of freedom in the plasma while $g_{s}$ denotes the effective number of relativistic degrees of freedom associated with in the entropy density. In the standard situation where $g_{s,\, r}= g_{\rho,\, r} = 106.75$ and $g_{s,\, eq}= g_{\rho,\, eq} = 3.94$ we obtain ${\mathcal C}(g_{s}, \, g_{\rho})\to 0.75$. }
\begin{equation}
\Omega_{\sigma}(a_{eq}) = \frac{8 \pi}{3 \, {\mathcal C}(g_{s}, g_{\rho})}\, \biggl(\frac{M}{M_{P}}\biggr)^2 \, \biggl(\frac{H_{m}}{H_{r}}\biggr)^{\frac{3 (1 - \delta)}{2 (\delta +1)}}\, \sqrt{\frac{H_{m}}{H_{r}}},
\label{DM2}
\end{equation}
where, as before, $H_{r}$ denotes the expansion rate at the onset of the radiation stage 
while $H_{m} = {\mathcal O}(m)$.  The result of Eq. (\ref{DM2}) 
depends on the duration of the intermediate post-inflationary stage and on its expansion rate.
From the results of Eqs. (\ref{NAD4a}) and (\ref{OSC5}) the non-adiabatic curvature inhomogeneities can be estimates after equality (but before decoupling). For the timelines of Figs. \ref{FIGURE2} and \ref{FIGURE3}
the evolution of curvature inhomogeneities obeys: 
\begin{equation}
\frac{\partial {\mathcal R}}{\partial \ln{a}} = \frac{8 \pi}{3} \,\biggl(\frac{M}{M_{P}}\biggr)^2 \,
\biggl(\frac{H_{m}}{H_{r}}\biggr)^{\frac{2- \delta}{1 + \delta}} \biggl(\frac{a}{a_{r}}\biggr) \, 
\frac{\chi_{\sigma}(k)}{M}.
\label{DM3}
\end{equation}
We can now integrate once the expression of Eq. (\ref{DM3}) and obtain 
the value of curvature inhomogeneities at equality that is the appropriate 
reference time since the initial conditions of the temperature inhomogeneities 
are set around equality (and before photon decoupling):
\begin{equation}
{\mathcal R}(k, \, a_{eq}) = {\mathcal R}_{ad}(k) + \frac{8 \pi}{3} \,\,{\mathcal C}(g_{\rho}, g_{s})\,\,\biggl(\frac{M}{M_{P}}\biggr)^2 \,\,
\biggl(\frac{H_{m}}{H_{r}}\biggr)^{\frac{3 ( 1- \delta)}{2 ( 1 + \delta)}} \,\,
\sqrt{\frac{H_{m}}{H_{eq}}} \,\, \frac{ \chi_{\sigma}(k)}{M}.
\label{DM4}
\end{equation}
In Eq. (\ref{DM4}) the identity $a_{eq}/a_{r} = {\mathcal C}(g_{\rho}, g_{s}) \sqrt{H_{r}/H_{eq}}$ has been employed. We recall now that $\chi_{\sigma}(k) = H_{k}/(2\pi)$ and this observation ultimately 
implies that the ratio between the non-adiabatic and the adiabatic contributions reads:
\begin{eqnarray}
Z(M,\,m,\,r_{T},\, \delta,\, H_{r}) &=& \frac{{\mathcal R}_{nad}(k, a_{eq})}{{\mathcal R}_{ad}(k)}
\nonumber\\
&=& \frac{\sqrt{ \pi \, r_{T}}}{3} \, \biggl(\frac{M}{M_{P}}\biggr) \, \biggl(\frac{m}{H_{r}}\biggr)^{\frac{3(1-\delta)}{2(\delta +1)}} \, {\mathcal C}(g_{\rho}, g_{s})\, \sqrt{\frac{m}{H_{eq}}}.
\label{DM5}
\end{eqnarray}
Thanks to Eq.(\ref{DM5}) we can also compute the ratio between the corresponding 
power spectra, namely
\begin{equation}
\frac{P_{nad}(k)}{P_{ad}(k)} = Z^2(M,\,m,\,r_{T},\, \delta,\, H_{r}) \biggl(\frac{k}{k_{p}}\biggr)^{n_{{\mathcal R}}},\qquad n_{{\mathcal R}} = 4 \,\epsilon_{k} - 2 \,\overline{\eta}_{k},
\label{DM6}
\end{equation}
The presence of non-adiabatic modes is notoriously constrained by current data; therefore
we must require, in a conservative perspective, that $Z^2(M,\,m,\,r_{T},\, \delta,\, H_{r}) < {\mathcal O}(10^{-2})$ \cite{FIVEa,FIVEb,FIVEc,FIVEd,FIVEe}. We can therefore consider simultaneously the bounds on $\Omega_{\sigma}(a_{eq})$ and on $Z(M,\,m,\,r_{T},\, \delta,\, H_{r})$: this step constrains 
the post-inflationary evolution. Before focussing on the explicit forms of the constraints it is however relevant to examine 
the independent limits coming from the tensor modes of the geometry in the high frequency range.

\renewcommand{\theequation}{5.\arabic{equation}}
\setcounter{equation}{0}
\section{Non-adiabatic spectra, dark matter and tensor modes}
\label{sec5}
The constraints stemming from the non-adiabatic curvature inhomogeneities and from the dark matter fraction are supplemented by the limits associated with the high frequency gravitons that are produced when the post-inflationary evolution is not dominated by radiation. At the present time, the spectral energy density of the relic gravitons in critical units [denoted hereunder by $\Omega_{gw}(\nu,\tau_{0})$] extends between the aHz range and the THz domain. The shape, the slopes and the amplitudes of $\Omega_{gw}(\nu,\tau_{0})$ generally bear the marks of the post-inflationary expansion rate \cite{LIGO3}. When a  conventional inflationary stage is followed by a radiation-dominated epoch $\Omega_{gw}(\nu,\tau_{0})$ is nearly scale-invariant in the high frequency range while at low frequencies it scales as $\nu^{-2}$ \cite{FOURg,FOURh,FOURi,FOURl}. Conversely if the post-inflationary epoch includes a stage whose expansion rate is either faster 
or slower than radiation $\Omega_{gw}(\nu,\tau_{0})$ is, respectively, either a decreasing or an increasing function of the comoving frequency\footnote{Similarly if there are many phases expanding at different rates the spectral energy density contains a sequence of peaks. Finally it is always true that $\Omega_{gw}(\nu,\tau_{0}) \propto r_{T}$ in the aHz region, but at 
higher frequencies the scaling of $\Omega_{gw}(\nu,\tau_{0})$ is more complicated (see, in this respect, the discussion reported after Eq. (\ref{APB6}) in appendix \ref{APPB}).}(see Eqs. (\ref{APB5})--(\ref{APB6}) and discussion therein). In the initial part of this section the focus is on the physical 
bounds related to the tensor modes of the geometry; to avoid lengthy digressions, few relevant considerations have been included in the appendix \ref{APPB}. In the bulk of this section the constraints coming from the non-adiabatic curvature inhomogeneities and  from the critical dark matter fraction are combined with the limits stemming from the relic gravitons. 

\subsection{Low frequency and high frequency spectra of the tensor modes}
The timelines of the post-inflationary evolutions affect the tensor to scalar ratio and the examples illustrated in Figs.  \ref{FIGURE2} and \ref{FIGURE3} (for a given choice of the inflationary potential) lead to values of $r_{T}$ that are larger or smaller than the ones inferred from a decelerated stage dominated by radiation. The bounds associated with $r_{T}(k)$  involve the CMB wavenumbers $k= {\mathcal O}(k_{p})$ and these scales correspond to the low-frequency tail of the graviton spectrum since $\nu_{p} = k_{p}/(2\pi) = {\mathcal O}(3) \, \mathrm{aHz}$ where $1\,\, \mathrm{aHz} = 10^{-18} \, \mathrm{Hz}$.
In the aHz domain the signal of the tensor modes is comparatively smaller when the post-inflationary expansion rate is slower than radiation but the converse is true in the high frequency domain where $\Omega_{gw}(\nu,\tau_{0})$ gets enhanced. There actually exists an absolute limit on the high frequency 
domain of the spectrum implying that the maximal frequency must not exceed the THz domain \cite{EIGHT1}. Indeed, the spectral energy density 
of the produced gravitons can be written, at the present time, as\footnote{Throughout this investigation units $\hbar=c =1$ are employed; in these units $ 2\pi \nu = k$. If we would restore the dependence upon the 
Planck constant we would discover that $\Omega_{gw}(\nu,\tau_{0}) \propto \hbar^2$ 
\cite{THIRTEENb} showing that the quantum mechanical nature of the underlying effect.}
\begin{equation}
\Omega_{gw}(\nu,\tau_{0}) = \frac{1}{\rho_{c}} \frac{d \,\overline{\rho}_{gw}}{d \ln{k}} = 
\frac{128\, \pi^3}{3} \frac{\nu^4}{H_{0}^2 M_{P}^2}\, \overline{n}_{\nu},
\label{TM1}
\end{equation}
where $\rho_{c} = 3 \, H_{0}^2 \, \overline{M}_{P}^2$ is the current value of the 
critical energy density and $\overline{n}_{\nu}$ is the averaged multiplicity 
of the produced gravitons. The amplification of the tensor and scalar modes of the geometry actually corresponds to the production of gravitons \cite{ONEa,ONEb, FOURa,FOURb,FOURc,FOURd,LIGO3} and curvature phonons, as originally argued in Refs. \cite{THIRTEENa1,THIRTEENa2} even before the formulation of the inflationary paradigm. The second equality in Eq. (\ref{TM1})
 follows from the first one since the averaged energy density of the gravitons can be generally expressed 
as $d \overline{\rho}_{gw} = 2 \, k\, \overline{n}_{k}\, d^{3} k/(2\pi)^3$ where 
the factor $2$ accounts for the two tensor polarizations (see e.g. \cite{THIRTEENc}).  
With these specifications the maximal frequency of the spectrum corresponds to the production of a single pair of gravitons with opposite three-momenta. Since
the averaged multiplicity is smoothly suppressed at high frequencies (see appendix \ref{APPB} and discussion therein) we could also require 
that only few pairs of gravitons are produced at $\nu\to \nu_{max}$  (i.e. $\overline{n}_{\nu_{max}} = {\mathcal O}(1)$) and nothing would change in the final estimate. Thus, from Eq. (\ref{TM1}), we have 
\begin{equation}
\Omega_{gw}(\nu_{max},\tau_{0}) = 
\frac{128\, \pi^3}{3} \frac{\nu_{max}^4}{H_{0}^2 M_{P}^2}\, \overline{n}_{\nu_{max}} \to
\frac{128\, \pi^3}{3} \frac{\nu_{max}^4}{H_{0}^2 M_{P}^2}.
\label{TM2}
\end{equation}
In the case of the timeline of Fig. \ref{FIGURE1} where the decelerated timeline is dominated by radiation, the maximal frequency of the spectrum is
\begin{equation}
\overline{\nu}_{max} = \frac{ ( 2 \Omega_{R0})^{1/4}}{2 \pi} {\mathcal C}(g_{\rho}, g_{s}) \, \sqrt{H_{1}H_{0}},
\label{TM3}
\end{equation}
where $\Omega_{R0}$ is the critical fraction in radiation; in the conventional lore, depending 
upon the values of the various parameters, the value of $\overline{\nu}_{max}$ given in Eq. (\ref{TM2}) ranges between $200$ and $300$ MHz. When Eq. (\ref{TM3}) is inserted into Eq. (\ref{TM2})  $h_{0}^2 \Omega_{gw}(\overline{\nu}_{max}, \tau_{0})$ 
becomes
\begin{equation}
h_{0}^2 \Omega_{gw}(\overline{\nu}_{max}) = \frac{r_{T}}{3} \,{\mathcal A}_{{\mathcal R}} \, 
h_{0}^2 \, \Omega_{R\,0}\, {\mathcal C}(g_{\rho}, g_{s}) < {\mathcal O}(10^{-16}),
\label{TM3a}
\end{equation}
where $h_{0}$ is the current indetermination on the Hubble rate and $h_{0}^2 \, \Omega_{R\,0} = 4.15 \, \times 10^{-5}$; note that $H_{1}$ appearing 
in Eq. (\ref{TM3}) has been estimated from $(H_{1}/M_{P}) \simeq (H_{k}/M_{P}) = \sqrt{\pi \, r_{T}\, {\mathcal A}_{\mathcal R}}/4$. Furthermore, the late-time sources of damping (like 
the neutrino free-streaming \cite{FOURTEENa,FOURTEENb,FOURTEENc,FOURTEENd,FOURTEENe}) have been neglected so that the estimate of Eq. (\ref{TM3a}) should be further reduced 
a factor ${\mathcal O}(0.08)$. The result of Eq. (\ref{TM3a}) shows that, in the conventional lore of Fig. \ref{FIGURE1}, $h_{0}^2 \Omega_{gw}(\nu, \tau_{0})$ remains unconstrained for frequencies larger than $\nu_{bbn}$ 
\begin{eqnarray}
\nu_{bbn} &=& 8.17 \times 10^{-33} g_{\rho,\, bbn}^{1/4} \, T_{bbn} \biggl(\frac{h_{0}^2 \Omega_{R0}}{4.15\times 10^{-5}} \biggr)^{1/4}
\nonumber\\
&=& {\mathcal O}(2) \times 10^{-2} \biggl(\frac{g_{\rho,\, bbn}}{10.75}\biggr)^{1/4} \biggl(\frac{T_{bbn}}{\,\,\mathrm{MeV}}\biggr) 
\biggl(\frac{h_{0}^2 \Omega_{R0}}{4.15 \times 10^{-5}}\biggr)^{1/4}\,\,\mathrm{nHz},
\label{TM3b}
\end{eqnarray}
where $g_{\rho,\, bbn}$ is the effective number of relativistic species appearing in the 
energy density at BBN; Eq. (\ref{TM3b}) is approximately of the order of the BBN expansion rate redshifted at the present epoch. In connection with the range $\nu\geq \nu_{bbn}$, we must make sure that the energy density of the produced gravitons does not excessively increase the expansion rate at the time of BBN. It is actually well known that the relativistic species (like the gravitons) speed up the expansion rate at the time when the light nuclei are formed. The abundances of the light elements (particularly of the $^{4}\mathrm{He}$) are actually sensitive to the expansion rate at the BBN time, the presence of extra-relativistic degrees of freedom is strongly constrained through the 
so-called BBN bound (see, e.g. \cite{FIFTEENa,FIFTEENb,FIFTEENc,FIFTEENd,FIFTEENe,FIFTEENf}): 
\begin{equation}
h_{0}^2  \int_{\nu_{bbn}}^{\nu_{max}}
  \Omega_{gw}(\nu,\tau_{0}) d\ln{\nu} = 5.61 \times 10^{-6} \Delta N_{\nu} 
  \biggl(\frac{h_{0}^2 \Omega_{\gamma\,0}}{2.47 \times 10^{-5}}\biggr),
\label{TM4}
\end{equation}
where $\Omega_{\gamma\,0}$ is the critical fraction of photons at the 
present time and $\Delta N_{\nu}$ effectively parametrize the supplementary 
relativistic species that have been traditionally identified with neutrino species. However the supplementary species do not need to be fermionic since also the bosons (like gravitons) 
can increase the expansion at the nucleosynthesis epoch. The integral appearing in Eq. (\ref{TM4}) runs 
over all the frequencies larger than $\nu_{bbn}$; a
comparison between Eqs. (\ref{TM3})--(\ref{TM3a}) and (\ref{TM4}) shows that, in conventional 
lore (see Fig. \ref{FIGURE1} and discussions therein) no supplementary constraints are expected
since $h_{0}^2 \Omega_{gw}(\nu, \tau_{0})$ is always decreasing and smaller 
than ${\mathcal O}(10^{-16})$ in the highest frequency range.

Having discussed the conventional evolution of Fig. \ref{FIGURE1}, we now come the timelines of Figs. \ref{FIGURE2} and \ref{FIGURE3}. In this case the explicit expressions of maximal frequency of the spectrum 
can be written in terms of $\overline{\nu}_{max}$ defined in Eq. (\ref{TM3}):
\begin{equation}
\nu_{max} = (H_{r}/H_{1})^{\frac{\delta -1}{2(\delta+1)}} \, \overline{\nu}_{max}.
\label{TM6}
\end{equation}
Since $H_{r} < H_{1}$,  $\nu_{max}> \overline{\nu}_{max}$ when $\delta< 1$ (i.e. when the post-inflationary expansion rate is slower than radiation); conversely if $\delta> 1$ (and the expansion rate is faster than radiation) $\nu_{max} < \overline{\nu}_{max}$. Moreover, thanks to the results of appendix \ref{APPB}
(see in particular Eqs. (\ref{APB1})--(\ref{APB1b})), the spectral energy density increases with frequency when $\delta <1$ while it decreases 
when $\delta>1$.  This observation implies that when $\delta < 1$ the integral appearing of Eq. (\ref{TM4}) is dominated by $\nu_{max}$ while for $\delta > 1$ the same integral is dominated by $\nu_{bbn}$ and it is automatically negligible in the context of the standard lore. The concrete form of the constraint implied by Eq. (\ref{TM4}) is 
\begin{equation}
h_{0}^2 \Omega_{gw}(\nu_{max}, \tau_{0}) = \frac{16}{3\pi} \,\,h_{0}^2 \Omega_{R0}\,\, {\mathcal C}^4(g_{\rho}, g_{s}) \,\,\biggl(\frac{H_{r}}{M_{P}}\biggr)^{ 2 \frac{(\delta -1)}{\delta+1}}\,\, \biggr(\frac{H_{1}}{M_{P}}\biggl)^{\frac{4}{\delta +1}} < {\mathcal O}(10^{-6}).
\label{TM7}
\end{equation}
The maximal bound on $\nu_{max}$ (implying  $\nu_{max} < {\mathcal O}(\mathrm{THz})$) \cite{EIGHT1} can also be written as
\begin{equation}
\biggl(\frac{H_{r}}{H_{1}}\biggr)^{\frac{\delta-1}{2(\delta +1)}}\overline{\nu}_{max} < {\mathcal O}(10^{12}) \, \mathrm{Hz},
\label{TM8}
\end{equation}
where the value of $\overline{\nu}_{max}$ is explicitly given by:
\begin{equation}
\overline{\nu}_{max} = 271.93\,\, {\mathcal C}(g_{\rho}, g_{s}) \,\,
\biggl(\frac{{\mathcal A}_{{\mathcal R}}}{2.41\times 10^{-9}}\biggr)^{1/4} \, \biggl(\frac{r_{T}}{0.06}\biggr)^{1/4}\, \biggl(\frac{h_{0}^2 \Omega_{R0}}{4.15\times 10^{-5}}\biggr)^{1/4}\, \mathrm{MHz}.
\label{TM9}
\end{equation}
The conditions of Eqs. (\ref{TM7})--(\ref{TM8}) are in fact equivalent and we can  trade one for the other. More accurate strategies to implement the bound (\ref{TM4}) 
are briefly outlined in the appendix \ref{APPB} but they are not essential for the present ends.
\begin{figure}[!ht]
\centering
\includegraphics[height=7cm]{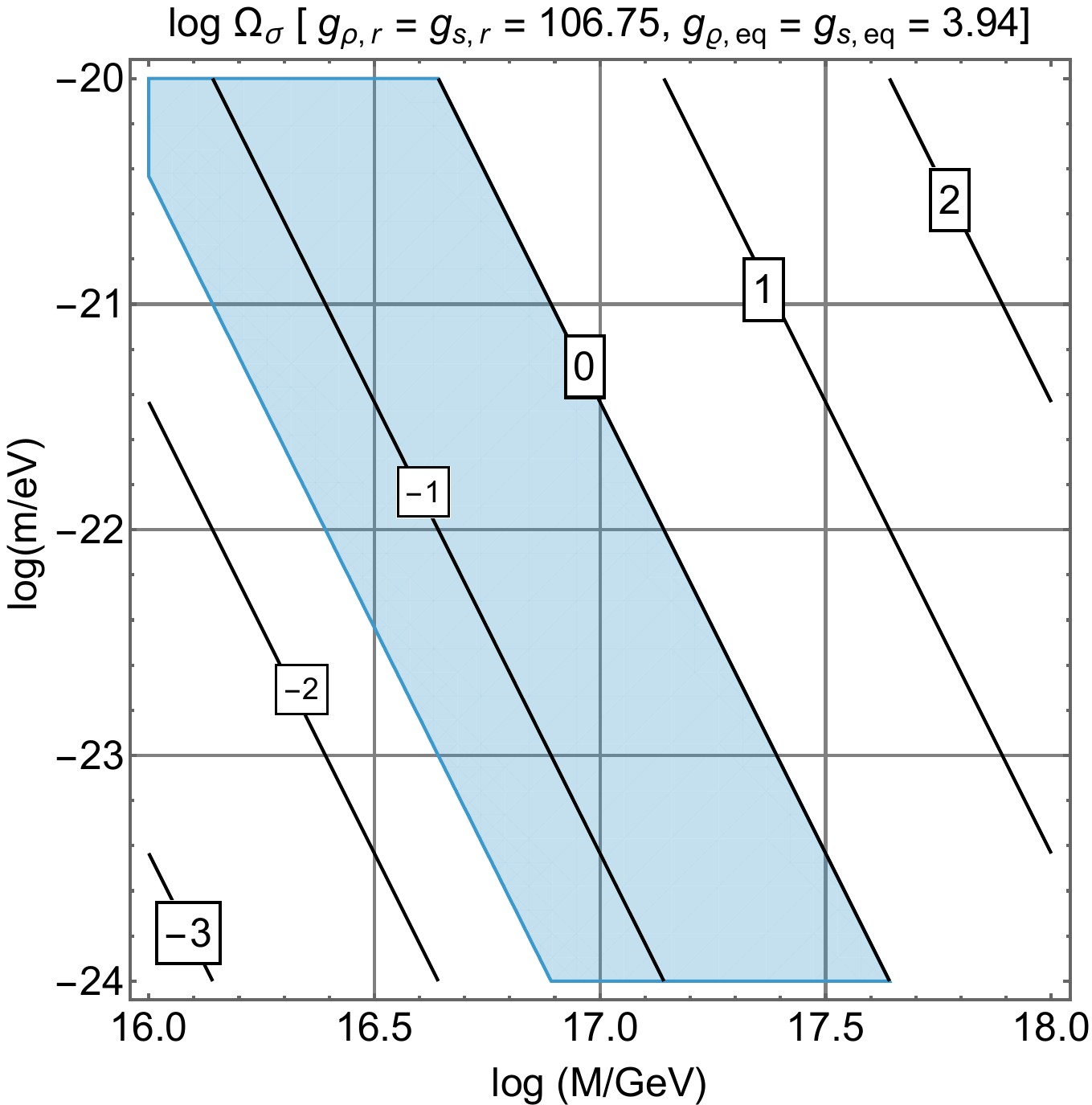}
\includegraphics[height=7cm]{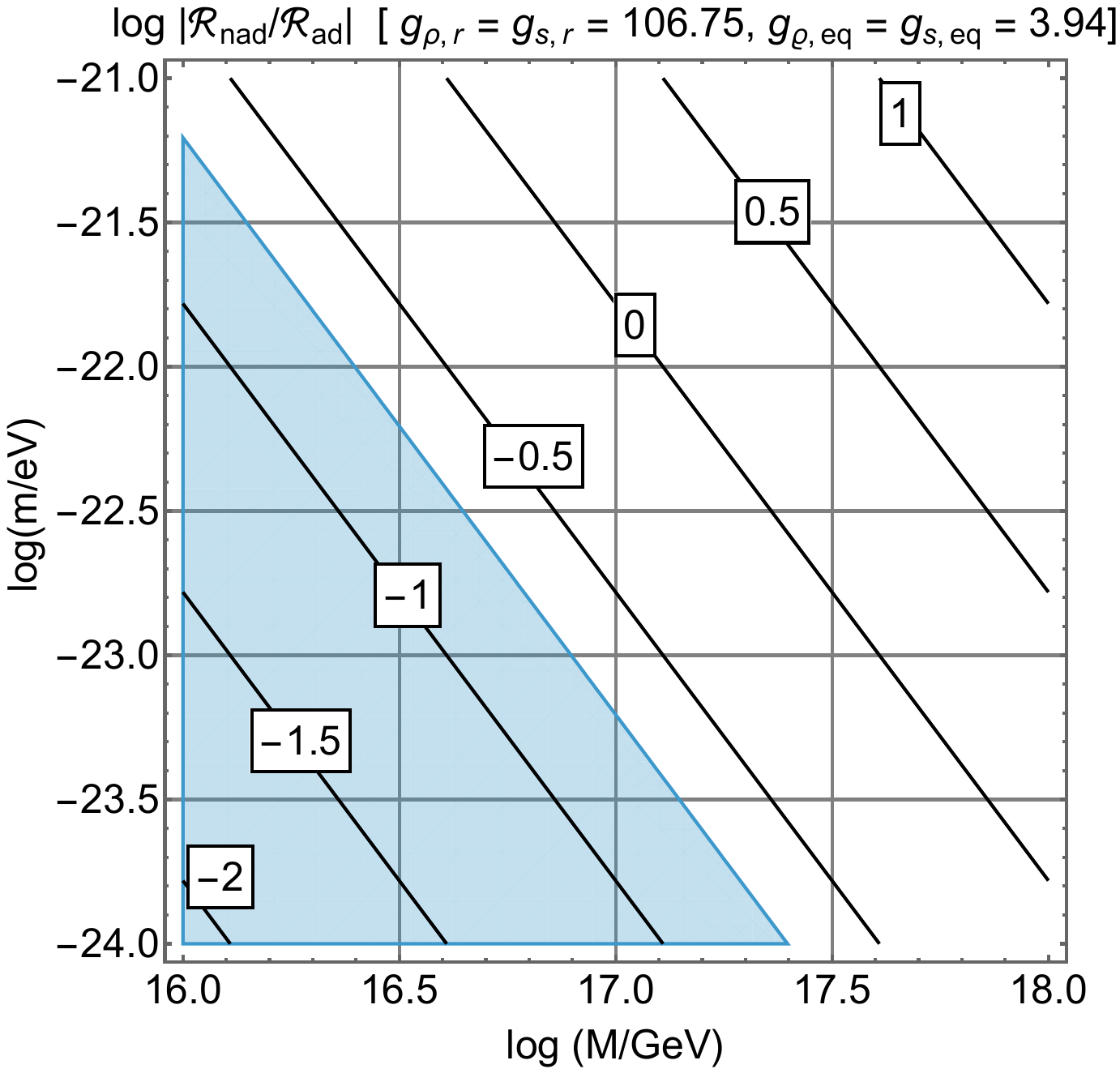}
\caption[a]{The range of variation of $\Omega_{\sigma}$ and of $|{\mathcal R}_{nad}/{\mathcal R}_{ad}|$ is independently illustrated. In the plot at the left the labels correspond to the common logarithms of $\Omega_{\sigma}$; in the right plot the labels correspond instead to the common logarithm of $|{\mathcal R}_{nad}/{\mathcal R}_{ad}|$. The selected value of $r_{T}$ coincides with $0.03$ and it is 
consistent with the actual limits (i.e. $r_{T} < {\mathcal O}(0.034)$).}
\label{FIGNA}      
\end{figure} 
\subsection{Timelines dominated by radiation}
The evolution of the ultralight scalars will now be analyzed in the light 
of the various sets of bounds deduced so far. As before we are going to consider, as a reference 
scenario,  the case of a post-inflationary timeline dominated by radiation (see Fig. \ref{FIGURE1}). In this situation, recalling the results obtained in the previous sections, the critical fraction of the ultralight scalar at the equality epoch 
\begin{equation}
\Omega_{\sigma} = \frac{8\pi}{3}  {\mathcal C}(g_{\rho}, g_{s}) \biggl(\frac{M}{M_{P}}\biggr)^2 \, \sqrt{\frac{m}{H_{eq}}}.
\label{TMR1}
\end{equation}
At the same reference time, the ratio between the non-adiabatic and the adiabatic inhomogeneities is estimated as follows:
\begin{equation}
\biggl|\frac{{\mathcal R}_{nad}}{{\mathcal R}_{ad}}\biggr|= \frac{\sqrt{\pi \, r_{T}}}{8} {\mathcal C}(g_{\rho}, g_{s}) \biggl(\frac{M}{M_{P}}\biggr)^2 \, \sqrt{\frac{m}{H_{eq}}}.
\label{TMR2}
\end{equation}
\begin{figure}[!ht]
\centering
\includegraphics[height=7cm]{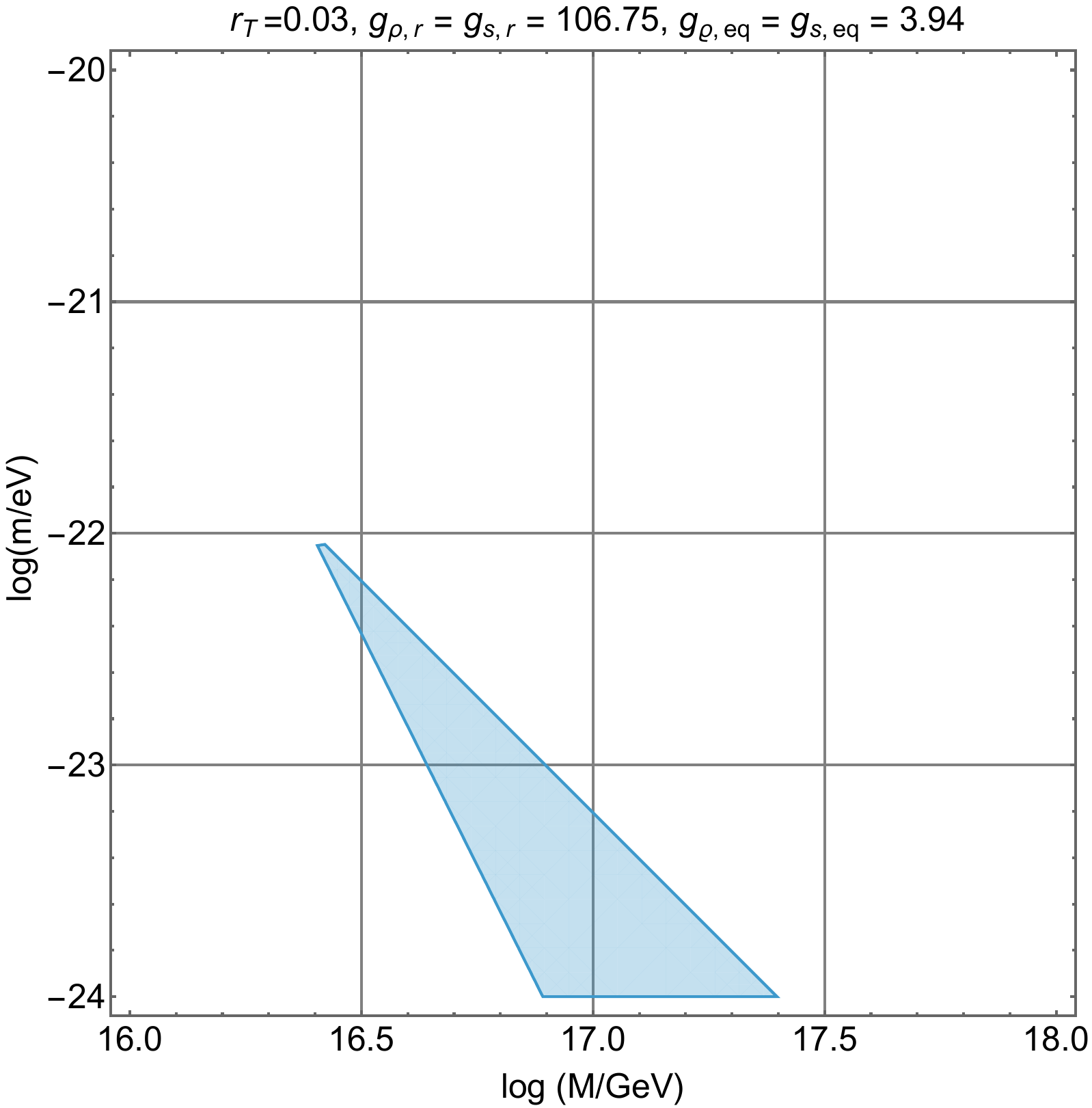}
\includegraphics[height=7cm]{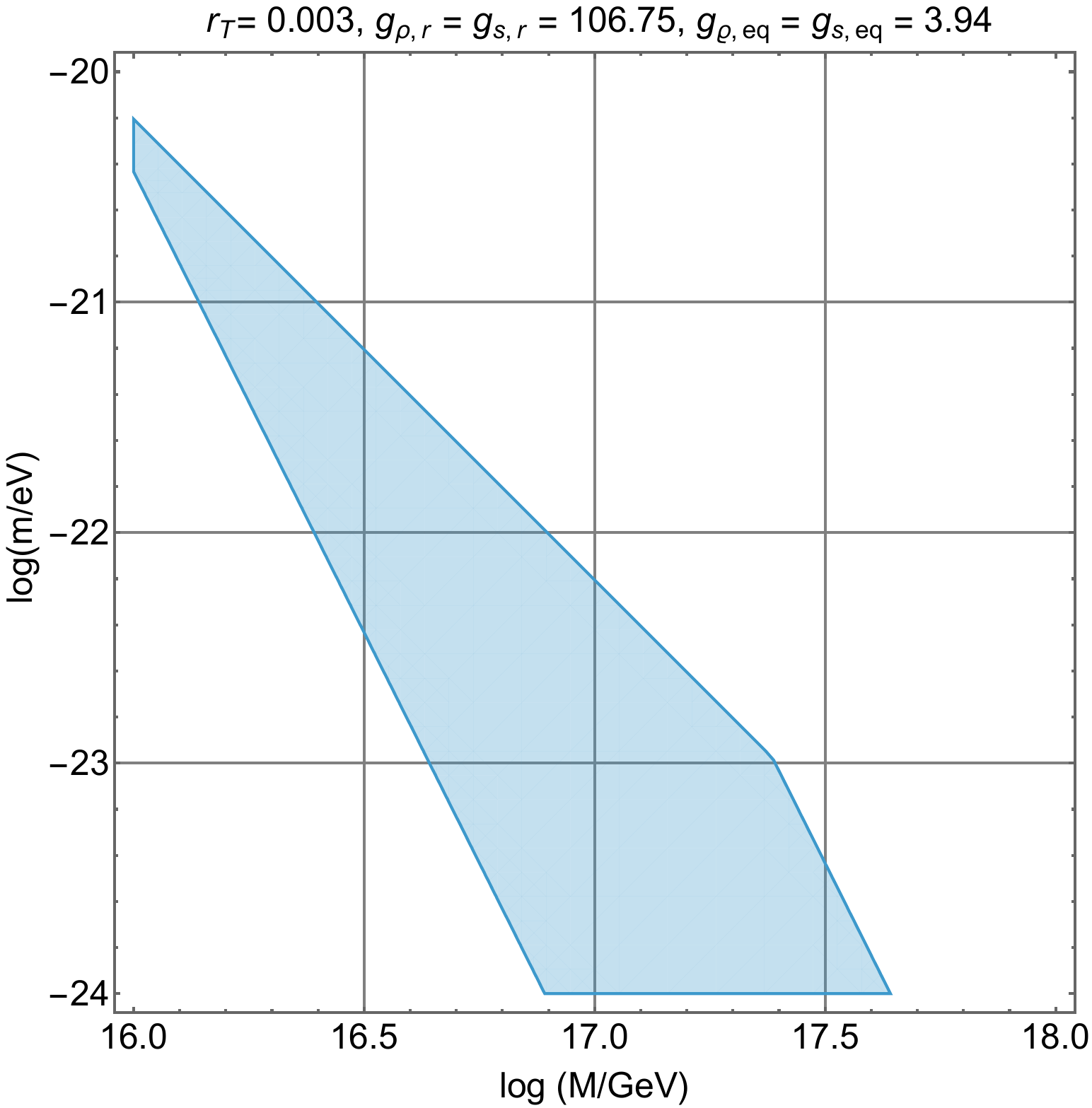}
\caption[a]{We simultaneously illustrate the constraints associated with Eqs. (\ref{TMR1}) and (\ref{TMR2}). 
In the plot at the left  we have chosen $r_{T}= 0.03$ while in the right plot $r_{T} =0.003$. When the values of $r_{T}$ 
decrease the shaded area describing the intersection between the different bounds gets wider.}
\label{FIGNB}      
\end{figure} 
If  Eqs. (\ref{TMR1}) and (\ref{TMR2}) are combined we can obtain a direct relation between $|{\mathcal R}_{nad}/{\mathcal R}_{ad}|$ and
$\Omega_{\sigma}$: 
\begin{equation}
\biggl|\frac{{\mathcal R}_{nad}}{{\mathcal R}_{ad}} \biggr|= \frac{\sqrt{r_{T} \, \pi}}{8} \biggl(\frac{M_{P}}{M}\biggr) \Omega_{\sigma}.
\label{TMR3}
\end{equation}
For a fully uncorrelated non-adiabatic mode in the matter sector the current limits would imply that the ratio of the power spectra should be smaller than ${\mathcal O}(0.038)$ \cite{FIVEa,FIVEb,FIVEc,FIVEd,FIVEe}. We are considering here the direct ratio between the non-adiabatic and the adiabatic components even if other authors prefer to bound the non-adiabatic 
spectra over the total inhomogeneity (i.e. adiabatic and non-adiabatic). If the non-adiabatic and the adiabatic components are correlated the bounds are quantitatively similar and slightly more 
stringent when the two components are anticorrelated. Since Eq. (\ref{TMR3}) involves the amplitudes (i.e. the square roots of the power spectra) we should demand $|{\mathcal R}_{nad}/{\mathcal R}_{ad}| < {\mathcal O}(0.2)$. 
Equation (\ref{TMR3}) also involves $r_{T}$ and this means that there is a further constraint on the right hand side corresponding to the current limit on the tensor to scalar ratio, i.e. $r_{T} < {\mathcal O}(0.034)$. Roughly speaking Eq. (\ref{TMR3}) is compatible with observational data provided $r_{T}$ is sufficiently small. 
\begin{figure}[!ht]
\centering
\includegraphics[height=6cm]{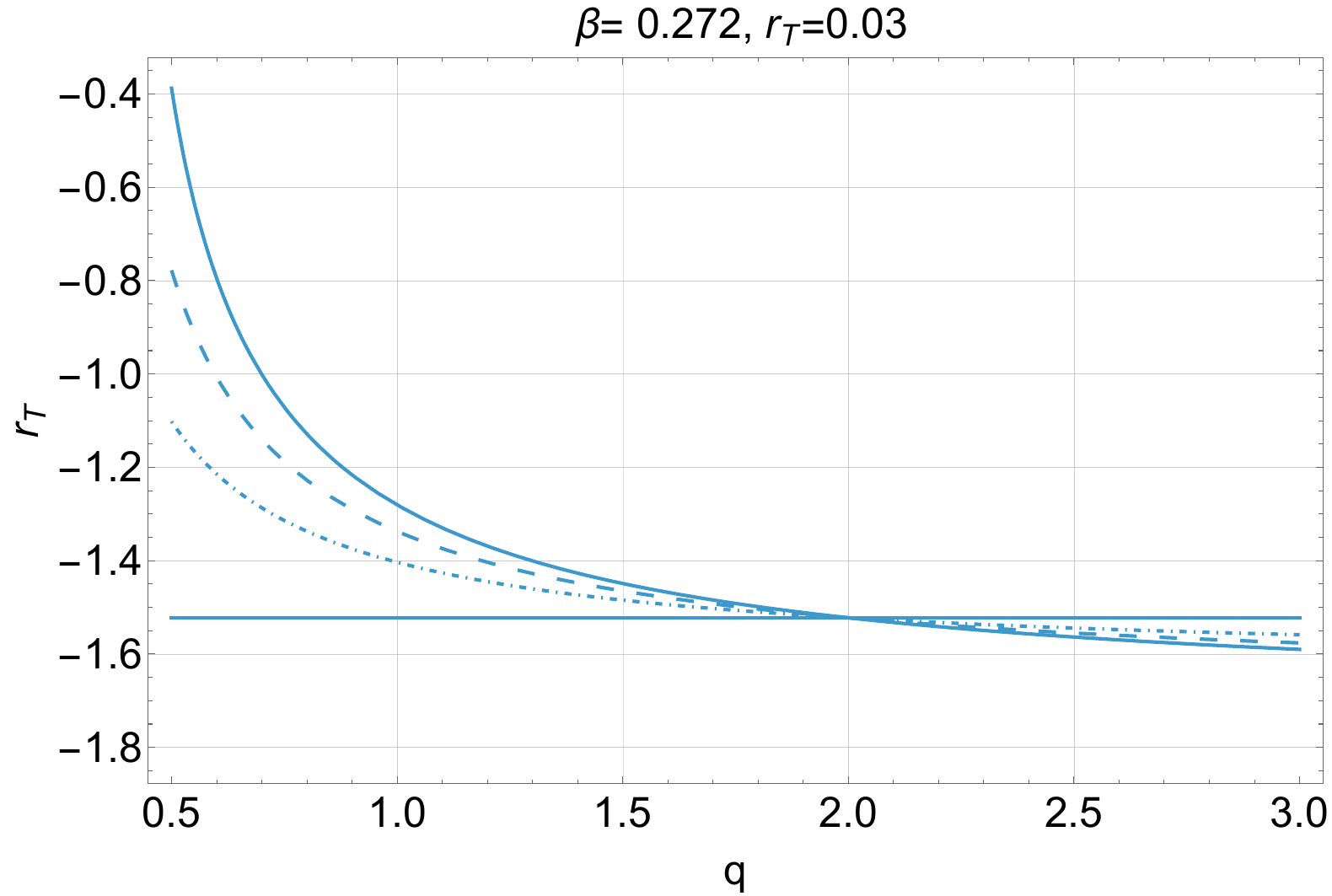}
\includegraphics[height=6cm]{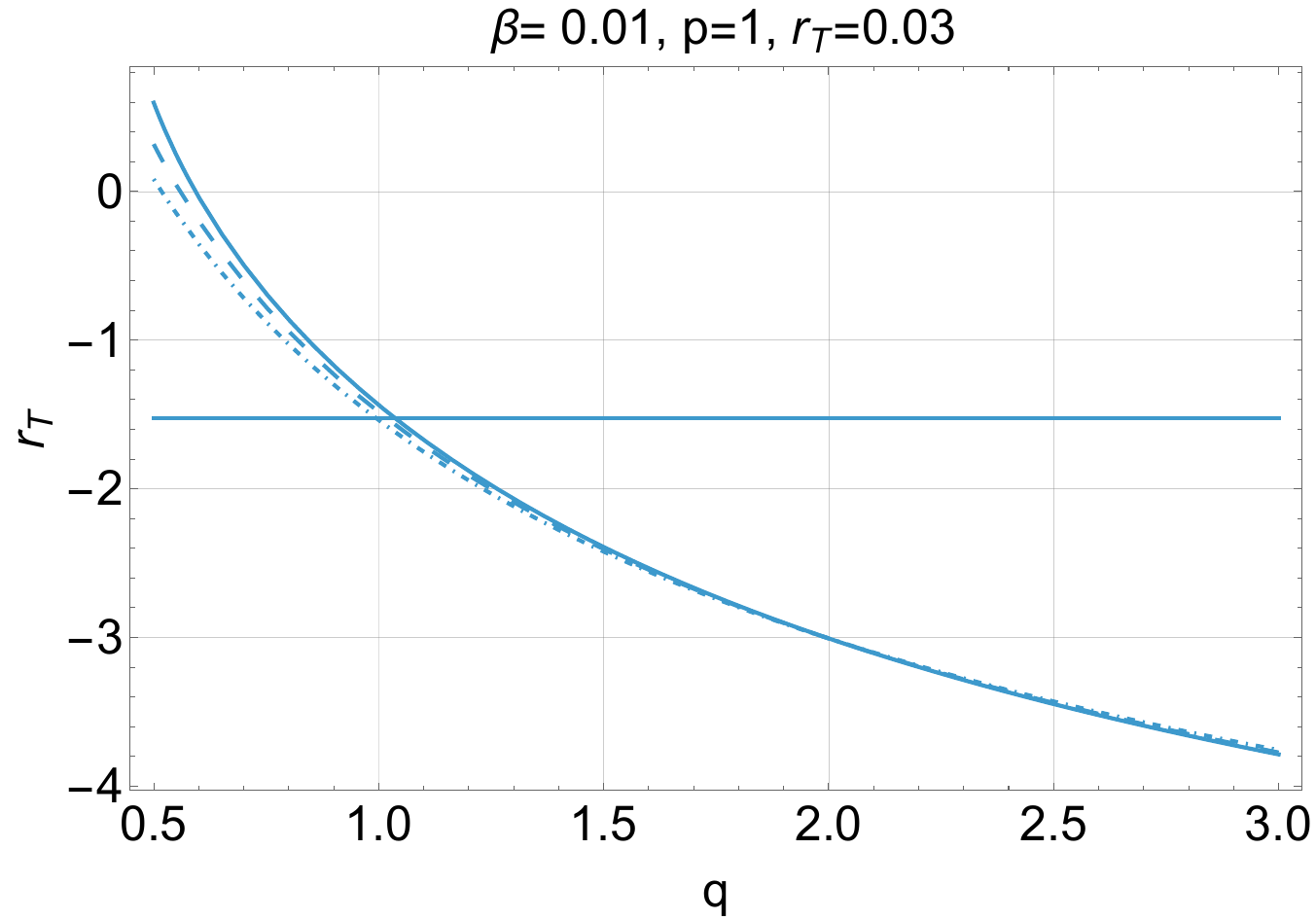}
\caption[a]{The common logarithm of $r_{T}(k)$ (always for $k = {\mathcal O}(k_{p})$) is illustrated as a function of $q$ for the two 
different potentials discussed in Eqs. (\ref{INF8})--(\ref{INF9}). The full, dashed and dot-dashed curves 
correspond to three different values of $(H_{r}/H_{1})$ ranging from $10^{-38}$ to $10^{-20}$ (dot-dashed).}
\label{FIGNC}      
\end{figure}

In Fig. \ref{FIGNA} we illustrate Eqs. (\ref{TMR2})--(\ref{TMR3}). In particular we require that $\Omega_{\sigma}$ ranges between
$0.03$ and $1$; this is the range corresponding to the shaded are in the left plot of Fig. \ref{FIGNA}. In the right plot of the same 
figure the shaded region illustrates instead the condition $|{\mathcal R}_{nad}/{\mathcal R}_{ad}| < {\mathcal O}(0.2)$, as implied by the limits on the entropic modes. The region obtained from the intersection of the two plots is very sensitive to the values of $r_{T}$. 
 In Fig. \ref{FIGNB} the intersection of the conditions described in Fig. \ref{FIGNA} is illustrated. In the plot at the left $r_{T} = 0.03$ while in the plot at the right $r_{T} = 0.003$. This shows 
that the dark matter interpretation and the bounds on the non-adiabatic modes 
are progressively more consistent when the values of $r_{T}$ are comparatively smaller than the current constraints stipulating that $r_{T} < 0.034$. The results of Figs. \ref{FIGNA} and \ref{FIGNB} hold when the post-inflationary timeline is dominated by radiation. In what 
follows, thanks to the analyses of the previous sections, we can relax the assumption of 
a unique radiation-dominated stage extending between the end of inflation and the equality time.

\subsection{Timelines expanding faster than radiation}
When the expansion rate is faster than radiation 
the total number of $e$-folds elapsed since the crossing of the scales $k = {\mathcal O}(k_{p})$ gets smaller than ${\mathcal O}(60)$. This observation implies, as explained at the end of sec. \ref{sec3}, that $r_{T}(k)$ is comparatively larger than in the case of radiation dominance and the constraints coming from the current data may exclude this possibility for a given form 
of the inflaton potential\footnote{The following stenographic notation $r_{T}(k) = r_{T}(N_{k})$ (where $k = {\mathcal O}(k_{p})$ are the CMB scales) will be adopted throughout. Since the conventional notations the $k$-dependence is generally omitted (i.e. $r_{T}= r_{T}(k)$) and we shall also neglect $k$-dependence when not strictly necessary.}. This possibility has been actually envisaged in order to increase the value of $r_{T}(k)$ computed from the monomial potentials with the purpose of accounting for the observational determinations of the $B$-mode polarization measured by the Bicep-2 experiment \cite{SIXb,SIXc,SIXd}. Unfortunately the results reported by the Bicep-2 experiment turned out to be seriously affected by the contaminations of late-time foregrounds \cite{SIXd} so that the current data are compatible with much smaller values of $r_{T}(k)$. Since the monomial potentials are now observationally disfavoured, we may consider the plateau-like potentials (see Eqs. (\ref{INF8})--(\ref{INF9}) and discussions therein) suggesting instead $r_{T}(k) = {\mathcal O}(1/N_{k}^2)$; barring for the numerical factors this scaling is potentially more consistent with the observational data.
From the examples of Eq. (\ref{INF9}) we would have, in particular, 
\begin{equation}
r_{T}(k) = \frac{8}{\beta^2 \, N_{k}^2}, \qquad r_{T}(k) = \frac{32 \, q^2 \,  \beta^{ - 2/(1+ 2q/p)}}{[ 4\, q\, ( p + 2 q)\, N_{k}/p]^{(p+ 4 q)/(p+ 2 q)}},
\label{TMF1}
\end{equation}
The first and second expressions of $r_{T}(k)$ in Eq. (\ref{TMF1}) come, respectively, from $v_{1}(\overline{\varphi})$ and 
$v_{2}(\overline{\varphi})$ of Eq. (\ref{INF9}). 
For example, when the post-inflationary evolution is dominated by radiation, the first expression of Eq. (\ref{TMF1}) would imply 
\begin{equation}
r_{T}(k) = \frac{1}{450 \beta^2} (N_{k}/60)^{-2} \leq {\mathcal O}(0.034).
\label{TMF2}
\end{equation}
If $N_{k} = {\mathcal O}(60)$ (as implied by a radiation timeline) the range of $\beta$ consistent with $r_{T}(k) = {\mathcal O}(0.03) < 0.034$ is $\beta = {\mathcal O}(0.27)$. 
If we now stick to the class of examples of Eq. (\ref{INF9}), the post-inflationary expansion rate 
in controlled by $q$. As established in Fig. \ref{FIGURE5} and in the related discussion, 
what matters is the expansion rate averaged over the oscillations; from Eqs. (\ref{OSC1})--(\ref{OSC1a}) the averaged rate of expansion after inflation when $v(\overline{\varphi}) = \overline{\varphi}^{2\, q}$ is given by 
\begin{equation}
\frac{1}{\overline{\delta}} = -1 + \overline{\epsilon} = \frac{ 2 q -1}{q + 1}.
\label{TMF3}
\end{equation}
In case the post-inflationary evolution is faster than radiation the number of $e$-folds 
elapsed since the crossing of the CMB scales gets smaller depending on the value of $q$. 
Recalling now the results of Eqs. (\ref{OSC1})--(\ref{OSC1a}) and (\ref{OSC2}) (see also Fig. \ref{FIGURE5}), 
the effective expansion rate is  $\overline{\delta} > 1$ (i.e. faster than radiation) provided $q < 2$.
This observation implies that 
\begin{equation}
N_{k} = {\mathcal O}(60) + [(2 - q)/(6\, q)] \ln{(H_{r}/H_{1})}.
\label{TMF4}
\end{equation}
In the left panel of Fig. \ref{FIGNC} we illustrate the common logarithm of $r_{T}$ as a function of $q$. For sufficiently small $q$ the values of $r_{T}$ exceed $\log{0.03} = -1.52$ that corresponds to the horizontal straight lines in both plots. In the right panel of the same figure we repeat the same exercise with the value of $r_{T}$ determined 
from the potential $v_{2}(\overline{\varphi})$ given in Eq. (\ref{INF9}). In both plots 
the full, dashed and dot dashed curves correspond, respectively, to the following three choices of $H_{r}/ H_{1}$: $(H_{r}/H_{1}) = 10^{-38}$ (full),
$(H_{r}/H_{1}) = 10^{-30}$ (dashed) and $(H_{r}/H_{1}) = 10^{-20}$ (dot-dashed).  
These examples of Fig. \ref{FIGNC} clarify the general trend: if a given potential leads to a value of $r_{T}$ compatible with the current data, a post-inflationary stage expanding faster than radiation simply decreases the total number of $e$-folds by simultaneously increasing the values of $r_{T}$. This means that, in practice, the post-inflationary stages 
expanding faster than radiation are excluded, at least in the context 
of the conventional inflationary scenarios based on plateau-like potentials. 

\subsection{Timelines expanding slower than radiation}
\begin{figure}[!ht]
\centering
\includegraphics[height=7cm]{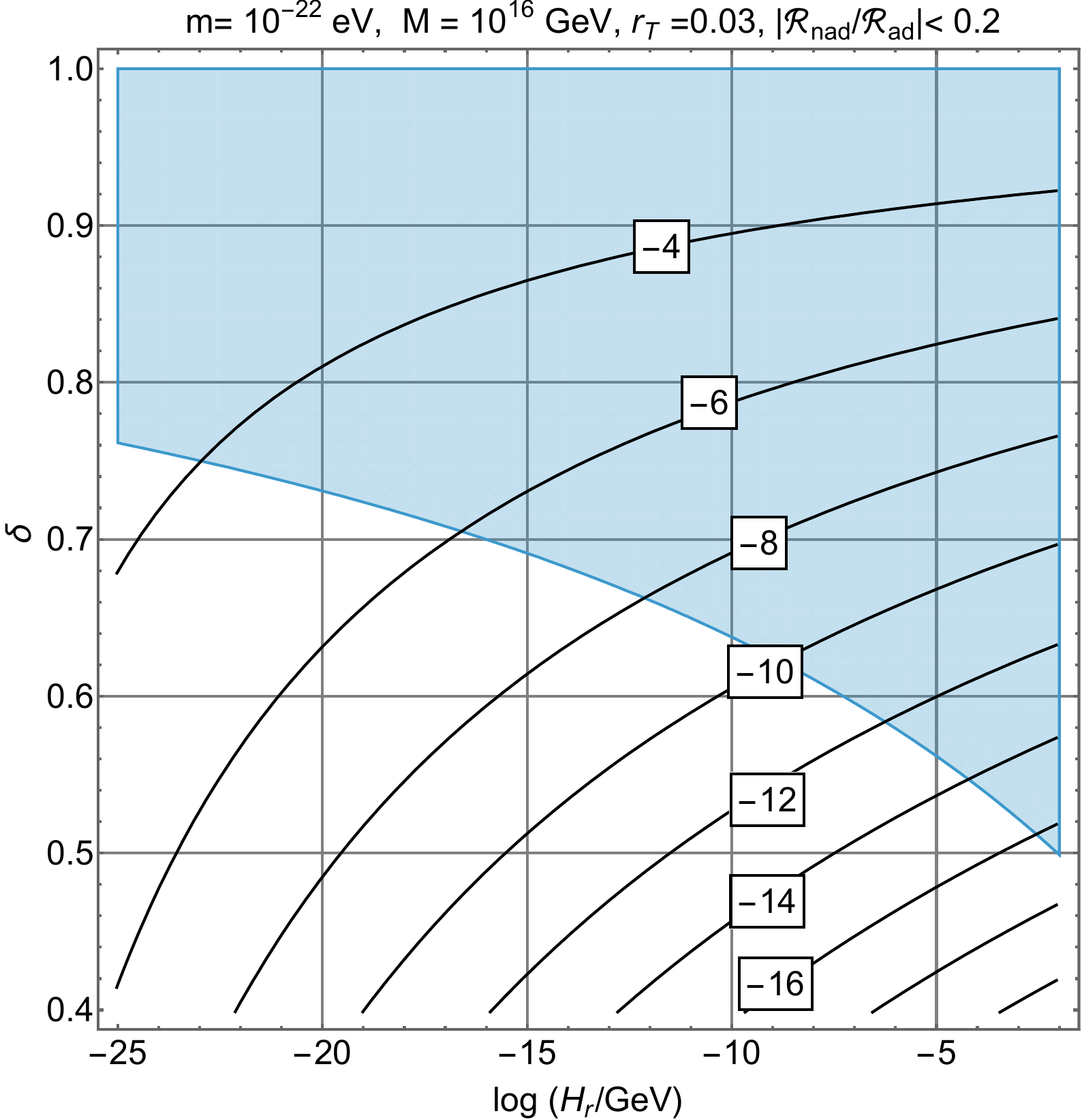}
\includegraphics[height=7cm]{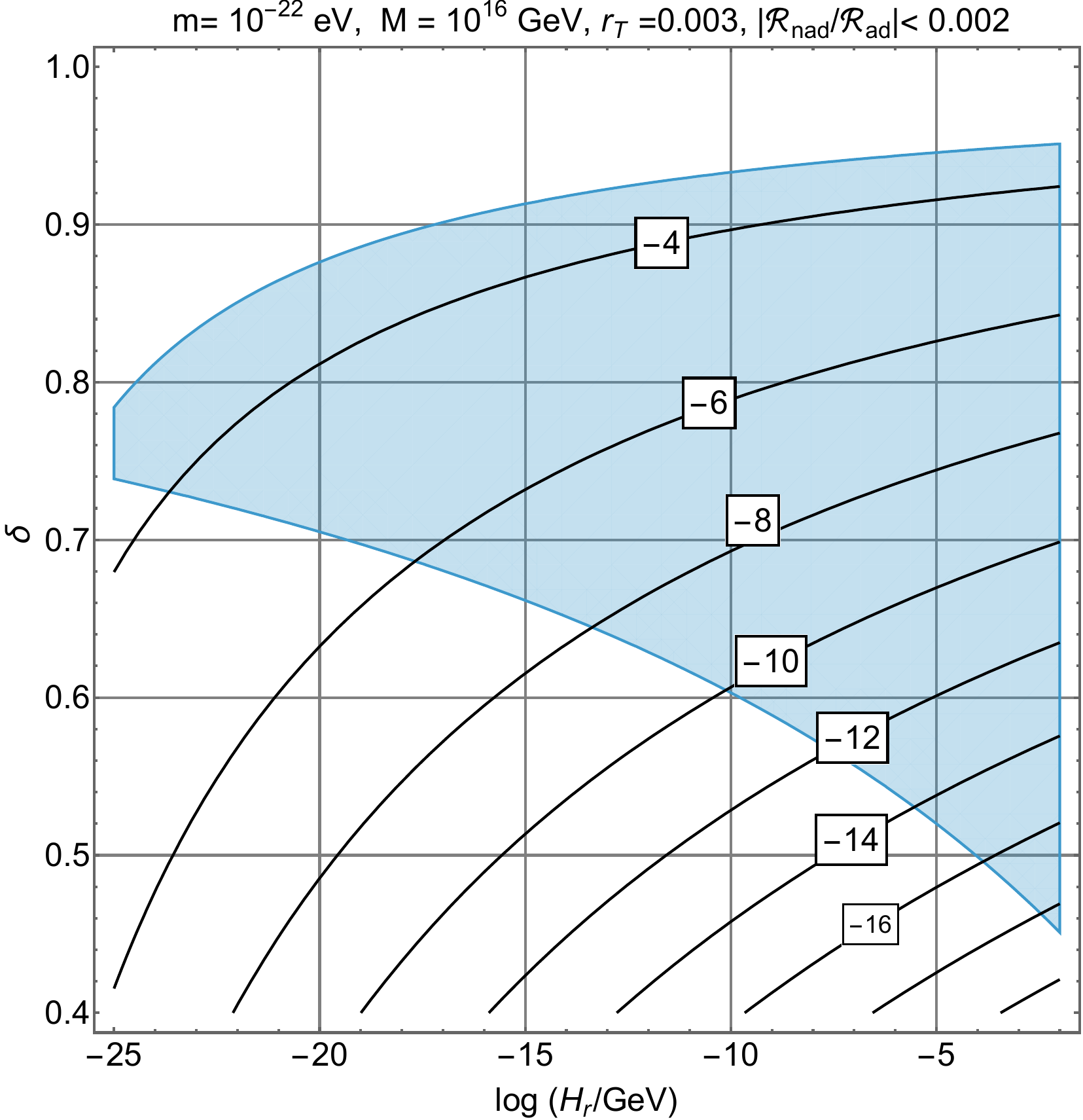}
\caption[a]{The constraints on a decelerated expansion rate slower than radiation are illustrated in the plane $(H_{r}, \, \delta)$. The remaining parameters are fixed as specified in each of the plots. The shaded area corresponds to the region where all the relevant constraints are satisfied. The lower limit on $H_{r}$ is $10^{-44} M_{P} = 
{\mathcal O}(10^{-25}) \, \mathrm{GeV}$ which is the value implied by the considerations 
related to BBN. The various labels correspond to the common logarithm of $\Omega_{\sigma}$.}
\label{FIGND}      
\end{figure}

A post-inflationary stage expanding at a rate slower than radiation is 
less constrained by the presence of ultralight scalar particles.
Furthermore, because the number of $e$-folds elapsed since the crossing of the CMB scales 
gets larger, the values of $r_{T}$ are suppressed  for a given shape 
of the inflaton potential. In this situation the spectrum of the relic gravitons is much less constrained in the 
aHz region than in the highest frequency domain (i.e. between the GHz and the THz).
The ratio between the non-adiabatic and the adiabatic components is affected by a decelerated expansion 
rate slower than radiation by two independent effects.
On the one hand a reduction of $r_{T}$ directly suppresses the $|{\mathcal R}_{nad}/{\mathcal R}_{ad}|$. On the other 
hand a sequence of stages expanding slower than radiation further decreases the non-adiabatic contribution, as 
previously computed at the end of sec. \ref{sec4}. For an arbitrary sequence of stages expanding at different 
rates we actually have that for typical scales $k = {\mathcal O}(k_{p})$ 
\begin{equation}
\biggl|\frac{{\mathcal R}_{nad}}{{\mathcal R}_{ad}}\biggr| = \frac{\sqrt{r_{T}\, \pi}}{3}
\,\, \biggl(\frac{M}{M_{P}}\biggr) \, \sqrt{\frac{m}{H_{eq}}} \, \prod_{i=1}^{n-1} \,\, \biggl(\frac{H_{i+1}}{H_{i}} \biggr)^{\frac{3 (1- \delta_{i})}{2(\delta_{i} + 1)}},
\label{TMS1}
\end{equation}
where $H_{i+1} < H_{i}$. The product appearing in Eq. (\ref{TMS1}) runs from $1$ to $(n-1)$;
we remind that the beginning of the $n$-th stage coincides with the onset of the radiation epoch (i.e. $H_{n} = H_{r}$). In general terms the various $\delta_{i}$ can be either larger or smaller than $1$ but in what 
follows we are going to consider the simplest physical situation where all the various $\delta_{i}$ collapse to a single value; in this case the product of the ratio of scales coincides with $(H_{r}/H_{1})$ and the single power is $3( 1- \delta)/[2(\delta+1)]$. In case 
then decelerated stage slower than radiation is sustained by the oscillations of the inflaton we must replace $\delta \to \overline{\delta}$ (see Eq. (\ref{TMF3})) with $ q> 2$.
In Fig. \ref{FIGND}  the constraints are illustrated in the plane $(\delta, \, H_{r})$ 
for fixed values of $m$, $M$ and $r_{T}$. In the left panel 
we consider the standard bounds on the non-adiabatic component 
(for $r_{T} =0.03$, i.e. close to the current limit); in the right panel 
we analyze instead the effect of a constraint on $|{\mathcal R}_{nad}/{\mathcal R}_{ad}|$ 
$100$ times more stringent than the current one together with a smaller value of $r_{T}$. 
The shaded areas in both plots correspond to the regions where 
the bounds coming from the tensor modes are satisfied (see, in particular, Eqs. (\ref{TM4})--(\ref{TM6})).
The left panel of Fig. \ref{FIGND} is dominated by the bounds coming from the high frequency 
gravitons. This means that the limits on the non-adiabatic component would be eventually relevant for 
constraining the post-inflationary expansion rate only provided the bounds on  $|{\mathcal R}_{nad}/{\mathcal R}_{ad}|$ would improve by (at least) two orders of magnitude (see the right panel in Fig. \ref{FIGND}); a reduction in $r_{T}$ is comparatively less relevant.
The ultralight scalars are, at the moment, 
less constrained if the expansion rate is slower than radiation provided the limit on $|{\mathcal R}_{nad}/{\mathcal R}_{ad}|$ is close to its current value (i.e. ${\mathcal O}(0.2)$). In case these constraints would dramatically improve, 
the limits on the expansion rate from the ultralight scalars could become crucial. The  different labels in Fig. \ref{FIGND} correspond to the common logarithm of $\Omega_{\sigma}$. 
Thus the presence of a stage expanding  slower than radiation not only relaxes all the constraints 
coming from the non-adiabatic modes but also reduces $\Omega_{sigma}$ unless 
$\delta$ gets closer to $1$. 
\begin{figure}[!ht]
\centering
\includegraphics[height=7cm]{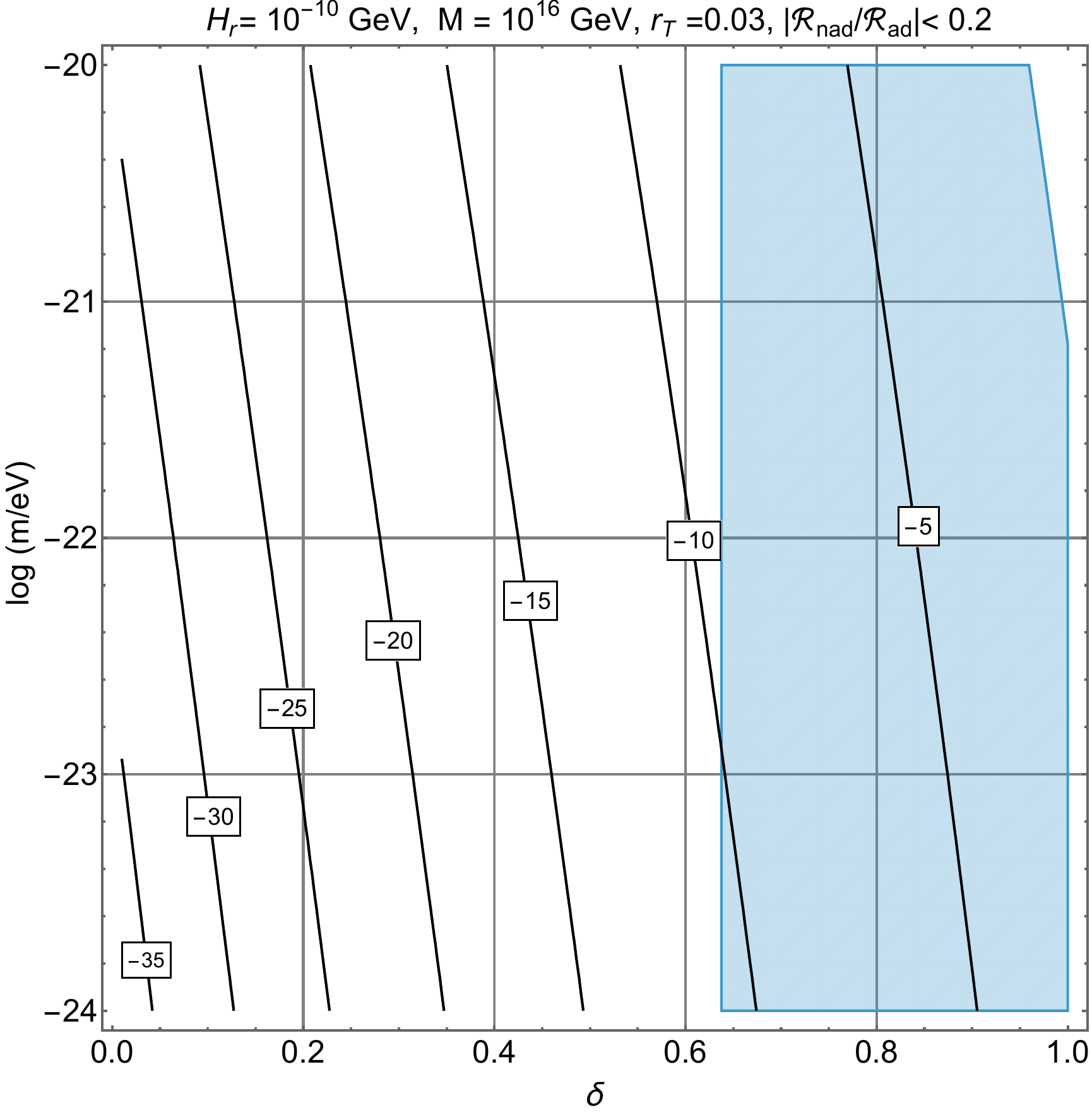}
\includegraphics[height=7cm]{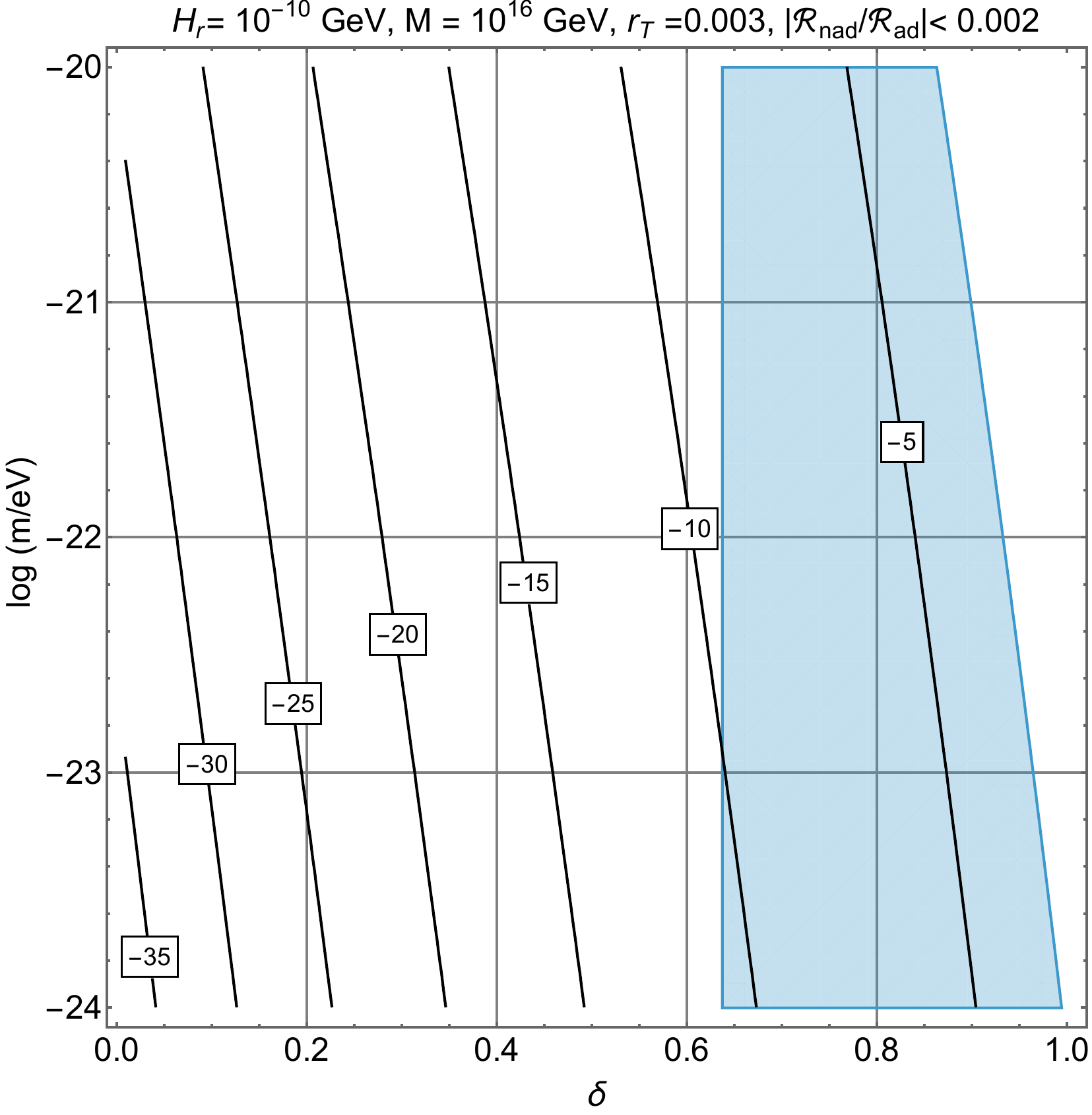}
\caption[a]{We illustrate the constraints in the plane $(m,\, \delta)$. The values of $M$ and $H_{r}$ are fixed 
and $r_{T}$ gets reduced by one order of magnitude between the left and the right plot. As in Fig. \ref{FIGND}, in both 
plots the various labels indicate the common logarithm of $\Omega_{\sigma}$.}
\label{FIGNE}      
\end{figure}
Similar conclusions follow from different sections of the parameter space:
for instance in Fig. \ref{FIGNE} we illustrate the constrains in the plane $(m,\, \delta)$ 
for fixed values of $M$, $H_{r}$ and $r_{T}$. In both plots we consider the case $M= 10^{-16}\, \mathrm{Gev}$ and $H_{r} = 10^{-10} \, \mathrm{GeV}$. The values of $H_{r}$ are suggested by Fig. \ref{FIGND} and they are compatible with the constraints from BBN requiring $H_{r} \geq 10^{-44} \, M_{P} = {\mathcal O}(10^{-25}) \mathrm{GeV}$. Also in this situation
the constraints on a post-inflationary stage with $\delta < 1$ 
do not come from the non-adiabatic fluctuations unless 
the constraints on ${\mathcal R}_{nad}$ would eventually become $100$ more stringent. 
The values of $\Omega_{\sigma}$ compatible with a dark matter interpretation 
correspond to $\delta \leq 1$ but in this case $r_{T}$ must also be very close 
to the current limit with tensor modes potentially observable in the aHz range.

\newpage
\renewcommand{\theequation}{6.\arabic{equation}}
\setcounter{equation}{0}
\section{Concluding Remarks}
\label{sec6}
The current version of the adiabatic paradigm suggests that the large-scale curvature 
inhomogeneities are Gaussian and predominantly adiabatic. As a result both 
the non-adiabatic inhomogeneities and the tensor modes of the geometry
are strongly constrained. This is why, when the timeline of the post-inflationary 
expansion rate is modified in comparison with the one customarily adopted within the standard lore,
two relevant physical questions must be independently analyzed. On the one hand the tensor to scalar ratio may be either reduced or increased, depending on the post-inflationary timeline. On the other hand an alteration of the decelerated evolution is plausible as long as the large-scale inhomogeneities are adiabatic but it is unclear what happens when non-adiabatic fluctuations are concurrently present. In this inestigation the attention has been focussed on the non-adiabatic modes induced by the ultralight scalar bosons that are one of the most interesting candidates also in the light of their potential dark matter interpretation.

After deducing the governing equations of the system (with particular attention to the dynamics of the adiabatic and non-adiabatic inhomogeneities) the evolution of the large-scale fluctuations has been carefully scrutinized in the physical frameworks dictated by different classes of post-inflationary histories. It turns out that, when the post-inflationary expansion rate is faster than radiation, the tensor to scalar ratio experiences a sharp enhancement that also increases 
the relative balance between the non-adiabatic and the adiabatic contribution. The 
reason behind this intriguing effect is that, ultimately, the total  number of $e$-folds elapsed 
since the exit of the CMB scales depends upon the nature and the properties of the decelerated rate of expansion after inflation. 

The situation is different when the timeline of the post-inflationary evolution is dominated by 
an expansion rate slower than radiation. In this case the tensor to scalar ratio inherits a 
further degree of reduction for a given form of the inflaton potential. In this way the non-adiabatic 
contributions are suppressed in two complementary ways. First a reduction 
of the tensor to scalar ratio decreases the ratio between the non-adiabatic and the adiabatic 
spectra. Second the non-adiabatic modes experience a direct dynamical 
suppression that is absent when the post-inflationary evolution is dominated by radiation. The most relevant constraints on the expansion rate come from the high frequency spectrum of the tensor modes of the geometry and from the big-bang nucleosynthesis bounds. The simultaneous analysis of the different classes of limits shows that the constraints on the non-adiabatic modes and on the tensor to scalar ratio are relaxed when the post-inflationary evolution expands at a rate slower than radiation well before the onset of the big-bang nucleosynthesis epoch. In an even more general perspective the discussion presented here also suggests that the bounds on the non-adiabatic contribution obtained from the CMB data analysis do depend upon the post-inflationary timeline.

\section*{Acknowledgements}
It is  a pleasure to thank A. Gentil-Beccot, L. Pieper, S. Rohr and J. Vigen of the CERN Scientific Information Service for their kind help during the preparation of this manuscript.

\newpage
\begin{appendix}
\renewcommand{\theequation}{A.\arabic{equation}}
\setcounter{equation}{0}
\section{Curvature inhomogeneities from two fields}
\label{APPA}
The first-order (relativistic) fluctuations of the geometry ($\delta^{(1)} g_{\mu\nu}(\vec{x}, \tau)$ in what follows) notoriously consist of $10$ independent components that can be decomposed in terms of $4$ scalar functions, $2$ solenoidal vectors, and $1$ rank-two 
tensor  in three spatial dimensions \cite{ONEa,ONEb} (see also \cite{EIGHT2,EIGHT3,EIGHT4}). The rank-two tensor is both traceless and solenoidal it carries two independent polarizations that correspond to the tensor modes of the geometry. Since the vector fluctuations do not play a specific role in the present discussion, the first-order fluctuations of the metric are 
$\delta^{(1)}\, g_{\mu\nu}(\vec{x},\tau) =\delta_{s}^{(1)}\, g_{\mu\nu}(\vec{x},\tau) + 
\delta_{t}^{(1)}\, g_{\mu\nu}(\vec{x},\tau)$ where the subscripts denote, respectively, the 
scalar and the tensor components contributions. In the case of the scalar fluctuations (that are 
directly related to the curvature inhomogeneities) the perturbed entries of the metric are:
\begin{equation}
\delta_{s}^{(1)} g_{00} = 2\, a^2 \, \phi, \quad \delta_{s}^{(1)} g_{0i} = - a^2 \partial_{i} B, 
\quad \delta_{s}^{(1)} g_{i\, j} = 2 \, a^2 \, \bigl( \psi\, \delta_{i\,j} - \partial_{i} \partial_{j} E\bigr)
\label{APA1}
\end{equation}
where all the four independent functions $\phi$, $B$, $\psi$ and $E$ depend 
upon $\vec{x}$ and $\tau$ while the scale factor is only time dependent [i.e. 
$a \equiv a(\tau)$]. The current data suggest that three-geometry is spatially flat \cite{FIVEa,FIVEb,FIVEc,FIVEd,FIVEe}; this means that the spatial derivatives appearing in Eq. (\ref{APA1}) are just ordinary derivative. In spite of this the decomposition (\ref{APA1}) can be easily generalized to the case where the three-geometry is curved by promoting the spatial derivatives 
to the status of covariant derivatives with respect to the metric of the sub-manifold. The infinitesimal coordinate transformations preserving the scalar nature of the fluctuations are given by $x^{\mu} \to \widetilde{x}^{\mu} = x^{\mu} + \epsilon^{\mu}$ where $\epsilon^{\mu} = (\epsilon^{0}, \partial^{i} \epsilon)$. Under these transformations the $4$ functions introduced in Eq. (\ref{APA1}) transform as:
\begin{eqnarray}
\phi \to \widetilde{\phi} &=& \phi - {\mathcal H} \epsilon_{0} - \epsilon_{0}^{\prime}, \qquad 
B \to \widetilde{B} = B + \epsilon_{0} - \epsilon^{\prime},
\label{APA2}\\
\psi\to \widetilde{\psi} &=& \psi + {\mathcal H} \, \epsilon_{0}, \qquad E\to \widetilde{E} = E -  \epsilon.
\label{APA3}
\end{eqnarray}
From Eqs. (\ref{APA2})--(\ref{APA3}) it is possible to construct two well known 
gauge-invariant combinations \cite{EIGHT2,EIGHT3} that have been already introduced in sec. \ref{sec2} (see, in particular, Eqs. (\ref{PP4})--(\ref{PP6})):
\begin{equation}
\Psi = \psi - {\mathcal H} (B - E^{\prime}), \qquad \Phi = \phi + 
{\mathcal H}\, (B - E^{\prime}) + (B - E^{\prime})^{\prime}.
\label{APA4}
\end{equation}
The variables $\Psi$ and $\Phi$ (customarily referred to as Bardeen potentials) are invariant under the transformations of Eqs. (\ref{APA2})--(\ref{APA3}). Recalling now the global variables $\rho_{t}$ and $p_{t}$ introduced in Eq. (\ref{BB1}),  their 
corresponding scalar fluctuations (i.e. $\delta_{s}^{(1)} \rho_{t}$ and $\delta_{s}^{(1)} p_{t}$) also change for infinitesimal coordinate transformations; when using $\Phi$ and $\Psi$ as pivotal variables the corresponding gauge-invariant fluctuations of the sources must 
be consistently employed and they are:
\begin{equation}
\delta^{(g)} \rho_{t} = \delta_{s}^{(1)} \rho_{t} + \rho_{t}^{\prime} \, (B- E^{\prime}), 
\qquad \delta^{(g)} p_{t} = \delta_{s}^{(1)} p_{t} + p_{t}^{\prime} \, (B- E^{\prime}).
\label{APA4a}
\end{equation}
For the present ends the two most relevant gauge-invariant variables are associated with  the fluctuations of spatial curvature on comoving orthogonal hypersurfaces (denoted by ${\mathcal R}$) and with the density contrast on uniform curvature hypersurfaces (denotes by $\zeta$). Since they are invariant under infinitesimal coordinate transformations,  both  quantities are expressible in terms of $\Phi$ and $\Psi$ in the following manner:
\begin{equation}
{\mathcal R} = - \Psi - {\mathcal H} ({\mathcal H} \Phi + \Psi^{\prime})/({\mathcal H}^2 - {\mathcal H}^{\prime}), \qquad \zeta = - \Psi - {\mathcal H}\,\,\delta^{(g)} \rho_{t}/\rho_{t}^{\prime}.
\label{APA5}
\end{equation}
From the Hamiltonian constraint of Eq. (\ref{PP4}) the explicit connection between ${\mathcal R}$ and $\zeta$ becomes 
\begin{equation}
{\mathcal R} - \zeta = - 2\nabla^2 \Psi/[3 \, \ell_{P}^2\,  a^2 ( p_{t} + \rho_{t})].
\label{APA6}
\end{equation}
Equation (\ref{APA6}) implies that, in Fourier space,  ${\mathcal R} = \zeta + {\mathcal O}(k^2 \tau^2)$ and this means that ${\mathcal R}$ and $\zeta$ can be used interchangeably in the large-scale limit. 
Since ${\mathcal R}$ and $\zeta$ are both gauge-invariant they can be computed in the most convenient coordinate system. For infinitesimal coordinate transformations the fluctuations 
of $\varphi$ and $\sigma$ (i.e. $\delta \varphi = \chi_{\varphi}$ and $\delta \sigma = \chi_{\sigma}$) transform as
\begin{equation}
\chi_{\varphi} \to \widetilde{\chi}_{\varphi} = \chi_{\varphi} - \varphi^{\prime} \epsilon_{0}, 
\qquad  \chi_{\sigma} \to \widetilde{\chi}_{\sigma} = \chi_{\sigma} - \sigma^{\prime} \epsilon_{0}.
\label{APA7}
\end{equation}
Therefore, in the two-field case, a particularly useful gauge is $\psi=0$ and $E =0$. In this coordinate system $\Psi = - {\mathcal H}\, B$ and $\Phi = \phi + {\mathcal H}\, B + B^{\prime}$. From Eq. (\ref{APA5}) the curvature inhomogeneities assume a particularly simple expression:
\begin{equation}
{\mathcal R} = - \frac{{\mathcal H}^2 \, \phi}{{\mathcal H}^2 - {\mathcal H}^{\prime}}  = 
- \frac{{\mathcal H}(\varphi^{\prime}\, \chi_{\varphi} + \sigma^{\prime} \, \chi_{\sigma})}{\varphi^{\prime \, 2} + \sigma^{\prime\, 2}}.
\label{APA8}
\end{equation}
Since the momentum constraint and the background equations 
imply:
\begin{equation}
\phi =\ell_{P}^2 [ (\varphi^{\prime}/{\mathcal H}) \, \chi_{\varphi} + 
(\sigma^{\prime}/{\mathcal H})\chi_{\sigma}]/2, \qquad {\mathcal H}^2 - {\mathcal H}^{\prime} = \ell_{P}^2(\varphi^{\prime\, 2} + \sigma^{\prime\, 2})/2,
\label{APA9}
\end{equation}
the second equality in Eq. (\ref{APA8}) follows.
In the gauge $\psi = E=0$ the evolution equations of the system can be 
directly reduced to the dynamics of $\chi_{\varphi}$ and $\chi_{\sigma}$; for this 
purpose we start from the perturbed Klein-Gordon equations in curved background 
\begin{eqnarray}
&& \chi_{\varphi}^{\prime\prime} + 2 {\mathcal H} \,\chi_{\varphi}^{\prime} - \nabla^2 \chi_{\varphi} + 
a^2 \,V_{,\,\varphi\varphi}\, \chi_{\varphi}  + 2 \,\phi \,a^2\,V_{,\, \varphi}  - 
\varphi^{\prime} \phi^{\prime} - \varphi^{\prime} \nabla^2 B=0,
\label{APA10}\\
&& \chi_{\sigma}^{\prime\prime} + 2 {\mathcal H} \,\chi_{\sigma}^{\prime} - \nabla^2 \chi_{\sigma} + 
a^2\, U_{,\,\sigma\sigma} \,\chi_{\sigma}  + 2 \,\phi \,a^2 U_{,\,\sigma}- 
\sigma^{\prime} \phi^{\prime} - \sigma^{\prime}\,\nabla^2 B=0.
\label{APA11}
\end{eqnarray}
Although Eqs. (\ref{APA10})--(\ref{APA11}) are both coupled to $\phi$ and $\nabla^2 B$, the formal dependence upon the metric inhomogeneities is totally eliminated by using 
the perturbed components of the Einstein equations.  
In the gauge $\psi = E =0$ the $(i\neq j)$, $(00)$ components of the perturbed 
Einstein equations lead, respectively, to:
\begin{eqnarray}
&& B' + 2 {\mathcal H} B + \phi =0,
\label{APA12}\\
&& {\mathcal H} \nabla^2 B + ({\mathcal H}^{\prime}+ 2 {\mathcal H}^2) \phi = - \frac{\ell_{P}^2}{2}\biggl[ \varphi^{\prime} \chi_{\varphi}^{\prime}+ \sigma^{\prime} \chi_{\sigma}^{\prime}
+V_{,\,\varphi} a^2 \chi_{\varphi} + U_{,\,\sigma} a^2 \chi_{\sigma}\biggr].
\label{APA13}
\end{eqnarray}
The $(0\,i)$ component of the perturbed Einstein equations 
correspond to the first relation already appearing in Eq. (\ref{APA9}). 
After inserting 
Eqs. (\ref{APA9}), (\ref{APA12}) and (\ref{APA13}) into Eqs. (\ref{APA10})--(\ref{APA11}) the dependence upon $B$ and 
$\phi$ (i.e. the fluctuations of the metric) can be totally eliminated.  Indeed, Eqs.  (\ref{APA9}), (\ref{APA12}) and (\ref{APA13}) imply
\begin{eqnarray}
&& \phi = \frac{\ell_{P}^2}{2} \biggl( \frac{\varphi'}{{\mathcal H}} \chi_{\varphi} + \frac{\sigma'}{{\mathcal H}} \chi_{\sigma}\biggr),
\label{APA14}\\
&& \phi' = \frac{\ell_{P}^2}{2}\biggl[  \frac{\varphi'}{{\mathcal H}} \chi_{\varphi}' + \frac{\sigma'}{{\mathcal H}} \chi_{\sigma}' +
\biggl( \frac{\varphi'}{{\mathcal H}}\biggr)' \chi_{\varphi} + \biggl(\frac{\sigma'}{{\mathcal H}}\biggr)' \chi_{\sigma}\biggr],
\label{APA15}\\
&& \nabla^2 B = - \frac{\ell_{P}^2}{2} \biggl\{ ({\mathcal H}' + 2 {\mathcal H}^2)\biggl[ \frac{\varphi'}{{\mathcal H}} \chi_{\varphi} + \frac{\sigma'}{{\mathcal H}} \chi_{\sigma}
\biggr] +  \frac{\varphi'}{{\mathcal H}} \chi_{\varphi}' + \frac{\sigma'}{{\mathcal H}} \chi_{\sigma}' +
\frac{\partial W}{\partial\varphi} \frac{a^2}{{\mathcal H}} \chi_{\varphi} +
\frac{\partial W}{\partial\sigma} \frac{a^2}{{\mathcal H}} \chi_{\sigma}\biggr\}.
\label{APA16}
\end{eqnarray}
Equation (\ref{APA15}) is the derivative of Eq. (\ref{APA14}) and Eq. (\ref{APA13}) follows from Eq. (\ref{APA13})--(\ref{APA14}).
Using now Eqs. (\ref{APA14})--(\ref{APA16}) to eliminate $\phi$, $\phi'$ and $\nabla^2 B$ from
 Eqs. (\ref{APA10}) and (\ref{APA11}) the result of Eqs. (\ref{NAD4}) and (\ref{NAD5}) 
is readily obtained. Not all the terms are simultaneously dominant during the various dynamical stages of the evolution; for instance from  Eqs. (\ref{NAD4})--(\ref{NAD5}) and (\ref{NAD8}) the evolution of the Fourier modes of $\overline{\chi}_{\varphi}$  during the inflationary stage obeys: 
\begin{equation}
\overline{\chi}_{\varphi}^{\prime\prime} + \biggl[k^2 - \frac{\nu_{\varphi}^2 - 1/4}{\tau^2} \biggr] \overline{\chi}_{\varphi} =0, \qquad \overline{\chi}_{\sigma}^{\prime\prime} + \biggl[k^2 - \frac{\nu_{\sigma}^2 - 1/4}{\tau^2} \biggr] \overline{\chi}_{\sigma} =0, 
\label{NAD8b}
\end{equation}
where $\overline{\chi}_{\varphi}(k,\tau) = a(\tau) \chi_{\varphi}(k,\tau)$ and $\overline{\chi}_{\sigma}(k,\tau) = a(\tau) \chi_{\sigma}(k,\tau)$. The results of Eq. (\ref{NAD8b}) follow by recalling that $a\, H = - 1/[(1 -\epsilon) \tau]$ (see also Eq. (\ref{NAD8a}) and discussion therein). After some lengthy but straightforward calculations the Bessel indices $\nu_{\varphi}$ and $\nu_{\sigma}$ appearing in Eq. (\ref{NAD8b}) are given by:
\begin{equation}
\nu_{\varphi} = 3/2 + 3 \epsilon - \overline{\eta} + {\mathcal O}(\epsilon^2) + {\mathcal O}(\overline{\eta}^2) + 
{\mathcal O}(\overline{\eta} \, \epsilon), \qquad \nu_{\sigma} = 3/2 + \epsilon + {\mathcal O}(\epsilon^2),
\label{NAD8c}
\end{equation}
where $\epsilon$ and $\overline{\eta}$ are the slow-roll parameters defined within the 
notations introduced in Eq. (\ref{INF3}). With the appropriate boundary conditions 
dictated by quantum mechanics, the solutions of Eq. (\ref{NAD8b}) are:
\begin{equation}
\chi_{\varphi}(k,\tau) = \frac{{\mathcal N}_{\varphi}}{a\, \sqrt{2 k}} \,\sqrt{- k \tau} \, H_{\nu_{\varphi}}^{(1)}(- k\tau), \qquad \chi_{\sigma}(k,\tau) = \frac{{\mathcal N}_{\sigma}}{a\, \sqrt{2 k}} \,\sqrt{- k \tau} \, H_{\nu_{\sigma}}^{(1)}(- k\tau),
\label{NAD8d}
\end{equation}
where ${\mathcal N}_{\sigma}$ and ${\mathcal N}_{\varphi}$ only differ by a phase, i.e. $|{\mathcal N}_{\sigma}| = |{\mathcal N}_{\varphi}| = \sqrt{\pi/2}$. The curvature inhomogeneities now follow after inserting Eq. (\ref{NAD8d}) into Eq. (\ref{NAD4a}):
\begin{equation}
{\mathcal R}_{k}(\tau) = \frac{1}{a\,\sqrt{2 k}} \sqrt{- k\tau} \biggl(\frac{{\mathcal H}}{\varphi^{\prime}}\biggr) \biggl[ 
{\mathcal N}_{\varphi} H_{\nu_{\varphi}}^{(1)}(- k\tau)+ \biggl(\frac{\sigma^{\prime}}{\varphi^{\prime}}\biggr){\mathcal N}_{\sigma} H_{\nu_{\sigma}}^{(1)}(- k\tau)\biggr]. 
\label{NAD8e}
\end{equation}
The value of $(\sigma^{\prime}/\varphi^{\prime})= \dot{\sigma}/\dot{\varphi}$ is actually negligibly small: since 
$H \dot{\sigma}$ and $m^2 \sigma$ are comparable we can estimate that $\dot{\sigma}/\dot{\varphi} = (m^2/H^2) \, (M/\overline{M}_{P})/\sqrt{2\epsilon}$ which is ${\mathcal O}(10^{-90})$ for $m = {\mathcal O}(10^{-22}) \, \mathrm{eV}$. It follows 
that the spectrum of curvature inhomogeneities matches with the standard result already mentioned earlier on, in particular 
\begin{equation}
P_{{\mathcal R}}(k,\tau) = \frac{k^3}{2\pi^2} \bigl| {\mathcal R}_{k}(\tau) \bigr|^2 = \frac{k^3}{2\pi^2} \biggl|\frac{H}{\dot{\varphi}}\biggr|^2 \bigl| \chi_{\varphi}(k,\tau)\bigr|^2.
\label{NAD8f}
\end{equation}
After plugging the result of Eqs. (\ref{NAD8d})--(\ref{NAD8e}) into Eq. (\ref{NAD8f}) we can expand the Hankel function in its small argument limit $| k \tau | \ll 1$ (i.e. $k < a\, H$ ) so that, ultimately, the explicit expression of the power spectrum of ${\mathcal R}$ is:
\begin{equation}
P_{{\mathcal R}}(k,\tau) = \frac{H^2}{\pi \, \epsilon \, M_{P}} \biggl(\frac{k}{a H}\biggr)^{n_{s} -1}, \qquad 
n_{s} -1 = 3 - 2 \nu_{\varphi} = 2 \overline{\eta} - 6 \epsilon.
\label{NAD8g}
\end{equation}
As repeatedly discussed in the bulk of the paper, the slow-roll parameters are evaluated when the CMB scales $k = {\mathcal O}(k_{p})$ crossed the comoving Hubble radius. In this case $n_{s} \to n_{s}(k) = 
1 + 2 \overline{\eta}_{k} - 6 \epsilon_{k}$.
Even if the spectrum of $\overline{\chi}_{\sigma}$ does not contribute, in practice, to the spectrum 
of the adiabatic mode of Eq. (\ref{NAD8g}), the fluctuations of $\sigma$ become relevant 
later on and lead eventually to a non-adiabatic pressure fluctuation. For this reason 
we are also interested in the spectrum of $\overline{\chi}_{\sigma}$ whose general expression is:
\begin{equation}
P_{\sigma}(k,\tau) = \frac{k^3}{2 \pi^2 a^2} \bigl| \overline{\chi}_{\sigma}(k,\tau)\bigr|^2 = 
\frac{k^2}{4\pi^2 a^2} \bigl| {\mathcal N}_{\sigma}\bigr|^2 (- k \tau) \bigl|H_{\nu_{\sigma}}^{(1)}(- k \tau)\bigr|^2. 
\label{NAD8h}
\end{equation}
After expanding the Hankel functions in the small argument limit we obtain:
\begin{equation}
P_{\sigma}(k,\tau) = \frac{H^2}{\pi^3} 2^{2 \nu_{\sigma} -3} \Gamma^2(\nu_{\sigma}) \, \biggl(\frac{k}{a\, H}\biggr)^{3 - 2 \nu_{\sigma}},
\label{NAD8i}
\end{equation}
which coincides with Eq. (\ref{INF13}) of sec. \ref{sec2}. We finally stress that the discussion conducted in the gauge $\psi= E =0$ can be easily reformulated in gauge-invariant terms by recalling that the momentum constraint can be written as 
 ${\mathcal H} \Phi + \Psi^{\prime} = \ell_{P}^2[ \varphi^{\prime} \, X_{\varphi} + 
 \sigma^{\prime} X_{\sigma}]$, 
 where $X_{\varphi} = \chi_{\varphi} + \varphi^{\prime}\,(B- E^{\prime})$
 and $X_{\sigma} = \chi_{\sigma} + \sigma^{\prime}(B- E^{\prime})$ are the
 gauge-invariant variables associated with the scalar field fluctuations. In terms 
 of $X_{\varphi}$ and $X_{\sigma}$ we can construct two further variables\footnote{The evolution of $V_{\varphi}$ and $V_{\sigma}$ formally coincides with Eqs. (\ref{NAD4})--(\ref{NAD5}) by replacing $\chi_{\varphi} \to V_{\varphi}$ 
and $\chi_{\sigma} \to V_{\sigma}$. The same is true for the curvature inhomogeneities 
whose explicit expression in terms of $V_{\varphi}$ and $V_{\sigma}$ now becomes 
${\mathcal R} = - {\mathcal H}[ \varphi^{\prime} V_{\varphi} + \sigma^{\prime} V_{\sigma}]/( \varphi^{\prime\, 2} + \sigma^{\prime\, 2})$ which is exactly 
the result of Eq. (\ref{APA8}) with $\chi_{\varphi} \to V_{\varphi}$ 
and $\chi_{\sigma} \to V_{\sigma}$.}   $V_{\varphi} = X_{\varphi} + (\varphi^{\prime}/{\mathcal H}) \Psi$ and $ V_{\sigma} = X_{\sigma} + (\sigma^{\prime}/{\mathcal H}) \Psi$.

\renewcommand{\theequation}{B.\arabic{equation}}
\setcounter{equation}{0}
\section{Complements on the constraints from the tensor modes}
\label{APPB}
The tensor modes of the geometry constrain the post-inflationary evolution, 
in the low and in the high frequency ranges. The low frequency limits correspond to the  CMB scales $k= {\mathcal O}(k_{p})$ and are associated with comoving frequencies ${\mathcal O}(\mathrm{aHz})$; conversely the high frequency bounds involve a domain ranging between the GHz and the THz, i.e. roughly comparable with the maximal frequency range $\nu_{max}$ that cannot exceed the THz region  \cite{EIGHT1} (see Eqs. (\ref{TM3}) and (\ref{TM6}) and discussions therein). From Eq. (\ref{TM2})   
$\nu_{max} < {\mathcal O}(10^{-2}) \sqrt{H_{0} M_{P}}$ implying
 imply  $\nu_{max} < {\mathcal O}(\mathrm{THz})$ since $H_{0} = {\mathcal O}(\mathrm{aHz})$ and $M_{P} = {\mathcal O}(10^{43}) \mathrm{Hz}$.
 A more accurate form of the high frequency limit follows from the observation 
that at the maximal frequency only one (or few) pairs of gravitons are effectively
produced, i.e. $\overline{n}_{\nu} = {\mathcal O}(1)$ for $\nu \to \nu_{max}$. Indeed, 
 the averaged multiplicity is exponentially suppressed for $\nu > \nu_{max}$ 
while it goes like a power for $\nu < \nu_{max}$:
\begin{equation}
\overline{n}_{\nu} = \gamma\, \frac{x^{\,n_{T} - 3}}{e^{\gamma\, x} -1},\qquad 
n_{T} = \frac{32 -4\,r_{T}}{16 - r_{T}} - 2 \delta,
\label{APB1}
\end{equation}
where $\gamma$ is a numerical factor ${\mathcal O}(1)$, 
$x =(\nu/\nu_{max})$ and $n_{T}$ indicates the tensor spectral index at high frequencies. 
In Eq. (\ref{APB1}) we consider, for simplicity, only the situation where the gravitons are produced from the vacuum since these are initial data compatible with a long inflationary stage where the total number of $e$-folds exceeds $N_{k}$. When $\delta \to 1$ Eq. (\ref{APB1}) together with Eq. (\ref{TM6}) imply $n_{T} \to - r_{T}/8 + {\mathcal O}(r_{T}^2)$ for $ \nu_{max} \to \overline{\nu}_{max} = {\mathcal O}(300) \, \mathrm{MHz}$: this limit accounts for the nearly scale-invariant limit of the spectrum where the maximal frequency is much smaller than the THz. When the post-inflationary expansion rate is slower than radiation (i.e. $\delta < 1$) we have instead 
\begin{equation}
n_{T} = 2 - 2 \delta - r_{T}/8 + {\mathcal O}(r_{T}^2), \qquad \nu_{max} \gg \overline{\nu}_{max}.
\label{APB1b}
\end{equation}
Since $\delta < 1$ the spectral index is positive (i.e. $n_{T} >0$)  and the ${\mathcal O}(r_{T})$ correction 
is immaterial for an approximate evaluation. When the post-inflationary evolution is slower 
than radiation Eq. (\ref{TM6}) implies that $\nu_{max} \gg \overline{\nu}_{max}$. Because the radiation must dominate before BBN we should require  $H_{r} \geq {\mathcal O}(10^{- 44}) M_{P}$ (i.e. $H_{r} \geq {\mathcal O}(10^{-16}) \, \mathrm{eV}$). For instance,  form Eq. (\ref{TM6}) we would have $\nu_{max} = {\mathcal O}(4) \mathrm{THz} \gg \overline{\nu}_{max}$ when $H_{r} = 10^{-31} \, M_{P}$ and $\delta \to 1/2$.
The same logic can be applied to a post-inflationary expansion rate faster 
than radiation (i.e. $\delta > 1$); also in this case  $n_{T} = 2 - 2\delta + {\mathcal O}(r_{T})$ but this time the spectral index is negative $n_{T}<0$. Similarly $\nu_{max} \ll \overline{\nu}_{max}$; for example when $H_{r} = 10^{-31} \, M_{P}$ and $\delta \to 1$ we would get, from Eq. (\ref{TM6}), $\nu_{max} = {\mathcal O}(20)\, \mathrm{kHz} \ll \overline{\nu}_{max}$. Thanks to Eq. (\ref{APB1}) BBN bound can be reformulated in the following manner:
\begin{equation}
3.66 \times 10^{-37} \biggl(\frac{\nu_{max}}{\mathrm{kHz}}\biggr)^4\, 
{\mathcal I}(n_{T}, \, \gamma) \leq 5.61  \times 10^{-6}\, \Delta N_{\nu} \,\biggl(\frac{h_{0}^2 \,\Omega_{\gamma0}}{2.47 \times 10^{-5}}\biggr).
\label{APB2}
\end{equation}
In Eq. (\ref{APB2}) ${\mathcal I}(n_{T}, \, \gamma)$ indicates the following definite integral: 
\begin{equation}
{\mathcal I}(n_{T}, \, \gamma) =  \, \gamma \, \int_{x_{bbn}}^{\infty} \, \frac{x^{m_{T}}}{e^{\gamma x} -1} \, d x,\qquad x = \nu/\nu_{max}, \qquad x_{bbn} = \nu_{bbn}/\nu_{max}.
\label{APB3}
\end{equation}
The integral of Eq. (\ref{APB3}) is convergent in the limit $x \to \infty$ (i.e. $\nu\to \infty$) whereas  limit of integration leads to a potential divergence 
for $n_{T}\to 0$ but this never occurs since $n_{T}$ never goes to zero. In the standard lore $n_{T} \simeq - r_{T}/8 \ll 1$ but the exact scale-invariant case is never reached. When $n_{T} >0$ we can ignore 
the contribution of the lower bound of integration.
 Equations (\ref{APB2})--(\ref{APB3}) imply $\nu_{max} \leq  {\mathcal O}(10^{11}) \,[\Delta N_{\nu}/{\mathcal I}(n_{T}, \gamma)]^{1/4} \, \mathrm{Hz} < \mathrm{THz}$
and $[\Delta N_{\nu}/Q(m_{T}, \gamma)]< {\mathcal O}(10)$. The actual bounds on $\Delta N_{\nu}$ range from $\Delta N_{\nu} \leq 0.2$ to $\Delta N_{\nu} \leq 1$ so that the integrated spectral density must range, at most, between  $10^{-6}$ and $10^{-5}$.

It follows from Eq. (\ref{APB1}) that the slope of $\Omega_{gw}(\nu,\tau_{0})$ bears the mark of the post-inflationary expansion rate. For a single post-inflationary stage expanding with an arbitrary 
rate $\Omega_{gw}(\nu,\tau_{0})$ consists of three spectral regions.
In the high frequency region $\Omega_{gw}( \nu, \tau_{0})$ follows directly 
after inserting Eq. (\ref{APB1}) into Eq. (\ref{TM1}); thus for $\nu \leq \nu_{max}$
the explicit expression of the spectral energy density is
\begin{equation}
\Omega_{gw}(\nu,\tau_{0}) = \Omega_{gw}(\nu_{max},\tau_{0}) (\nu/\nu_{max})^{n_{T}}
\label{APB5}
\end{equation}
where  $n_{T}$ has been already given in Eq. (\ref{APB1}). Recalling the results 
of Eq. (\ref{APB1b}), Eq. (\ref{APB5}) shows that  $\Omega_{gw}(\nu,\tau_{0})$ increases with the comoving 
frequency when the post-inflationary expansion rate is slower than radiation while it decreases if the decelerated expansion rate is faster than radiation, as originally argued in \cite{SEVENTEEN1}. The high frequency normalization appearing in Eq. (\ref{APB5}) is 
\begin{equation}
\Omega_{gw}(\nu_{max}, \tau_{0}) = \frac{\Omega_{R0}}{6} \, r_{T}\, {\mathcal A}_{{\mathcal R}}\, {\mathcal C}(g_{\rho}, g_{s}) \, \biggl(\frac{H_{r}}{H_{1}}\biggr)^{2 \frac{(\delta -1)}{\delta+1}}.
\label{APB6}
\end{equation}
According to Eq. (\ref{APB6}) the high frequency normalization does not simply 
scale as $r_{T}$ but rather as $\Omega_{gw}(\nu_{max}, \tau_{0}) \propto r_{T}^{2/(\delta +1)}$; this is because  $H_{1}$ scales as $\sqrt{r_{T}}$ and implies 
a dependence upon $\delta$ in the final result. The results of Eqs. (\ref{APB5})--(\ref{APB6}) 
hold in the high frequency region of the spectrum, i.e. for $ \nu_{r} < \nu < \nu_{max}$ 
where $\nu_{r} = \sqrt{H_{r}/H_{1}}$ is the transition frequency between the intermediate region and the high frequency domain. When $\nu< \nu_{r}$ the spectral energy density 
is quasi-flat. Finally, in the low frequency region (i.e. $\nu < \nu_{dec} = {\mathcal O}(100) \, \mathrm{aHz}$ we have $ \Omega_{gw} \propto (\nu/\nu_{eq})^{-2}$. 
 We finally remark that if there are different stages of expansion that are either faster  or slower than radiation the spectral energy density is characterized by a sequence of peaks and through that can be accurately predicted and may also be relevant, in specific scenarios, for frequencies below the Hz \cite{SEVENTEEN2}.
\end{appendix}


\newpage

\begin{thebibliography}{99}

\itemsep -1pt

\bibitem{ONEa} S. Weinberg, {\it Cosmology} (Oxford University Press, Oxford 2008).

\bibitem{ONEb}  P.~J.~E.~Peebles, {\it Principles of Physical Cosmology}, (Princeton University Press, Princeton, NJ 1993).

\bibitem{TWOa} P.~J.~E.~Peebles and J.~T.~Yu,  Astrophys.\ J.\  {\bf 162} 815 (1970).

\bibitem{TWOb} S.~Weinberg, Phys. Rev. D {\bf 67}, 123504 (2003).

\bibitem{THREEa} D.~N.~Spergel \textit{et al.} [WMAP collaboration], Astrophys. J. Suppl. {\bf 148}, 175 (2003).

\bibitem{THREEb} A.~Kogut \textit{et al.} [WMAP], Astrophys. J. Suppl. {\bf 148}, 161 (2003).

\bibitem{THREEc} D.~N.~Spergel \textit{et al.} [WMAP], Astrophys. J. Suppl. {\bf 170}, 377 (2007).

\bibitem{THREEd} L.~Page \textit{et al.} [WMAP], Astrophys. J. Suppl. {\bf 170}, 335 (2007).

\bibitem{THREEe} G.~Hinshaw \textit{et al.} [WMAP], Astrophys. J. Suppl. {\bf 208}, 19 (2013)

\bibitem{FOURa}  L.~Parker,  Phys.\ Rev.\ Lett.\  {\bf 21}, 562 (1968).

\bibitem{FOURb} L.~Parker,  Phys.\ Rev.\  {\bf 183}, 1057 (1969).

\bibitem{FOURc} L.~P.~Grishchuk,  Sov.\ Phys.\ JETP {\bf 40}, 409 (1975)   [Zh.\ Eksp.\ Teor.\ Fiz.\  {\bf 67}, 825 (1974)].

\bibitem{FOURd} L.~P.~Grishchuk,  Annals N.\ Y.\ Acad.\ Sci.\  {\bf 302}, 439 (1977).

\bibitem{FOURe}  L. H. Ford and L. Parker, Phys.\ Rev.\ D {\bf 16}, 245 (1977).

\bibitem{FOURf} B. L. Hu and L. Parker, Phys. Lett. A {\bf 63}, 217 (1977).

\bibitem{FOURg}  A. A. Starobinsky, JETP Lett. {\bf 30}, 682 (1979) [Pis'ma Zh. Eksp. Teor. Fiz. {\bf 30}, 719 (1979)].

\bibitem{FOURh} V. A. Rubakov, M. V. Sazhin, and A. V. Veryaskin, Phys. Lett. B {\bf 115}, 189 (1982).

\bibitem{FOURi} L.~F.~Abbott and M.~B.~Wise, Nucl.\ Phys.\ B {\bf 244}, 541 (1984).
   
\bibitem{FOURl} S.~W.~Hawking,  Phys.\ Lett.\  {\bf 150B}, 339 (1985).

\bibitem{FIVEa}  N.~Aghanim {\it et al.} [Planck Collaboration], Astron. Astrophys. {\bf 641}, A6 (2020).

\bibitem{FIVEb}  Y.~Akrami {\it et al.} [Planck Collaboration], Astron. Astrophys. {\bf 641}, A10 (2020).

\bibitem{FIVEc} S.~Aiola \textit{et al.} [Atacama Cosmology Telescope], JCAP {\bf 12}, 047 (2020).

\bibitem{FIVEd} P.~Ade {\it et al.} [BICEP and Keck], Phys. Rev. Lett. {\bf 127}, 151301 (2021).

\bibitem{FIVEe} E.~Calabrese \textit{et al.} [Atacama Cosmology Telescope], JCAP {\bf 11}, 063 (2025).

\bibitem{SIXa}  M.~Giovannini, Phys. Rev. D {\bf 108},  123508 (2023).

\bibitem{SIXb}  L. Dai, M. Kamionkowski, J. Wang Phys. Rev. Lett.  {\bf 113}, 041302 (2014).

\bibitem{SIXc} J.~B.~Munoz and M.~Kamionkowski, Phys. Rev. D {\bf 91}, 043521 (2015). 

\bibitem{SIXd} P. A. R. Ade et al. (BICEP2 Collaboration), Phys. Rev. Lett. {\bf 112}, 241101 (2014).

\bibitem{SEVENa} R. D. Peccei and H. R. Quinn, Phys. Rev. Lett. {\bf 38}, 1440  (1977).

\bibitem{SEVENb}  R. D. Peccei and H. R. Quinn,  Phys. Rev. D {\bf 16}, 1791 (1977).

\bibitem{SEVENc} S. Weinberg, Phys. Rev. Lett. {\bf 40}, 223 (1978).

\bibitem{SEVENd} F. Wilczek, Phys. Rev. Lett. {\bf 40}, 279 (1978).

\bibitem{SEVENe} J. Kim, Phys. Rep. {\bf 150}, 1 (1987).

\bibitem{SEVENf} L.~Hui, J.~P.~Ostriker, S.~Tremaine and E.~Witten, Phys. Rev. D {\bf 95}, 043541 (2017).

\bibitem{SEVENg} D.~J.~E.~Marsh, D.~Grin, R.~Hlozek and P.~G.~Ferreira, Phys. Rev. Lett. {\bf 113}, 011801 (2014).

\bibitem{LIGO0}  B.~Abbott {\it et al.} [LIGO Collaboration], Phys.\ Rev.\ D {\bf 69}, 122004 (2004); Phys.\ Rev.\ Lett.\  {\bf 95}, 221101 (2005).  

\bibitem{LIGO0a} J.~Aasi {\it et al.} [LIGO/Virgo Collaboration], Phys.\ Rev.\ Lett.\  {\bf 113}, 231101 (2014); Phys.\ Rev.\ D {\bf 91},  022003 (2015).

\bibitem{LIGO0b} B.~P.~Abbott {\it et al.} [LIGO/Virgo Collaboration],  Phys.\ Rev.\ Lett.\  {\bf 118}, 121101 (2017)  Erratum: [Phys.\ Rev.\ Lett.\  {\bf 119}, 029901 (2017)]; Phys.\ Rev.\ D {\bf 100}, 061101(R) (2019).

\bibitem{LIGO1} R.~Abbott {\it et al.} [KAGRA, Virgo and LIGO Scientific], Phys. Rev. D {\bf 104}, 022004 (2021).

\bibitem{LIGO1a} A.~G.~Abac {\it et al.} [Virgo, KAGRA and LIGO Scientific], Phys. Rev. D {\bf 114},  042001 (2026)

\bibitem{LIGO1b} A.~G.~Abac \textit{et al.} [LIGO Scientific, VIRGO and KAGRA], [arXiv:2608.23477 [gr-qc]].

\bibitem{LIGO3} M.~Giovannini, Prog. Part. Nucl. Phys. {\bf 112}, 103774 (2020).

\bibitem{NANO1} Z. Arzoumanian {\it et al.}, Astrophys. J. Lett. {\bf 905}, L34 (2020).

\bibitem{NANO2} G.~Agazie \textit{et al.} [NANOGrav], Astrophys. J. Lett. {\bf 951},  L8 (2023).

\bibitem{PPTA1} B. Goncharov {\it et al.} Astrophys. J. Lett. {\bf 917}, L19 (2021).

\bibitem{PPTA2} D.~J.~Reardon, {\it et al.}, Astrophys. J. Lett. {\bf 951},  L6 (2023).

\bibitem{PPTA2a}M.~Giovannini, Eur. Phys. J. C {\bf 84},  67 (2024).

\bibitem{EIGHT1} M.~Giovannini, Phys. Lett. B {\bf 854}, 138769 (2024).

\bibitem{EIGHT2} J. Bardeen, Phys. Rev. D {\bf 22}, 1882 (1980).

\bibitem{EIGHT3} J. Bardeen, P. Steinhardt, and M. Turner, Phys. Rev. {\bf D28}, 679 (1983).

 \bibitem{EIGHT4} J.~A.~Frieman and M.~S.~Turner, Phys.\ Rev.\ D {\bf 30}, 265 (1984).
 
 \bibitem{NINE1} J.~Valiviita, M.~Savelainen, M.~Talvitie, H.~Kurki-Suonio and S.~Rusak,  Astrophys.\ J.\  {\bf 753}, 151 (2012)

\bibitem{NINE2} R.~Keskitalo, H.~Kurki-Suonio, V.~Muhonen and J.~Valiviita, JCAP {\bf 0709}, 008 (2007)

\bibitem{NINE3} H.~Kurki-Suonio, V.~Muhonen and J.~Valiviita, Phys.\ Rev.\ D {\bf 71}, 063005 (2005).

\bibitem{NINE4} J.~Valiviita and V.~Muhonen,  Phys.\ Rev.\ Lett.\  {\bf 91}, 131302 (2003).

\bibitem{NINE5} M.~Giovannini, Class. Quant. Grav. {\bf 23}, 4991 (2006); 
Phys. Rev. D {\bf 90}, 123513 (2014)

\bibitem{NINE6} A.~R.~Liddle and S.~M.~Leach, Phys. Rev. D {\bf 68}, 103503 (2003)

\bibitem{TEN1}  M. Abramowitz and I. A. Stegun, {\it Handbook of Mathematical Functions} (Dover, New York, 1972).

\bibitem{TEN2}  A. Erdelyi, W. Magnus, F. Obehettinger, and F. Tricomi,  {\it Higher Trascendental Functions} (Mc Graw-Hill, New York, 1953).  

\bibitem{ELEVENa} M. S. Turner, Phys. Rev. D {\bf 28}, 1243 (1983).

\bibitem{ELEVENb}  L. H. Ford, Phys. Rev. D {\bf 35}, 2955 (1987).

\bibitem{ELEVENc} J. D. Barrow, Phys. Rev. D {\bf 48}, 1585 (1993).

\bibitem{TWELVEa} A. Liddle, P. Parsons, and J.D.  Barrow,   Phys. Rev. D {\bf 50}, 7222 (1994).

\bibitem{TWELVEb}  A. Liddle and D. Lyth, D. H. (1992) Phys. Lett. B {\bf 291}, 391 (1992).

\bibitem{TWELVEc} L.~Boubekeur and D.~H.~Lyth, JCAP {\bf 07}, 010 (2005)

\bibitem{TWELVEd} N.~K.~Stein and W.~H.~Kinney, JCAP {\bf 03}, 027 (2023).

\bibitem{TWELVEe} W. J. Wolf, Phys. Rev. D {\bf 110}, 043521 (2024).

\bibitem{TWELVEf} H. Motohashi and A. A. Starobinsky, JCAP {\bf 11}, 025 (2019).

\bibitem{TWELVEg} A. Mohammadi, T. Golanbari, S. Nasri and K. Saaidi, Phys. Rev. D {\bf 101}, 123537 (2020).

\bibitem{TWELVEh} S.~Weinberg, Phys. Rev. D {\bf 67}, 123504 (2003).

\bibitem{TWELVEh2} S.~Weinberg, Phys. Rev. D {\bf 70}, 043541 (2004)

\bibitem{TWELVEi} C. P. Ma and E. Bertschinger, Astrophys. J. {\bf 455}, 7 (1995).

\bibitem{TWELVEl} W. Press and E. Vishniac, Astrophys. J. {\bf 239}, 1 (1980); {\bf 236}, 323 (1980).

\bibitem{THIRTEENa1} A. D. Sakharov,  Sov. Phys. JETP {\bf 22}, 241 (1966) [Zh. Eksp. Teor. Fiz. {\bf 49}, 345 (1965)].

\bibitem{THIRTEENa2}  V.~N.~Lukash,  Sov.\ Phys.\ JETP {\bf 52}, 807 (1980) [Zh. Eksp. Teor. Fiz. {\bf 79}, 1601 (1980)].

\bibitem{THIRTEENb} L.~M.~Krauss and F.~Wilczek, Phys. Rev. D {\bf 89},  047501 (2014).

\bibitem{THIRTEENc} M.~Giovannini, JCAP {\bf 05}, 056 (2023).

\bibitem{FOURTEENa}  S. Weinberg, Phys. Rev. D {\bf 69}, 023503 (2004).

\bibitem{FOURTEENb} D.~A.~Dicus and W.~W.~Repko,  Phys.\ Rev.\  D {\bf 72}, 088302 (2005).
 
\bibitem{FOURTEENc} H.~X.~Miao and Y.~Zhang, Phys.\ Rev.\ D {\bf 75}, 104009 (2007).

\bibitem{FOURTEENd}  B.~A.~Stefanek and W.~W.~Repko, Phys.\ Rev.\ D {\bf 88},  083536 (2013).

\bibitem{FOURTEENe}  K.~W.~Ng, Phys.\ Rev.\ D {\bf 86}, 103510 (2012).

\bibitem{FIFTEENa}  V.F. Schwartzman, Pis'ma Zh. Eksp. Teor. Fiz. {\bf 9}, 315 (1969) [JETP Lett. {\bf 9}, 184 (1969)].
  
\bibitem{FIFTEENb} C.J. Copi, D.N. Schramm, and M.S. Turner, Phys. Rev. D {\bf 55}, 3389 (1997).

\bibitem{FIFTEENc}  S. Burles, K.M. Nollett, J.W. Truran, and M.S. Turner, Phys. Rev. Lett. {\bf 82}, 4176 (1999).

\bibitem{FIFTEENd} R. Cyburt, B. Fields, and K. Olive, Astropart. Phys. {\bf 17}, 87 (2002).

\bibitem{FIFTEENe} M.~Giovannini, H.~Kurki-Suonio and E.~Sihvola, Phys.\ Rev.\  D {\bf 66}, 043504 (2002).

\bibitem{FIFTEENf} R. Cyburt, B.~D.~Fields, K.~A.~Olive and T.~H.~Yeh,  Rev.\ Mod.\ Phys.\  {\bf 88}, 015004 (2016).

\bibitem{SEVENTEEN1} M.~Giovannini, Phys. Rev. D {\bf 58}, 083504 (1998); 
Phys. Rev. D {\bf 60}, 123511 (1999); Class. Quant. Grav. {\bf 16}, 2905 (1999).

\bibitem{SEVENTEEN2} M.~Giovannini, Phys. Rev. D {\bf 105}, 103524 (2022); Eur. Phys. J. C {\bf 82}, 828 (2022).

\end{thebibliography}
\end{document}